\documentclass[11pt,a4paper]{article}

\usepackage{jheppub}

\usepackage{amssymb}
\usepackage{amsmath}
\usepackage{siunitx}
\numberwithin{equation}{section}
\usepackage{autobreak}
\usepackage{physics}
\usepackage{mathtools}
\usepackage{slashed}
\usepackage{empheq}
\usepackage{lmodern}
\usepackage{cleveref}
\usepackage{mathrsfs}
\usepackage{cancel}
\usepackage{enumitem}
\usepackage{upgreek}
\usepackage[version=4]{mhchem}
\normalsize
\usepackage{booktabs}
\usepackage{isotope}
\usepackage{multirow}
\usepackage{footnote}
\usepackage{bbold}
\usepackage{scalerel}

\usepackage[b]{esvect}
\usepackage{graphicx}
\usepackage{dcolumn}
\usepackage{bm}
\usepackage{textcomp}
\usepackage[latin1]{inputenc}
\usepackage{wasysym}
\usepackage{cases}

\AtBeginDocument{\RenewCommandCopy\qty\SI}

\usepackage[italic]{hepnicenames}
\makeatletter\def\@shiftlen@anti@gen@bar{0mu}\makeatother
\usepackage[svgnames]{xcolor}
\usepackage[compat=1.1.0]{tikz-feynhand}
\definecolor{electron}{HTML}{1f77b4}
\definecolor{crimson}{HTML}{dd143c}
\definecolor{newgreen}{HTML}{009900}

\newcommand{\del}{\partial}
\renewcommand{\d}{\mathrm{d}}
\newcommand{\eps}{\varepsilon}

\newcommand{\had}{\mathcal{H}}

\renewcommand{\to}{\rightarrow}

\DeclareSIUnit\barn{b}

\makeatletter
\renewcommand*\env@matrix[1][\arraystretch]{%
	\edef\arraystretch{#1}%
	\hskip -\arraycolsep
	\let\@ifnextchar\new@ifnextchar
	\array{*\c@MaxMatrixCols c}}
\makeatother

\makeatletter
\newcommand{\biggg}{\bBigg@\thr@@}
\newcommand{\Biggg}{\bBigg@{3.5}}

\renewcommand*{\thesection}{\arabic{section}}

\renewcommand*{\p@subsection}{}

\renewcommand*{\p@subsubsection}{}
\makeatother

\newsavebox\foobox 
\newlength{\foodim}
\newcommand{\slantbox}[2][0]{\mbox{%
		\sbox{\foobox}{#2}%
		\foodim=#1\wd\foobox
		\hskip \wd\foobox
		\hskip -0.5\foodim
		\pdfsave
		\pdfsetmatrix{1 0 #1 1}%
		\llap{\usebox{\foobox}}%
		\pdfrestore
		\hskip 0.5\foodim
}}
\def\lgr{\slantbox[-.45]{$\mathscr{L}$}}

\makeatletter
\newcommand{\dalembertian}{\mathop{\mathpalette\dalembertian@\relax}}
\newcommand{\dalembertian@}[2]{%
  \begingroup
  \sbox\z@{$\m@th#1\square$}%
  \dimen0=\fontdimen8
    \ifx#1\displaystyle\textfont\else
    \ifx#1\textstyle\textfont\else
    \ifx#1\scriptstyle\scriptfont\else
    \scriptscriptfont\fi\fi\fi3
  \makebox[\wd\z@]{%
    \hbox to \ht\z@{%
      \vrule width \dimen0
      \kern-\dimen0
      \vbox to \ht\z@{
        \hrule height \dimen0 width \ht\z@
        \vss
        \hrule height 2\dimen0
      }%
      \kern-2.5\dimen0
      \vrule width 2.5\dimen0
    }%
  }%
  \endgroup
}
\makeatother

\newcommand*\xbar[1]{%
	\hbox{%
		\vbox{%
			\hrule height 0.6pt 
			\kern0.3ex
			\hbox{%
				\kern-0.05em
				\ensuremath{#1}%
				\kern-0.05em
			}%
		}%
	}%
}

\newcommand{\ad}{A_\text{\tiny{D}}}
\newcommand{\asm}{A_\text{\tiny{SM}}}
\newcommand{\fd}{F_\text{\tiny{D}}}
\newcommand{\fsm}{F_\text{\tiny{SM}}}
\newcommand{\ap}{{A^\prime}}
\newcommand{\fp}{F^\prime}

\newcommand{\bh}{\text{\tiny{BH}}}
\newcommand{\vcs}{\text{\tiny{VCS}}}

\definecolor{electron}{HTML}{1f77b4}
\definecolor{kern}{HTML}{2ca02c}
\definecolor{orange}{HTML}{FF7F00}
\definecolor{kern}{HTML}{2ca02c}
\definecolor{vpurple}{HTML}{9467bd}
\definecolor{pseudo}{HTML}{9400d3}

\makeatletter
\g@addto@macro\bfseries{\boldmath}
\makeatother

\let\vec\vv

\let\Gamma\varGamma

\allowdisplaybreaks

\graphicspath{{./figs/}}
\usepackage{subfiles}

\preprint{BONN-TH-2026-09, MPP-2026-47}

\title{Theory Calculations for LDMX and LOHENGRIN beyond Coherent Bethe-Heitler Scattering}

\author[a]{Martin Sch\"urmann,}
\author[a]{Herbi K. Dreiner,}
\author[b]{Rhorry Gauld}

\affiliation[a]{Bethe Center for Theoretical Physics \& Physikalisches Institut der Universit\"at Bonn, \\ Nu{\ss}allee 12, 53115 Bonn, Germany}
\affiliation[b]{Max Planck Institute for Physics, Boltzmannstra{\ss}e 8, 85748 Garching, Germany}

\emailAdd{mar-schuermann@uni-bonn.de}
\emailAdd{dreiner@uni-bonn.de}
\emailAdd{rgauld@mpp.mpg.de}

\abstract{
The Light Dark Matter eXperiment (LDMX), DarkSHINE, and \textsc{Lohengrin} are proposed new experiments. 
They aim to search for missing momentum signals sourced by the direct production of dark photons with masses in the MeV-GeV range in bremsstrahlung processes, in which an electron beam of a few GeV scatters off a fixed target. 
So far, the signal characteristics, \textit{i.e.} the behavior of the recoiling electron, have mostly been studied in coherent Bethe-Heitler electron-nucleus scattering with a dark photon that couples only to the Standard Model charged leptons. 
In this work, we present the calculations of the differential cross sections of all contributing real emission processes up to third order in the electromagnetic fine structure constant and fourth order in the kinetic mixing parameter associated with the dark photon. 
We consider a dark photon coupling to both the beam electron and the hadronic target and we take into account the scattering off both the target nucleus and its nuclear constituents.
Besides real emission processes, we also discuss virtual dark photon contributions and their relevance for the signal prediction.
After discussing the different phase space regions and constraints
emerging from the experimental setups, we show numerical results of the
cross sections and differential distributions, including the signal and 
dominant background.
Within our framework, we find that the \textsc{Lohengrin} experiment will require an extension of its HCAL to effectively veto background processes originating from  diffractive scattering.
Apart from that, the contributions beyond coherent Bethe-Heitler scattering, in the presence of realistic experimental selections, have only a limited effect on the predicted signal and background in the relevant dark photon mass range.
}

\begin{document}
\keywords{Light Dark Matter, Fixed Target}
\maketitle
\flushbottom


\section{Introduction}
\label{sec:intro}
The particle nature of dark matter (DM) is an unsolved mystery of 
physics \cite{Bertone:2016nfn}. Many attempts have been made to detect 
DM particles in laboratory experiments, without success so far 
\cite{battaglieri2017cosmic}. The Light 
Dark Matter eXperiment (LDMX) \cite{LDMX:2018cma, LDMX:2019gvz, 
Akesson:2022vza, LDMX:2023zbn, LDMX:2025pkp}, as well as DarkSHINE 
\cite{DarkSHINE:2022mak} and \textsc{Lohengrin} \cite{Bechtle:2024atq},
are new recently proposed fixed target experiments that aim to 
explore the parameter space of light DM particles feebly coupling to 
the electron. The search strategy of these types of experiments (which, following the chronology of the proposals, we refer to as LDMX-like) relies on 
missing momentum in the final state: an electron beam of a few 
GeV scatters off a heavy nucleus in a thin target to produce new dark 
sector particles in bremsstrahlung-like processes. For sufficiently 
thin targets, the final state electron will be observable. It typically
carries only a small fraction of the beam energy and a sizable 
transverse momentum w.r.t. the beam direction, if the produced dark 
sector particle is sufficiently heavy. In contrast to reappearance 
experiments, the dark sector particles are assumed to not decay into 
visible Standard Model (SM) particles but into other DM particles, such
that their potential decay chain cannot be observed by
any detector. Then, the main observable is the momentum of the 
recoiling electron.

The dark photon extension of the SM provides a great physics motivation for the existence of feebly coupling particles \cite{Okun:1982xi, HOLDOM1986196, 
Boehm:2003hm, Pospelov:2008zw}. 
It is a potentially massive vector 
particle stemming from a new (spontaneously broken) $\text{U}(1)_ 
\text{\tiny{D}}$ gauge symmetry in a dark sector. It can couple to SM 
fermions by kinetically mixing with the photon or hypercharge boson 
\cite{Filippi:2020kii, Fabbrichesi:2020wbt, Dreiner:2013mua}; the 
resulting couplings are suppressed by the kinetic mixing parameter, 
$\eps \ll 1$. Minimal dark sector extensions including a dark photon $A'$
and dark matter scalars/fermions $\chi$ charged under $\text{U}(1)_\text{\tiny{D}}$ with light masses in the MeV--GeV range can explain the observed DM relic abundance 
through the WIMPless miracle \cite{Feng:2008ya, Krnjaic:2025noj, 
Battaglieri:2017aum, Feng:2017drg}.
It is assumed that the dark sector was in thermal equilibirum with the SM bath, such that the DM relic abundance is produced through freeze-out.
If the mass hierarchy in the dark sector is $m_{A'} > 2m_\chi$, the DM particles annihilate directly into SM particles.
Such a DM freeze-out predicts a well-defined
target in the dark sector parameter space that is both consistent with the 
observed DM relic abundance but also falsifiable through 
laboratory experiments~\cite{Battaglieri:2017aum}. LDMX-like 
experiments can reach the sensitivities required to explore this part 
of the parameter space and can thus detect or rule out these classes of DM
models \cite{LDMX:2018cma, DarkSHINE:2022mak, Bechtle:2024atq}.
For the mass hierarchy $m_{\ap} > 2 m_\chi$, the dark photon will dominantly decay into invisible dark matter particles, given that the dark gauge coupling $g_\text{\tiny{D}}$ is much bigger than the coupling strength of the dark photon to SM states induced by kinetic mixing.
Several constrains have already been set on such dark sector models with an invisibly decaying dark photon through missing momentum/energy/mass experiments such as BaBar \cite{BaBar:2017tiz}, E137
\cite{Batell:2014mga}, LSND \cite{deNiverville:2011it, Batell:2009di}, MiniBooNE
\cite{MiniBooNEDM:2018cxm}, as well as NA64
\cite{Banerjee:2019pds}, see also Refs.~\cite{Alexander:2016aln,  Battaglieri:2017aum, Izaguirre:2014bca, Izaguirre:2015yja, Izaguirre:2017bqb}.
Additional constrains can be derived from astrophysical mechanisms and cosmological precision measurements, see \textit{e.g.} 
Refs.~\cite{Dent:2012mx, Dreiner:2013tja, Dreiner:2013mua, An:2013yfc, 
Redondo:2013lna} and Ref.~\cite{Sabti:2019mhn}, respectively.

The signal and background characteristics of LDMX-like experiments are 
determined by the four momenta of the final state particles, especially by the properties of the electron after the
emission of a dark photon (signal) and QED photon (background). So far, these signatures have been studied in 
coherent Bethe-Heitler scattering \cite{Bethe:1934za}, where the 
incident beam electron interacts with the charge distribution of the 
target nucleus via one-photon-exchange (OPE) to radiate a dark photon. 
In perturbation theory, this process exists at order $\mathcal{O} 
(\alpha^3\eps^2)$ in the squared amplitude, where $\alpha$ is the electromagnetic fine 
structure constant. There are additional contributions at this order
which have not yet been studied explicitly, and we will generalize 
the scattering formalism to account for them and achieve a more 
consistent treatment at $\mathcal{O}(\alpha^3\eps^2)$. Some calculations
also make use of the Weizs\"acker-Williams (WW) method 
\cite{vonWeizsacker:1934nji, Williams:1934ad, LDMX:2018cma, 
Blinov:2020epi, Eichlersmith:2022bit}, a phase space approximation which
treats the full $2\to3$ process as a $2\to2$ Compton process with 
on-shell photons emitted from the nucleus 
\cite{Tsai:1986tx}. The validity of the WW method has been questioned 
for this particular application in Refs.~\cite{Liu:2016mqv, Liu:2017htz,
Gninenko:2017yus}, since it produces relative deviations of $\mathcal 
{O}(1)$ in the predicted cross sections, especially -- as we found -- in
phase space regions relevant to the detection strategy of the novel 
experiments, \textit{i.e.} where the final state electron carries only a
small fraction of the beam energy.

\begin{figure}
    \centering
    \includegraphics[width=\textwidth]{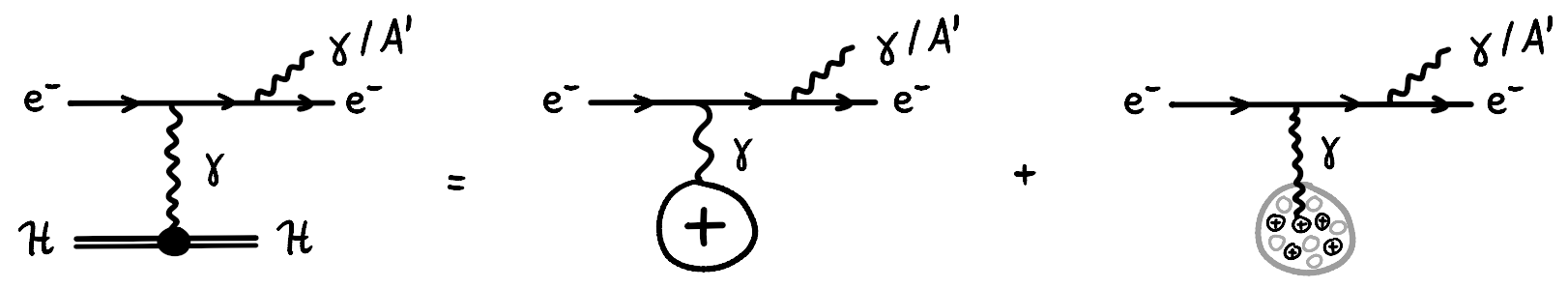}
    \caption{Generalization of the hadronic current in Bethe-Heitler scattering to account for both the interaction with the nuclear charge distribution and the nucleon charges and magnetic moments. $e^-, \had, \gamma$ and $A'$ represent the electron, hadronic system, photon and dark photon, respectively.}
    \label{fig:generalization_cartoon_1}
\end{figure}
\begin{figure}
    \centering
    \includegraphics[width=\textwidth]{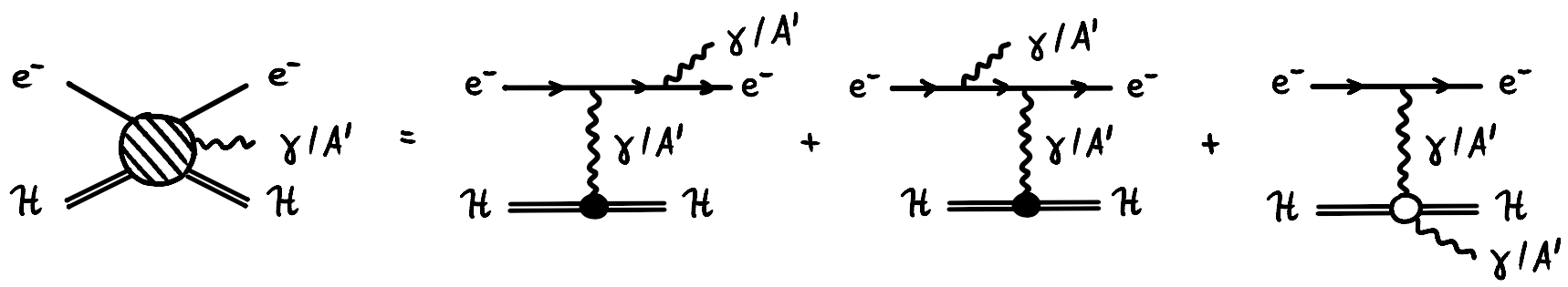}
    \caption{Generalization of dark photon reactions (to include hadronic systems) in the process we consider and addition of virtual Compton scattering.}
    \label{fig:generalization_cartoon_2}
\end{figure}

In this paper, we present the calculation framework and cross section results after accounting for additional contributions up to $\mathcal{O}(\alpha^3 \eps^2)$.
This includes the following aspects:
\begin{enumerate}[label=(\alph*)]
    \item In $2\to 3$ scattering:
    \begin{enumerate}[label={\arabic*.}]
    \item We consider not only coherent scattering off the nuclear charge distribution, but also incoherent contributions from diffractive scattering on individual nucleons, which are characterized by their electric charges and magnetic moments. This is schematically presented in Fig. \ref{fig:generalization_cartoon_1}.
    \item Fundamentally, the dark photon also couples to quarks, and 
    consequently to composite hadronic systems. It can then 
     appear as the virtual exchange particle between 
    leptonic and hadronic currents in the scattering (see 
    Fig.~\ref{fig:generalization_cartoon_2}).
    \item Besides Bethe-Heitler scattering there are further contributions from virtual Compton scattering, where 
    the (dark) photon is radiated from the hadronic current (see 
    Fig.~\ref{fig:generalization_cartoon_2}).
    \end{enumerate}
    \item In $2\to 2$ scattering we include
    contributions where no real dark photon is produced, but where it
    contributes as virtual exchange particle to cross sections up to
    $\mathcal{O}(\alpha^3\eps^2)$.
\end{enumerate}
These generalizations allow to study various 
possible BSM search strategies, that will be the target of LDMX-like experiments, in better detail.
Parts of the adopted formalism have been borrowed from the 
framework used to study neutrino trident scattering, see for example 
Ref.~\cite{Ballett:2018uuc}.

The paper is organized as follows. In Sec.~\ref{sec:physics_input} we
review the physical models entering the calculations, \textit{i.e.} the 
dark photon extension of the SM, as well as the electromagnetic 
interactions with the hadronic target. Sec.~\ref{sec:phasespace} 
provides a discussion of the phase space regions 
relevant for fixed target scenarios similar to LDMX. 
A discussion of BSM physics effects for scattering processes that include the elastic $2\to 2$ scattering line are the focus of Sec.~\ref{sec:2to2}, while detailed studies of $2\to 3$ real emission processes are provided in  Sec.~\ref{sec:2to3}. Sec.~\ref{sec:conclusions} provides some concluding
remarks and an outlook.
App. \ref{app:phasespace_integral} gives details about the phase space implementation.
App. \ref{app:coherent_compton_tensor}, \ref{app:dark_goldstone} and  \ref{app:renormalization_counterterms} provide some technical details about the dark photon extension, namely the dark photon Compton tensor in coherent scattering, the goldstone boson accompanying the dark photon as well as the renormalization procedure in the leptophilic limit, from which the renormalization counterterms are obtained.
Finally, App. \ref{app:supp_figures} provides supplementary figures.

\section{Calculational Framework}
\label{sec:physics_input}

As introduced in Sec.~\ref{sec:intro}, the scattering process of
interest here is
\begin{equation}
	e^-(p) + \had(P) \longrightarrow e^-(p^\prime) + X ,
	\label{eq:scatter}
\end{equation}
through which an electron scatters upon a nuclear target $\had$, and the
momentum of the outgoing electron is measured. This scattering process 
is typically dominated by the coherent scattering process through which 
a virtual photon is exchanged between the projectile and the
target nucleus, potentially in association with other radiation such as 
a photon or light DM particles.
The sensitivity of LDMX-like experiments benefits twofold from a precise theoretical modelling of both signal and background processes. Precise modelling allows for the selection of a more powerful signal region in the first place and allows to reduce the theory uncertainties that enter in the calculation of limits that the experiments will be able to set.
In this section we introduce the theoretical framework used for the description 
of the scattering process defined in Eq.~\eqref{eq:scatter} in both the SM 
and a dark sector extension of the SM.
We begin by introducing the considered dark sector extension, before discussing the scattering formalism and its eventual implementation 
in a computational framework for numerical evaluation.

\subsection{Dark Sector Extension of the Standard Model}
\label{sec:dark_photon_basics}

The dark photon is a spin-1 vector boson arising in a minimal dark sector
extension of the Standard Model. The model contains a $U(1)_\text
{\tiny{D}}$ gauge field $\ad^\mu$ with associated field strength tensor
$F^{\mu\nu}_{\text{\tiny{D}}} = \del^\mu\ad^\nu -\del^\nu\ad^\mu$, as 
well as DM scalars or fermions, which constitute a current $J^\mu_\text 
{\tiny{D}}$.

In general, one introduces the interactions of $A^\mu_\text{\tiny{D}}$ to
the SM via the kinetic mixing of the $U(1)_\text{\tiny{D}}$ field 
strength tensor with that of the $U(1)_\text{\tiny{Y}}$ hypercharge gauge
boson, $B^{\mu\nu} ~=~\partial^\mu B^\nu-\partial^\nu B^\mu$ 
\cite{Fabbrichesi:2020wbt}. The relevant terms in the Lagrangian may then
be written as,
\begin{equation}
		\lgr \supset -\frac{1}{4} {F_\text{\tiny{D}}}_{\mu\nu}F_\text{\tiny{D}}^{\mu\nu} + \frac{1}{2} m_{\ad}^2 {\ad}_\mu \ad^\mu - \frac{\sin\varepsilon_Y}{2} F_\text{\tiny{D}}^{\mu\nu} B_{\mu\nu} - g_\text{\tiny{D}} {\ad}_\mu J_\text{\tiny{D}}^\mu\, ,
\end{equation}
where $\eps_Y$ is the kinetic mixing parameter.
We assume that the $U(1)_\text{\tiny{D}}$ gauge boson has acquired 
a mass $m_{\ad}$, for instance through an abelian Higgs mechanism in the 
dark sector. After rotating away the kinetic mixing and performing 
electroweak (EW) symmetry breaking, one finds a $3\times 3$ mass matrix 
for linear combinations of the $A_\text{\tiny{D}}, B$ and $W^3$ bosons, 
which can be diagonalized while keeping only terms of order $\eps_Y^2$, 
\textit{cf.} Ref.~\cite{Dreiner:2013mua}. The interaction terms of the 
resulting mass eigenstates $A^\prime_\mu$ and the modified $Z^0$-boson 
$\tilde{Z}_\mu$ are,
\begin{equation}
		\lgr_\text{int} = -A_\mu^\prime \left( \sum_{f} g^{A^\prime}_f\, \xbar{f}\gamma^\mu f + g^{A^\prime}_\text{\tiny{D}}\, J_\text{\tiny{D}}^\mu \right) - \tilde{Z}_\mu \left( \sum_{f} g^{\tilde{Z}}_f\, \xbar{f}\gamma^\mu f + g^{\tilde{Z}}_\text{\tiny{D}}\, J_\text{\tiny{D}}^\mu \right)\, ,
\end{equation}
where the sums run over all chiral SM fermions $f$, and $A'_\mu$ 
denotes the dark photon after EW symmetry breaking. The $\tilde{Z}$ 
coupling constants have the form $g^{\tilde{Z}}_f=g^Z_f+\mathcal{O}(\eps 
_Y^2)$, where $g^Z_f$ are the usual SM couplings. Up to $\mathcal{O}( 
\eps_Y^2)$, the mass eigenvalue $m_{A^\prime}$ is related to the model 
parameter $m_{A_\text{D}}$ through,
\begin{equation}
    m_{A^\prime}^2 \approx m_{A_\text{D}}^2 \left( 1 + \eps_Y^2 \frac{m_Z^2 \cos^2\theta_W - m_{A_\text{D}}^2}{m_Z^2 - m_{A_\text{D}}^2} \right),
\end{equation}
where $\theta_W$ is the EW mixing angle and $m_Z^2$ satisfies the usual
relation, $m_Z^2 = \frac{v^2}{4}\left(g^2+g^{\prime\:2}\right)$ with the 
Higgs vev $v$ and $SU(2)_L$, $U(1)_Y$ coupling constants $g$, $g'$ respectively.
Due to the mixing of $\ad^\mu$ with the $SU(2)_L$ gauge
boson ${W^3}^\mu$, the dark photon couples differently to left and right 
chiral fermions: for example, the charged lepton couplings are
\begin{align}
    g^{\ap}_{\ell_L} &=  \frac{e \varepsilon_Y}{2\cos\theta_W}\frac{2m_Z^2\cos^2\theta_W- m_{\ad}^2}{m_Z^2-m_{\ad}^2} \, ,\\
    g^{\ap}_{\ell_R} &= \frac{e \varepsilon_Y}{\cos\theta_W}\left( 1- \frac{m_Z^2\sin^2\theta_W}{m_Z^2-m_{\ad}^2} \right).
\end{align}

LDMX-like experiments will operate at electron beam energies of a few $\SI{}{\giga 
\electronvolt}$, which is far below the EW scale. The experimentally 
accessible dark photon masses $m^2_{\ap} \simeq m^2_{\ad}$ are thus much 
smaller than $m_Z^2 \propto v^2$. Hence, we will apply the approximation $m_{\ad}
^2/m_Z^2 \ll 1$ to simplify analytical and numerical calculations. In 
this approximation, the charged fermion couplings reduce to,
\begin{equation}
    g^{A'}_{f} \simeq - Q_f e \varepsilon_Y \cos\theta_W \equiv - Q_f e \eps \,.
\end{equation}
The dark photon interactions are the same for left and right chrial 
fields in this approximation and the interaction Lagrangian reduces to a 
generic vector interaction. The same interaction terms are obtained when 
using the minimal (or `vanilla') kinetic mixing, where the dark sector 
field strength tensor mixes directly with the SM/QED photon field 
strength tensor, \textit{i.e.}, 
\begin{equation}
    \lgr_{\text{\tiny{SM}} \otimes \text{\tiny{D}}} = -\frac{\eps}{2} {\fsm}_{\mu\nu} \fd^{\mu\nu} \, .
\end{equation}
In presence of this mixing operator, the kinetic terms get diagonalized at first order in $\mathcal{\eps}$ by the field redefinition,
\begin{equation}
    A^\mu_\text{\tiny{SM}} \approx A^\mu - \eps \ap^\mu ,
    \label{eq:vanilla_field_rotation}
\end{equation}
where $A^\mu$ is the massless QED photon and $A^{\prime\,\mu}$ is the
physical dark photon state, which couples to electromagnetic currents, as
the QED photon, but with strength rescaled by $\varepsilon$.

For completeness, let us also state the dark photon propagator and 
polarization sum. Regardless of the imposed kinetic mixing operator, 
in unitary gauge the propagator of the dark photon mass eigenstate 
$\ap$ has the form,
\begin{align}
	\label{eq:DP_propagator}
	i \varPi_{\mu\nu}^\ap&= \frac{i}{q^2 - m_\ap^2} \left( -g_{\mu\nu} + \frac{q_\mu q_\nu}{m_\ap^2} \right)\,.
\end{align}
The polarization sum for a massive vector boson is,
\begin{align}
	\label{eq:DP_polsum}
	\sum_\text{pol}\, \varepsilon^{\ast}_\mu(k) \: \varepsilon_\nu(k) = -g_{\mu\nu} + \frac{k_\mu k_\nu}{m_\ap^2}\,.
\end{align}
To avoid confusion among the two vector bosons in Feynman diagrams, we 
always display dark photon lines in red, while all QED/SM 
parts are in greyscale.

The above SM extension does not yet describe the dark photon couplings to
composite hadronic systems, but in our case these can be obtained in a 
straightforward way. Since we are experimentally constrained to
searches for dark photons in the MeV to GeV energy range, \textit{i.e.} 
$m_{A^\prime}^2\simeq m_{A_\text{D}}^2 \ll m_Z^2$, there is a natural 
suppression of the EW mixing effects. Then we effectively have two spin-1
bosons (one massless, one massive) that couple to the same 
electromagnetic vector-like quark currents, only with relative rescaled
strength. The low-energy interactions of a dark photon at this mass-scale
will thus be identical to those of the QED photon, with the appropriate
$\varepsilon$ rescaling; only a potential propagator is modified.

Formally, the electromagnetic behavior of a composite hadronic system may
be described by a low-energy effective Lagrangian. There, one can then 
impose the field redefinition Eq.~\eqref{eq:vanilla_field_rotation} to formally derive the 
low energy couplings of the composite systems to the dark photon. The same 
procedure was also applied to the electromagnetic pion-nucleon Lagrangian in 
Ref.~\cite{Curtin:2023bcf}. In
the following section, we state the resulting amplitudes 
relevant for LDMX-like experiments.

\subsection{Scattering Regimes, Hadronic Amplitudes and Form Factors}
\label{sec:amplitudes_regimes}

In general, the differential scattering rate for the considered process $e^- + \had \longrightarrow e^- + X$ takes the form
\begin{equation} \label{eq:dsigma}
    \d\sigma_{e\had\to eX} = \frac{\overline{\sum}{|\mathcal{M}_{e\had\to eX}|^2}}{F} \,{\d}\varPhi_{eX} \,,
\end{equation}
with the scattering amplitude $\mathcal{M}$ (being squared, and summed/averaged), the flux factor $F$, and final-state phase
space measure ${\d}\varPhi$ taking their usual meaning.
The presence of a dark photon can modify this scattering rate as compared to the SM
expectation in several ways. For example: the exchange of a virtual dark 
photon can modify the scattering rate
of the interaction that involves only external SM particles; while the 
emission of a dark photon leads to a clear signature of missing momentum.
In principle, depending on the precision of the experiment, both of these 
signal types can be observed given a reliable theoretical description of the 
scattering process.

In either case, given a particular final state $X$, the main theoretical challenge is to provide an accurate approximation of the full (squared) scattering amplitude.
This is achieved in several steps. 
Firstly, we always work perturbatively in the electromagnetic coupling of the (dark) photon with matter such that the scattering amplitude is approximated as
\begin{equation}
\mathcal{M}_{e\had\to eX} \approx \mathcal{M}_{e\had\to eX}^{\rm tree} + \mathcal{M}^{\rm 1-loop}_{e\had\to  eX} + ... \,.
\end{equation}
Secondly, we organise/approximate the scattering process in which a (dark) photon 
is exchanged with a hadronic target into the following distinct regimes: coherent,
diffractive, and deep inelastic scattering. In these regimes the scattering process
is described through the exchange of a (dark) photon with atomic nuclei, bound 
nucleons, or quarks, respectively. In each case, the interaction of the photon with
the target can be parameterized in terms of hadronic currents, where the 
latter provide an effective description of the target 
which depends on the momentum exchange $Q^2$. These currents may be described by 
electromagnetic form factors or in terms of parton distribution functions at large 
$Q^2$ values. 
Qualitatively, the importance of the different scattering regimes can be summarised as:

\begin{itemize}
    \item At small momentum exchange, $Q^2 \lesssim {\rm GeV}^2$, the incoming electron resolves the photon field of the target nucleus (\textit{i.e.} the electric charge distribution of the whole nucleus) and the coherent scattering process between the electron and the nucleus provides the dominant contribution to the scattering process.
    \item At moderate momentum exchange, $Q^2 \sim {\rm GeV}^2$, the photon field of bound nucleons within the nucleus are resolved and the diffractive scattering process is numerically important. In this case, the nucleon may be ejected from the nucleus depending on whether $Q^2$ is relatively large compared to the binding energy of the nucleus.
    \item At large momentum exchange, $Q^2 \gtrsim {\rm GeV}^2$, the deep inelastic scattering process becomes dominant. 
\end{itemize}
At the electron beam energies in question (a few GeV), we assume the deep 
inelastic contribution to be sufficiently suppressed, so we will only consider the 
coherent and diffractive contributions to determine the signal yield and the 
leading QED photon background.

Under these assumptions, the scattering rate can be approximated as the sum of coherent and diffractive contributions~\cite{Ballett:2018uuc}:
\begin{equation}
\d \sigma = \d\sigma_\text{coh} + \d\sigma_\text{dif}.
\end{equation}

In this work, we will consider only contributions to the scattering amplitude that result from the exchange of a
single virtual vector boson $V=\{\gamma, A'\}$ between the incoming 
electron and the hadronic system. Consequently, the amplitudes in 
the scattering regime $\mathcal{R}$ = \{coh, dif\} can always be 
factorized into terms containing a leptonic 
current $L^\mu$ and a hadronic current $H^\nu_\text{\tiny{R}}$ (which can
be constructed independently), connected through a vector boson 
propagator $i\varPi_{\mu\nu}^V$, \textit{i.e.},
\begin{equation}
    i \mathcal{M}_\mathcal{R}(e^- \mathcal{H} \rightarrow e^- X) = \sum L^\mu i\varPi_{\mu\nu}^V H^\nu_\mathcal{R}\,.
\end{equation}
For our purposes, we can further categorise the scattering amplitudes to the case of either $2\to 2$ or $2\to 3$ processes.

%
\paragraph{The $2\to 2$ Scattering Case.} 
In this case, we simply have $X = \had$ and we further 
distinguish in the behaviour of the dark photon $A'$:
\begin{enumerate}[label = \alph*)]
    \item The general case: Due to the kinetic mixing portal, the dark 
    photon couples to any electromagnetic current, including those of
    composed hadronic systems. Then, the dominant dark sector 
    contribution to $2\to2$ scattering will be the interference of 
    tree-level graphs at $\mathcal{O}(\alpha^2\eps^2)$ in the cross section.
    \item The leptophilic case: If the hadronic interactions are neglected, the dominant dark sector contributions in $2\to 2$ scattering are interferences of tree level QED graphs with one-loop graphs in which the dark photon appears as a virtual particle in the leptonic current component $L^\mu$ of the amplitude.
    The contributions of this form are of order $\mathcal{O}(\alpha^3 \eps^2)$ and we consider this special case to study their connection with the corresponding real $A'$ emission, which also exists at $\mathcal{O}(\alpha^3 \eps^2)$.
\end{enumerate}

%
\paragraph{The $2\to 3$ Scattering Case.} 
In this case, we have $X = \had + R$, where $R = \{ \gamma, A' \}$ is a real,
on-shell vector boson. In general, the amplitudes are then split into 
Bethe-Heitler (BH), and virtual Compton scattering (VCS) terms. In the
former, the real external boson is emitted by the electron, while in
the latter case it is emitted from the hadronic target. Thus, the total
matrix element in the scattering regime $\mathcal{R}$ will have the form,
\begin{equation}
    i\mathcal{M}_\mathcal{R} (e^-\, \had
    \longrightarrow e^-\, \had\, R) = L^\mu_\bh \:i\varPi^V_{\mu\nu}\: H^\nu_{\bh,\,\mathcal{R}}
    + L^\mu_\vcs \: i\varPi^V_{\mu\nu} \: H^\nu_{\vcs,\,\mathcal{R}}\,.
    \label{eq:M_BH+VCS}
\end{equation}

The cross section contributions outlined above require the knowledge of 
the hadronic amplitudes $H^\nu_\mathcal{R}$ in the regime $\mathcal{R}$. To that 
end, first note that the hadronic VCS piece can always be written as
the contraction of an external polarization vector $\varepsilon^ 
{\ast}_\mu(k)$ with a Compton tensor $C^{\mu\nu}_{\mathcal{R}; R,V}$, \textit{i.e.},
\begin{equation}
	H^{\nu}_{\vcs, \mathcal{R}} = \varepsilon^{\ast}_\mu(k) \: iC^{\mu\nu}_{\mathcal{R}; R,V},
\end{equation}
where we introduced subscripts $R$ and $V$ on the Compton tensor representing the relevant real and virtual vector boson.

The hadronic amplitude $H^{\nu}_{\vcs, \mathcal{R}}$ is described by low energy 
theorems (LETs), which have been derived using fundamental symmetries. 
The key result from generalized LETs is that up to linear order in the 
photon energy, the scattering amplitude for VCS is fully determined by 
the Born contribution, which only depends on the nucleon's 
electromagnetic form factors and static properties.
Ref.~\cite{Fearing:1996gs} (for spin-0) and Refs.~\cite{Guichon:1995pu, Scherer:1996ux} (for spin-1/2) demonstrated that the gauge-invariant Born-level contribution is obtained via the vertex functions in Eqs.~\eqref{eq:H_nu_BH_gamma} and \eqref{eq:nucleon_vertex_param}, respectively.
Terms of higher order are known and can be expanded in powers of small photon momenta.
They depend on the polarizabilities of the target, \textit{cf.} Ref.~\cite{Moinester:2019sew}.
The computation of cross sections containing the Born-level VCS terms requires then the numerical evaluation of the targets electromagnetic form factors at lightlike arguments $k^2=0$ (for real photon radiation) and at timelike arguments $k^2 = m_{A'}^2$ (for real dark photon radiation).
For beam energies of $\mathcal{O}(\SI{}{\giga\electronvolt})$, the timelike form factors are always evaluated in the so-called unphysical region where $k^2 = m_{A'}^2 < 4 m_\had^2$, in which their values cannot be directly extracted from experimental data; instead, one must rely on theoretical models.

We assume that the VCS terms given below provide an accurate and consistent physical description of the scattering process only in the regime of small virtualities of the intermediate nucleus/nucleon.
At larger virtualities, several additional effects and sources of uncertainty arise.
Firstly, resonances of the corresponding hadronic target will appear, which typically lead to an enhancement of the cross section.
Secondly, sizable uncertainties enter through the form factors used in the calculation.
For nucleon bremsstrahlung, the timelike form factors relevant for the emission of a massive dark photon can be modeled using the vector meson dominance approach; however, for heavy dark photons these form factors show some significant dependence on the choice of input parameters and fitting procedures.
Moreover, existing form factors are extracted from data involving on-shell states.
To account for off-shell effects, a phenomenological form factor is commonly introduced which effectively dampens contributions from large virtualities.
Such a form factor is known for nucleon bremsstrahlung, but it carries significant model dependence, which can induce variations in the predicted cross section of up to an order of magnitude.
To our current knowledge, neither the form of the timelike form factors nor of an off-shell damping form factor is available for (dark) bremsstrahlung emitted in the coherent regime through the recoiling nucleus as a whole.

A reliable and consistent treatment of the large-virtuality regime is therefore beyond the scope of this work and is left for a possible future study.
To work only in the small virtuality regime, we place a simple cutoff on the invariant 
mass of the joint system of $\had$ and $R$, \textit{i.e.}
\begin{equation}
    \sqrt{(P' + k)^2} < \text{Resonance energy scale},
\end{equation}
where $k$ and $P'$ are the momenta of the outgoing vector boson and the 
hadronic system, respectively. The numerical values for the resonance cutoff will be given below for each case.
Above the cutoff, we simply set the VCS 
amplitude to zero. This limits the amount of energy which can 
be carried away through VCS processes and the resulting cross section results should be interpreted as a lower bound.

\subsubsection{Coherent Regime}
\label{sec:amplitudes_regimes_coherent}
As previously noted, in the coherent regime the electron interacts via 
single-vector-boson-exchange with the whole nucleus, without resolving 
the individual nucleons within.
Tungsten, which is most commonly considered to be the target 
material in LDMX-like experiments, is naturally most abundant as three scalar
isotopes with $J^P = 0^+ (\ce{^{182}_{74}W}, \ce{^{184}_{74}W}, 
\ce{^{186}_{74}W}$) and a smaller fraction of a single fermionic isotope with 
$J^P = 
\frac{1}{2}^-$ ($\ce{^{183}_{74}W}$).
Contributions to the cross sections arising from the scattering off the fermionic component will, due to its magnetic moment,
differ from the scalar case.
This difference, however, is suppressed by the inverse square of the nucleus mass \cite{Thomson:2013zua}.
For tungsten, the mass is $\SI{171.16}{\giga\electronvolt}$, which is much larger than 
the typical energy scale of the proposed experiments, $\mathcal{O} 
(\SIrange{1}{10}{\giga\electronvolt})$.
We will thus always work in an approximation where 
the nucleus is represented by a spin-0 field, which comes
with a single form factor $F(q^2)$.
We denote the nucleus by the symbol 
$\varPhi^+$ and will henceforth always work with the isotope $\varPhi^+=
\ce{^{184}_{74}W}$, as it is the most abundant one.

The ingredients for the scattering off a scalar hadronic target are known.
Refs.~\cite{Fearing:1996gs, Lvov:2001zdg, Moinester:2019sew} discuss the general behavior
of spin-0 targets in Compton scattering and state how the electromagnetic 
interactions can be written in terms of a low-energy effective Lagrangian, which we briefly discuss in App.~\ref{app:coherent_compton_tensor}.
With this and the field redefinition
in Eq.~\eqref{eq:vanilla_field_rotation}, the Bethe-Heitler currents are,
\begin{equation}
    H^\nu_{\bh,\text{coh};\, \gamma} = \begin{tikzpicture}[baseline=-2.6]
    	\setlength{\feynhandblobsize}{3mm}
    	\begin{feynhand}
    		\vertex (veg) at (0,1.2) {$\nu$};
    		\vertex [grayblob] (vhg) at (0,0) {};
    		\vertex (hi) at (-1.3,0) {$\varPhi^+$};
    		\vertex (ho) at (1.3,0) {$\varPhi^+$};
    		\propag [photon, revmom'={$q$} ] (veg) to (vhg);
    		\propag [scalar, mom'={$P$} ] (hi) to (vhg);
    		\propag [scalar, mom'={$P^\prime$}] (vhg) to (ho);
    	\end{feynhand}
    \end{tikzpicture} = ie F(q^2) \left(2P^\nu - q^\nu\right) \,
    ,
	\label{eq:H_nu_BH_gamma}
\end{equation}
as well as,
\begin{equation}
    H^\nu_{\bh, \text{coh};\, \ap} = 
    \begin{tikzpicture}[baseline=-2.6]
        \setlength{\feynhandblobsize}{3mm}
        \begin{feynhand}
        \vertex (veg) at (0,1.2) {$\nu$};
        \vertex [grayblob] (vhg) at (0,0) {};
        \vertex (hi) at (-1.3,0) {$\varPhi^+$};
        \vertex (ho) at (1.3,0) {$\varPhi^+$};
        \propag [photon, color = crimson, revmom'={[arrow style = black] $q$} ] (veg) to (vhg);
        \propag [scalar, mom'={$P$} ] (hi) to (vhg);
        \propag [scalar, mom'={$P^\prime$}] (vhg) to (ho);
    \end{feynhand}
\end{tikzpicture} =  -\varepsilon H^\nu_{\bh,\, \text{coh};\, \gamma}\,.
\label{eq:H_nu_BH_Ap}
\end{equation}
The Compton tensor is to lowest order in the soft photon approximation
given by the Born contribution,
\begin{align}
    C^{\mu\nu}_{\text{coh};\, \gamma,\gamma} &= 
    \begin{tikzpicture}[baseline=-2]
    	\setlength{\feynhandblobsize}{5mm}
    	\begin{feynhand}
    		\vertex (veg) at (-1,1) {$\mu$};
    		\vertex (veg2) at (+1,1) {$\nu$};
    		\vertex [NEblob] (vhg) at (0,0) {};
    		\vertex (hi) at (-1.05,-1.05) {$\varPhi^+$};
    		\vertex (ho) at (1.05,-1.05) {$\varPhi^+$};
    		\propag [photon, revmom'={$q$} ] (veg2) to (vhg);
    		\propag [photon, color=black, revmom'={$k$} ] (veg) to (vhg);
    		\propag [scalar, mom'={$P$} ] (hi) to (vhg);
    		\propag [scalar, mom={$P^\prime$}] (vhg) to (ho);
    	\end{feynhand}
   \end{tikzpicture} \nonumber \\
    &\approx e^2 F(q^2) F(k^2) \left[2g^{\mu\nu} - \frac{(2P - k)^\mu (2P^\prime + q)^\nu}{(P - k)^2 - m_\varPhi^2} -  \frac{(2P - q)^\nu (2P^\prime + k)^\mu}{(P - q)^2 - m_\varPhi^2} \right].\nonumber
\end{align}
Including the dark photon then yields the two pieces:
\begin{align}
C^{\mu\nu}_{\text{coh};\, \gamma, \ap} &= 
\begin{tikzpicture}[baseline=-2]
	\setlength{\feynhandblobsize}{5mm}
	\begin{feynhand}
		\vertex (veg) at (-1,1) {$\mu$};
		\vertex (veg2) at (+1,1) {$\nu$};
		\vertex [NEblob] (vhg) at (0,0) {};
		\vertex (hi) at (-1.05,-1.05) {$\varPhi^+$};
		\vertex (ho) at (1.05,-1.05) {$\varPhi^+$};
		\propag [photon, color=crimson, revmom'={[arrow style = black]$q$} ] (veg2) to (vhg);
		\propag [photon, color=black, revmom'={[arrow style = black]$k$} ] (veg) to (vhg);
		\propag [scalar, mom'={$P$} ] (hi) to (vhg);
		\propag [scalar, mom={$P^\prime$}] (vhg) to (ho);
	\end{feynhand}
\end{tikzpicture} \label{eq:Cmunu_gAp}\\
&\approx  - \varepsilon e^2 F(q^2) F(0) \left[2g^{\mu\nu} - \frac{(2P - k)^\mu (2P^\prime + q)^\nu}{(P - k)^2 - m_\varPhi^2} -  \frac{(2P - q)^\nu (2P^\prime + k)^\mu}{(P - q)^2 - m_\varPhi^2} \right]\,, \nonumber 
\end{align}
as well as,
\begin{align}
    C^{\mu\nu}_{\text{coh}; \ap,\gamma} &= 
    \begin{tikzpicture}[baseline=-2]
        \setlength{\feynhandblobsize}{5mm}
        \begin{feynhand}
        \vertex (veg) at (-1,1) {$\mu$};
        \vertex (veg2) at (+1,1) {$\nu$};
        \vertex [NEblob] (vhg) at (0,0) {};
        \vertex (hi) at (-1.05,-1.05) {$\varPhi^+$};
        \vertex (ho) at (1.05,-1.05) {$\varPhi^+$};
        \propag [photon, color=black, revmom'={$q$} ] (veg2) to (vhg);
        \propag [photon, color=crimson, revmom'={[arrow style = black]$k$} ] (veg) to (vhg);
        \propag [scalar, mom'={$P$} ] (hi) to (vhg);
        \propag [scalar, mom={$P^\prime$}] (vhg) to (ho);
        \end{feynhand}
\end{tikzpicture} \nonumber \\
& \approx  - \varepsilon e^2 F(q^2) F(m_{A'}^2) \left[2g^{\mu\nu} - \frac{(2P - k)^\mu (2P^\prime + q)^\nu}{(P - k)^2 - m_\varPhi^2} -  \frac{(2P - q)^\nu (2P^\prime + k)^\mu}{(P - q)^2 - m_\varPhi^2} \right] \nonumber \\
& \quad  + 2\eps e^2 [F(q^2) - Z ] [F(m_{A'}^2) - Z ] \left( g^{\mu\nu} - \frac{q^\mu q^\nu}{q^2} \right). \label{eq:Cmunu_Apg}
\end{align}
Note that the effective Lagrangian gives rise to an additional term in the Compton tensor with an external real dark photon with nonzero mass; see App.~\ref{app:coherent_compton_tensor}.

There is also a Compton tensor with two dark photon legs, $C^{\mu\nu}_{\text{coh}; \,\ap, \ap} \propto \eps^2$.
However, at least one leg must connect to the electron line, which introduces an additional factor of $\eps$.
Consequently, contractions involving $C^{\mu\nu}_{\text{coh}; \,\ap, \ap} $ do not contribute to the squared amplitude at $\mathcal{O}(\eps^2)$, the leading order of interest here.
For an external QED photon we evaluate the form factor at zero momentum transfer, corresponding to the on-shell photon mass.
For an external dark photon, one must evaluate the form factor $F(m_{A'}^2)$.

For tungsten nuclei, resonances appear already at an energy scale of $\sim\mathcal{O}(\SI{100}{\kilo\electronvolt}-\SI{1}{\mega\electronvolt})$, see the level scheme in Ref.~\cite{MARTIN19771}.
To stay in the region where the VCS terms provide a reasonable physical description, we will work with a cutoff
\begin{equation}
    \sqrt{(P' + k)^2} < m_{\varPhi^+} + \SI{1}{\mega\electronvolt}\,.
    \label{eq:coh_res_cut}
\end{equation}
\paragraph{Form Factors.} Since we treat the target as a scalar, there is only one form factor, 
$F(q^2)$, corresponding to the electric charge distribution. In this 
study, we work in the space-like region (\textit{i.e.} $q^2 < 0$) with the analytic Woods-Saxon form factor given by Ref.~\cite{Ballett:2018uuc},
\begin{equation}
    F(q^2) = \frac{3 Z \pi a}{r_0^2 + \pi^2 a^2} \, \frac{\pi a \coth(\pi Q a) \sin(Qr_0) - r_0 \cos(Qr_0)}{Qr_0 \sinh(\pi Q a)}\quad \text{with} \quad Q = \sqrt{-q^2}\: .
\end{equation}
We take $r_0 = 1.126 A^{1/3}~\SI{}{\femto\meter}$ and $a = \SI{0.523}{\femto\meter}$ \cite{Ballett:2018uuc}.
To our current knowledge, the behaviour of the form factor in the timelike region has not been studied, which makes the exact evaluation of squared amplitudes stemming from the Compton tensor in Eq. \eqref{eq:Cmunu_Apg} impossible.
Still, to provide at least a qualitative statement about the relevance of such contributions, we will assume that $F(m_{A'})\sim \mathcal{O}(Z)$.
See also the results and discussions in Sec. \ref{sec:DP_signal_yield}.

\subsubsection{Diffractive Regime}
\label{sec:amplitudes_regimes_diffractive}
In the diffractive regime, the electron interacts with the constituents of the nucleus, the protons ($p$) and neutrons ($n$).
The matrix element, and thus the hadronic amplitude, for the electromagnetic interaction with a nucleon $N = \{p,n\}$ can be parameterized as,
\begin{align}
    H^\nu_{\bh, \text{dif};\, \gamma} &=
    \begin{tikzpicture}[baseline=-2.6]
    	\setlength{\feynhandblobsize}{3mm}
    	\begin{feynhand}
    		\vertex (veg) at (0,1.2) {$\nu$};
    		\vertex [grayblob] (vhg) at (0,0) {};
    		\vertex (hi) at (-1.2,0) {$N$};
    		\vertex (ho) at (1.2,0) {$N$};
    		\propag [photon, revmom'={$q$} ] (veg) to (vhg);
    		\propag [fermion, mom'={$P$} ] (hi) to (vhg);
    		\propag [fermion, mom'={$P^\prime$}] (vhg) to (ho);
    	\end{feynhand}
    \end{tikzpicture}\\
	&= \langle {N(P^\prime)} | J^\nu_\text{\tiny{EM}}(q) \ket{N(P)}\\
    &= e\, \overline{u}_N(P^\prime) \left[ \gamma^\nu F_1^N(Q^2) - \frac{i\sigma^{\nu\kappa} q_\kappa}{2m_N} F_2^N(Q^2)\right] u_N(P) \, \label{eq:nucleon_vertex_param}\\
    &= \, \overline{u}_N(P^\prime) \: \varGamma^\nu_N (q)\: u_N(P)\:,
\end{align}
where $F_{1,2}^N(Q^2)$ are the Dirac and Pauli form factors, $\varGamma^\nu 
_N(q)$ a short-hand notation for the vertex function, and $u_N(P)$ the Dirac
spinor of an external nucleon with momentum $P$. Since the dark photon also 
couples to the electromagnetic current, the same parametrization holds for its 
vertex function. Note however that in general, the interactions of the dark 
photon with quarks have the form of a $V-A$ theory, since the coupling to left 
and right chiral fermions is different, see the remarks in 
Sec.~\ref{sec:dark_photon_basics}. In the presence of an axial vector 
current, terms $\propto \gamma^5$ would enter 
Eq.~\eqref{eq:nucleon_vertex_param}, which carry an axial form factor and an 
induced pseudoscalar form factor. 
We expect these effects to be suppressed by 
the EW scale, because for the axial vector quark couplings $a_q$ we have, $a_q 
= g^{A^\prime}_{q_L} - g^{A^\prime}_{q_R} \propto m_{\ad}^2/m_Z^2 \approx 
m_{\ap}^2/m_Z^2 \ll 1$. So we only consider the well-known electromagnetic form
factors.

The effective field theory of electromagnetic interactions is described 
within the framework of chiral perturbation theory, where the form factors 
can formally be expressed in terms of low energy constants appearing in the 
effective Lagrangian \cite{Scherer:2012xha}. When imposing the field 
redefinition of Eq.~\eqref{eq:vanilla_field_rotation}, one thus 
straightforwardly obtains that the dark photon couples with the same form 
factors to the nucleon as the QED photon, but again with an overall rescaling 
of the vertex function by $-\eps$. So,
\begin{align}
	H^\nu_{\bh, \text{dif};\, \ap} =
	\begin{tikzpicture}[baseline=-2.6]
		\setlength{\feynhandblobsize}{3mm}
		\begin{feynhand}
			\vertex (veg) at (0,1.2) {$\nu$};
			\vertex [grayblob] (vhg) at (0,0) {};
			\vertex (hi) at (-1.2,0) {$N$};
			\vertex (ho) at (1.2,0) {$N$};
			\propag [photon, color=crimson, revmom'={[arrow style = black]$q$} ] (veg) to (vhg);
			\propag [fermion, mom'={$P$} ] (hi) to (vhg);
			\propag [fermion, mom'={$P^\prime$}] (vhg) to (ho);
		\end{feynhand}
	\end{tikzpicture} = \, \overline{u}_N(P^\prime) \: \Big(-\eps \varGamma^\nu_N (q)\Big)\: u_N(P)\,.
\end{align}
Again, the lowest order contribution is the Born-level Compton tensor, which is given in Refs.~\cite{Vanderhaeghen:2000ws, Djukanovic:2008skh},
\begin{align}
	C^{\mu\nu}_{\text{dif};\, \gamma,\gamma} &=
	\begin{tikzpicture}[baseline=-2]
		\setlength{\feynhandblobsize}{5mm}
		\begin{feynhand}
			\vertex (veg) at (-1,1) {$\mu$};
			\vertex (veg2) at (+1,1) {$\nu$};
			\vertex [NEblob] (vhg) at (0,0) {};
			\vertex (hi) at (-1,-1) {$N$};
			\vertex (ho) at (1,-1) {$N$};
			\propag [photon, revmom'={$k$} ] (veg) to (vhg);
			\propag [photon, color=black, revmom'={$q$} ] (veg2) to (vhg);
			\propag [fermion, mom'={$P$} ] (hi) to (vhg);
			\propag [fermion, mom={$P^\prime$}] (vhg) to (ho);
		\end{feynhand}
	\end{tikzpicture} \nonumber \\
	\label{eq:nucleon_compton_tensor}
	\\
	&\approx \overline{u}_N(P^\prime) \Bigg[  \Gamma^\nu_N(q) \frac{\slashed{P}-\slashed{k}+m_N}{(P-k)^2 - m_N^2} \Gamma^\mu_N(k) +  \Gamma^\mu_N(k) \frac{\slashed{P}-\slashed{q}+m_N}{(P-q)^2 - m_N^2} \Gamma^\nu_N(q) \bigg]  u_N(P)\,.\nonumber
\end{align}
Eq.~\eqref{eq:nucleon_compton_tensor} completely fixes the VCS amplitude up 
to next-to-leading order, \textit{i.e.} including linear terms, in 
the photon 
energy, see Ref. \cite{Djukanovic:2008skh} and citations therein. Imposing 
the vertex function rescaling then leads to:
\begin{align}
	C^{\mu\nu}_{\text{dif};\, \ap, \gamma} &= \begin{tikzpicture}[baseline=-2]
		\setlength{\feynhandblobsize}{5mm}
		\begin{feynhand}
			\vertex (veg) at (-1,1) {$\mu$};
			\vertex (veg2) at (+1,1) {$\nu$};
			\vertex [NEblob] (vhg) at (0,0) {};
			\vertex (hi) at (-1,-1) {$N$};
			\vertex (ho) at (1,-1) {$N$};
			\propag [photon, color=crimson, revmom'={[arrow style = black]$k$} ] (veg) to (vhg);
			\propag [photon, color=black, revmom'={$q$} ] (veg2) to (vhg);
			\propag [fermion, mom'={$P$} ] (hi) to (vhg);
			\propag [fermion, mom={$P^\prime$}] (vhg) to (ho);
		\end{feynhand}
	\end{tikzpicture} \approx -\eps\: C^{\mu\nu}_{\text{dif};\, \gamma,\gamma}\, ,
	\label{eq:nucleon_compton_tensor_DP1} \\
	C^{\mu\nu}_{\text{dif};\, \gamma, \ap} &= \begin{tikzpicture}[baseline=-2]
	\setlength{\feynhandblobsize}{5mm}
	\begin{feynhand}
		\vertex (veg) at (-1,1) {$\mu$};
		\vertex (veg2) at (+1,1) {$\nu$};
		\vertex [NEblob] (vhg) at (0,0) {};
		\vertex (hi) at (-1,-1) {$N$};
		\vertex (ho) at (1,-1) {$N$};
		\propag [photon, color=black, revmom'={[arrow style = black]$k$} ] (veg) to (vhg);
		\propag [photon, color=crimson, revmom'={[arrow style = black]$q$}] (veg2) to (vhg);
		\propag [fermion, mom'={$P$} ] (hi) to (vhg);
		\propag [fermion, mom={$P^\prime$}] (vhg) to (ho);
	\end{feynhand}
\end{tikzpicture} \approx -\eps\: C^{\mu\nu}_{\text{dif};\, \gamma,\gamma}\, ,
\label{eq:nucleon_compton_tensor_DP2}
\end{align}
where the only difference is the assignment of momenta and Lorentz indices.
We again omit the Compton tensor with two dark photon legs.

In case of nucleons we place a cutoff on the VCS contributions to ensure that only virtualities below the $\varDelta$ resonance region are probed,
\begin{equation}
    \sqrt{(P' + k)^2} < \SI{1.2}{\giga\electronvolt}\,.
    \label{eq:dif_res_cut}
\end{equation}
The total squared amplitude in diffractive scattering is the 
incoherent sum of proton and neutron contributions, scaled by the number of 
respective nucleons within the target nucleus,
\begin{equation}
    \overline{|\mathcal{M}|^2}_\text{dif} = Z \: \overline{|\mathcal{M}_p|^2} + (A-Z) \: \overline{|\mathcal{M}_n|^2} \, .
\end{equation}

\paragraph{Form Factors.}
In contrast to the coherent scattering case, the form factors required to evaluate the diffractive scattering amplitudes are reasonably well understood in both the space- and timelike regions; in the following we give some details about the parameterizations used in our computations.
In general, the Dirac and Pauli form factors may be translated into the electric and magnetic form factors $G_\text{\tiny{E,M}}^N(q^2)$, the so-called Sachs combinations.
In the space-like region and for small values of $-q^2$ they are well-desribed by the dipole parameterization $G_\text{\tiny{D}}(q^2)$ \cite{Thomson:2013zua}, \textit{i.e.},
\begin{equation}
    G_\text{\tiny{E,M}}^N(q^2) \propto G_\text{\tiny{D}}(q^2) = \left( 1 + \frac{-q^2}{\SI{0.71}{\giga\electronvolt\squared}} \right)^{-2} \, .
\end{equation}
We will operate also in regimes where contributions arise from phase space 
regions with $-q^2 \sim \SI{1}{\giga\electronvolt\squared}$ or higher, such 
that we will instead use the more robust Kelly parametrization, taken from 
Ref.~\cite{Kelly:2004hm}. The form factors $G^p_\text{\tiny{E}}$, $G^p_\text{\tiny{M}}$, and $G^n_\text{\tiny{M}}$ are then written in terms of polynomials in $\tau \equiv -q^2/(4m_N^2)$,
\begin{equation}
    \label{eq:Kelly_GK}
    G_\text{\tiny{K}}(q^2) = \frac{1 + a_1 \tau}{1 + \sum_{k=1}^{3} b_k \tau^k}\; ,
\end{equation}
while $G_\text{\tiny{E}}^n$ is well approximated by,
\begin{equation}
    \label{eq:kelly_GnE}
    G_\text{\tiny{E}}^n(q^2) = \frac{A\tau}{1+B\tau} G_\text{\tiny{D}}(q^2)\, ,
\end{equation}
where the Kelly coefficients $a_1, b_k, A$ and $B$ are obtained from fits to 
experimental data. The values used in this work are listed in 
Table~\ref{tab:Kellycoefficients}.
We then have the relations,
\begin{equation}
\label{eq:Kelly_NE}
G_\text{\tiny{E}}^N(q^2) = F_1^N(q^2) - \tau F_2^N(q^2) = \begin{cases}
G_\text{\tiny{E}}^n(q^2), \qquad &\text{if} \quad N = n\,, \\
G_{\text{\tiny{K}}}(q^2), \qquad &\text{if} \quad N = p\,, \\
\end{cases}
\end{equation}
and
\begin{equation}
\label{eq:Kelly_NM}
G_\text{\tiny{M}}^N(q^2) = F_1^N(q^2) + F_2^N(q^2) = \begin{cases}
\mu_n G_{\text{\tiny{K}}}(q^2), \qquad &\text{if} \quad N = n\,, \\
\mu_p G_{\text{\tiny{K}}}(q^2), \qquad &\text{if} \quad N = p\,, \\
\end{cases}
\end{equation}
where $\mu_n$ and $\mu_p$ are the nucleon magnetic moments of the neutron and proton, 
respectively, \textit{cf.} Eqs.~\eqref{eq:mu_p_value}--\eqref{eq:mu_n_value}. Given the above relations, it is 
straightforward to obtain the form factors $F^N_{1,2}(q^2)$ that appear in the 
(squared) amplitude expressions.

The form factors in the timelike region, $q^2 = m_{A'}^2 > 0$ can be described using the vector meson dominance (VMD) approach, which has been used in Refs.~\cite{Kling:2025udr, Foroughi-Abari:2021zbm, Gorbunov:2023jnx, Gorbunov:2024vrc}
specifically for the modelling of dark bremsstrahlung.
The VMD approach assumes that the nucleon form factors are described by a sum of iso-scalar, $F_i^\text{s} = F_{i,\omega} + F_{i,\phi}$, and iso-vector, $F_i^\text{v} = F_{i,\rho}$, form factors.
For the electromagnetic interaction, and similarly that induced by the dark photon, the nucleon form factors are then given by the sums
\begin{align}
    \label{eq:vmd}
    F_i^{p}(q^2) &= F_{i,\omega}(q^2) + F_{i,\phi}(q^2) + F_{i,\rho}(q^2)\,, \\
    F_i^{n}(q^2) &= F_{i,\omega}(q^2) + F_{i,\phi}(q^2) - F_{i,\rho}(q^2)\,.
\end{align}
The $F_{i,v}$ are modelled using a simple vector meson resonance ansatz.
With $n$ resonances, the form factors are written as a sum of Breit-Wigner functions,
\begin{equation}
    \label{eq:bw_ansatz}
    F_{i,v} = \sum_{j=0}^{n-1} a_{i, v_j} \text{BW}_{v_j}(q^2)\, ,
\end{equation}
where $v=\omega, \phi, \rho$ are the usual vector meson resonance families. 
For a resonance $v$ with mass $m_v$ and width $\varGamma_v$, the Breit-Wigner functions are given by,
\begin{equation}
    \label{eq:bw_form}
    \text{BW}_{v}(q^2) \equiv \frac{m_v^2}{m_v^2 - q^2 - im_v\varGamma_v} \, .
\end{equation}
The $3\times 2 \times n$ couplings $a_{i,v_j}$, corresponding to three resonance families and two form factors, are determined through either fitting or theoretical considerations.
The discussion of normalization conditions and analytic relations among the parameters is outlined in detail in Ref.~\cite{Kling:2025udr}.
This reference provides the best fit values of the remaining free parameters, which are $a_{1,v_0}$ as well as the masses and widths of the resonances.
We work with Model 1 presented in Ref.~\cite{Kling:2025udr}, in which the masses and widths of the highest resonances $\omega'', \rho'', \phi''$ are varied around their PDG values, while the parameters of the lower resonances $\omega, \omega', \rho, \rho', \phi, \phi'$ remain fixed to their PDG values.\footnote{The other models presented in Ref.~\cite{Kling:2025udr} allow for more variation of the masses and widths around their PDG values.
For dark photon masses in the relevant range of $\SIrange{1}{100}{\mega\electronvolt}$, all three presented models agree reasonably well with each other, so in this study we will not discuss any uncertainties caused by the different fit results.}

The form factors in the timelike region are complex functions.
We explicitly checked that the relevant squared amplitudes, specifically all terms arising in Eq. \eqref{eq:ME2_2to3_DarkPhoton}, only depend on real combinations of them, namely $\big|F_i^N(m_\ap^2)\big|^2$, $\Re\big\{ F_1^N(m_\ap^2) \, F_2^{N *}(m_\ap^2) \big\}$ as well as $\Re\big\{ F_i^N (m_\ap^2) \big\}$.

\paragraph{Pauli blocking.} Let us stress that we do not scatter off free nucleons, as the target nucleons are bound within a nucleus. To account for nuclear effects we implement the Pauli blocking mechanism as in Refs.~\cite{Ballett:2018uuc, Zhou:2019vxt}, which is believed to account for the dominant nuclear effects. Pauli blocking 
enforces that a nucleon struck by the intermediate photon does not scatter into an already occupied state.
The cross section, including this effect, is rescaled as
\begin{equation}
    \d \sigma \mapsto f(|\vv{q}|) \, \d\sigma\, ,
\end{equation}
where $f(|\vv{q}|)$ is a function which suppresses cross section contributions 
for small momentum transfers, with its argument, $|\vv{q}|$, being the absolute
value of the three-momentum transferred to the struck nucleon in the lab 
frame. According to Refs.~\cite{Bodek:2021trq, Ballett:2018uuc, Zhou:2019vxt}, 
the suppression factor in elastic scattering off a nucleon $N = \{ p, n\}$ 
reads,
\begin{equation}
    f_N(|\vv{q}|) = 
    \begin{dcases}
        \frac{3}{2} \frac{|\vv{q}|}{2 k^N_\text{\tiny{F}}} - \frac{1}{2} \left( \frac{|\vv{q}|}{2 k^N_\text{\tiny{F}}} \right)^3 , \qquad &\text{if} \quad |\vv{q}| < 2 k^N_\text{\tiny{F}}\, , \\
        1, &\text{if} \quad |\vv{q}| \geq 2 k^N_\text{\tiny{F}} \, ,
    \end{dcases}
\end{equation}
where $k^N_\text{\tiny{F}}$ is the Fermi momentum of nucleon $N$.
We treat the effect separately for protons and neutrons, since for large nuclei (like tungsten), the asymmetry of proton and neutron numbers becomes quite large, and thus the Fermi momenta will be different.
By treating the nucleus as an ideal Fermi gas we find the values $k_\text{\tiny{F}}^p \approx \SI{231}{\mega\electronvolt}$ and $k^n_\text{\tiny{F}} \approx \SI{263}{\mega\electronvolt}$ in tungsten.

\subsection{Tools}
\label{sec:amplitudes_tools}
A major goal of this work is to perform theoretical studies of signal and
background processes of the form of Eq.~\eqref{eq:scatter} in experimentally 
accessible regions of LDMX-like experiments.
To this end, we have developed a flexible numerical framework written in \texttt{C++} which implements the ingredients required to make differential scattering predictions in either the coherent or diffractive regimes discussed above.
In short, we implement the expressions for all terms that enter the differential scattering rate given in Eq.~\eqref{eq:dsigma}, \textit{i.e.} the fully differential phase-space as well as analytic expressions for the squared amplitudes.
This approach allows for the integration over the phase-space to be performed
numerically with arbitrary and often complicated selections on the outgoing
particles' momenta.
Our numerical framework is hereafter referred to as \texttt{Lohengrin++}.

\paragraph{Squared amplitude construction.}
To obtain the analytic expressions for the squared amplitudes appearing in Eq.~\eqref{eq:dsigma}, we first construct/calculate the various (sub)amplitudes which appear in the decomposition introduced in 
Sec.~\ref{sec:amplitudes_regimes}.
The construction of these amplitudes in the various scattering regimes was facilitated by the use of the following \texttt{Mathematica} packages: \texttt{FeynArts}~\cite{Hahn:2000kx}, \texttt{FormCalc}~\cite{Hahn:1998yk}, \texttt{FeynCalc}~\cite{Mertig:1990an, Shtabovenko:2016sxi, Shtabovenko:2020gxv} and \texttt{FeynRules}~\cite{Christensen:2008py, Alloul:2013bka}.
To obtain the amplitudes in the coherent regime, note that the lowest order scattering off the nucleus can be described with a simple effective field theory, where the nucleus is treated as a point-like particle associated to a complex scalar field $\varPhi^+$ charged under $U(1)_\text{\tiny{\tiny{EM}}}$, \textit{cf.} App. \ref{app:coherent_compton_tensor} and Eq. \eqref{eq:L_scalarQED}.
The Feynman rules involving the nucleus are then those from scalar QED, and the hadronic amplitudes emerging from this approach are the same as the ones given in Eqs.~\eqref{eq:H_nu_BH_gamma} -- \eqref{eq:Cmunu_gAp} 
after multiplying with the appropriate form factors and kinetic mixing parameter. 
We thus implemented a \texttt{FeynRules} model for scalar QED as well as the dark photon model described in Sec.~\ref{sec:dark_photon_basics} to allow for a straight-forward amplitude construction.
Further amplitude manipulations, for example the explicit tensor contractions involving for example the nucleon vertex function or the dark photon Compton tensor, were carried out using \texttt{FeynCalc}.
This approach was also used to construct one-loop Bethe-Heitler type amplitudes involving a dark photon (\textit{e.g.} a leptophilic dark photon). In that case we also made use of the external package \texttt{OneLOop}~\cite{vanHameren:2010cp}, to numerically evaluate scalar one-loop integrals.
Details on the renormalization procedure we implemented
for the one-loop calculation involving the dark photon are given in 
App.~\ref{app:renormalization_counterterms}. We 
follow in close analogy the procedure which has been detailed for the SM 
in Ref.~\cite{Denner:1991kt}, making the minor adaptations required for the 
dark photon case at hand.

\paragraph{Phase space integration.} The numerical 
integration is performed with the \texttt{Vegas} algorithm as implemented in 
the \texttt{Cuba} library; see Ref.~\cite{Hahn:2004fe}.
The expression(s) for the differential phase-space which were implemented in the code are summarized in App.~\ref{app:phasespace_integral}.
As already discussed, the motivation for this differential approach is that 
it allows to define arbitrary selections on the outgoing 
particles' momenta which is required to mimic the selections which must be applied in a realistic experimental setup.
In addition, we have included an option to generate events by making use of the Metropolis-Hastings algorithm as well as a feature to write event files in the LHE format.

\section{Final State Considerations: Experiment and Theory}

\label{sec:phasespace}

In this section, we go through a more detailed discussion of the different final state phase space regions in the considered experimental setups (\textit{i.e.} LDMX-like detectors). 
In the scattering process at hand,
\begin{equation}
    \label{eq:phasespace_process}
    e^-(p) + \had(P) \rightarrow e^-(p^\prime) + \had(P^\prime) ~[ + R(k)] \, ,
\end{equation}
we consider in this study for $R$ only the ordinary QED photon and a massive dark photon, which was introduced in Sec.~\ref{sec:dark_photon_basics}. 
However the following kinematical discussion applies to the emission of (pseudo)scalar or vector particles. 
In particular, Ref. \cite{Blinov:2020epi} already extended the spectrum of possible BSM particles detectable with LDMX and M$^3$ (Ref.~\cite{Kahn:2018cqs}) to renormalizable theories including any particles with spin $\leq 1$.

\subsection{Kinematic Variables}
\label{sec:phasespace_variables}

For LDMX-like experiments one of the key observable in the final state is the momentum $p'$ of (one of) the recoiling electron(s).
In the following, we work in the lab frame and place the right-handed coordinate system such that the target is at the origin, with the beam pointing in the positive $\hat{z}$-direction. 
The applied magnetic field shall point into the positive $\hat{y}$-direction.

For theoretical considerations, we find it convenient to work with the three spherical variables,
\begin{equation}
    (\xi_i,\theta_i,\phi_i)\, \qquad \text{with} \qquad i = e, \had, R\, ,
\end{equation}
where $\xi_i = E_i/E$ is the fraction of the beam energy $E$ carried by the final-state particle $i$, and $\theta_i$ and $\phi_i$ are its scattering angle with respect to the beam axis and its azimuthal angle.
In experimental contexts however, we will occasionally also work with the absolute value of the three momentum instead of $\xi_i$.
Instead of the scattering angles $\theta_i$, one could choose to work with the transverse momenta, but we prefer to completely disentangle the angle and energy information. 
In particular, the scattering angle is a very useful quantity since LDMX-like setups are forward detectors and thus cone-like with an opening pointing away from the interaction point.
In Sec.~\ref{sec:sig_bkg} we will also work with the absolute value of the electron momentum component perpendicular to the magnetic field, which we define as,
\begin{equation}
    p'_{\perp B} \equiv \sqrt{(p'_1)^2 + (p'_3)^2}\,.
\end{equation}

Another set of important kinematic quantities to consider are the (negative) squared momentum transfers between various external particles. 
For the different types of process being considered, these exchanged momenta can be associated with the momentum transfer through vector boson propagators $i\varPi_{\mu\nu}$.
We thus define,
\begin{subequations}
\begin{alignat}{2}
    Q^2 &\equiv -(p-p')^2 = -(P-P')^2 \qquad &&\text{for any}\: 2\to 2 \,\: \text{scattering},\\[2mm]
    Q^2_\bh &\equiv -(P-P')^2 = -(p-p'-k)^2 \qquad &&\text{for}\: 2\to 3\,\: \text{Bethe-Heitler scattering},\\[2mm]
    Q^2_\vcs &\equiv -(p-p')^2 = -(P-P'-k)^2\qquad &&\text{for}\: 2\to 3 \,\:  \text{virtual Compton scattering}.
\end{alignat}
\end{subequations}

Before discussing the available phase space regions for the different scattering processes, we find it informative to first discuss the capabilities of the proposed experimental setups, specifically that of LDMX as well as \textsc{Lohengrin}.
This will allow us to establish a set of realistic final-state selections for several potential experimental scenarios, which in turn have important implications for theory calculations and their range of validity.

\subsection{Measurable Final States}
\label{sec:phasespace_measurable}

The forward detector setups of LDMX-like experiments restrict the phase space in which the final state electron is accessible, \textit{i.e.} its tracks can be reconstructed. 
The form of the macroscopic electron trajectories is straightforward to calculate and they can be used to estimate the accessible portion in electron phase space. 
Furthermore, the detector technologies/capabilities place limitations on the resolution to which the various four-momentum components of the electron can be reliably measured.
The combination of this information in turn places restrictions on the experimentally accessible phase space regions.
This guides the definition of signal-regions where theory calculations can be utilised to assess which regions can enhance signal-to-background ratios, and ultimately discovery potential.

In the following we discuss two major detector concepts: LDMX with a beam energy of $E=\SI{8}{\giga\electronvolt}$, as well as \textsc{Lohengrin} with $E=\SI{3.2}{\giga\electronvolt}$.
In each case, we discuss their potential for reconstructing the electron kinematics and additionally discuss their potential for vetoing events with additional hadronic or photonic radiation.

\paragraph{Electron selections.}
The LDMX recoil tracker is assembled from silicon microstrip modules with a spacial resolution of $\SI{7}{\micro\meter}$ horizontally and $\SI{100}{\micro\meter}$ vertically, with the first tracking layer placed $\SI{1.5}{\centi\meter}$ behind the target \cite{LDMX:2025pkp}.
Hence, based only on simple geometric considerations, we estimate the order of magnitude of the angular resolution to be $\theta_\text {res} \sim\SI{1}{\milli\radian}$.\footref{fn:resolution}
Furthermore, it has been reported that electron tracks with momentum down to $\SI{50}{\mega \electronvolt}$, can be reconstructed \cite{LDMX:2025pkp}.
For the purpose of this study, we also estimate that the reconstruction efficiency for electrons with large scattering angles $\theta_e > \SI{0.5}{\radian}$ is low and is assumed to be zero here.
We note that the azimuthal angle $\phi_e$ also determines the trajectories traced by the electron in these detector setups, since the magnetic field breaks the rotational symmetry around the beam axis.
For small recoil electron energies and large scattering angles, the reconstruction efficiency for electrons in the final state will depend on the azimuthal angle $\phi_e$.
However, we found that this effect is numerically not important and therefore do not consider any $\phi_e$ selections.

A similar analysis can also be applied in the case of \textsc{Lohengrin}.
For \textsc{Lohengrin} the preliminary detector design implies that only electrons with $|\vec{p}'| > \SI{25}{\mega\electronvolt}$ and $\theta_e < \SI{0.25}{\radian}$ are accessible. 
In the \textsc{Lohengrin} case, the electrons from the primary beam are not directed into the ECAL. 
Concerning the angular resolution, it is planned to use DMAPS based on the TJ-Monopix2 design with pixel pitch of $\sim\SI{33} {\micro\meter}$ for the tracking system (\textit{cf.} Ref.~\cite{Bechtle:2024atq}), and a similar order of magnitude estimate of $\theta_\text {res} \sim\SI{1}{\milli\radian}$ can be applied.\footnote{Note that our resolution estimates are optimistic.
For the angle we only took into account the detector geometry, but imperfect hardware, measurement uncertainties and uncertainties from the track reconstruction algorithms will most certainly cause the angular resolution to be an order of magnitude bigger, $\theta_e \sim \SI{10}{\milli\radian}$.
Since we will need the angular cutoff only for conceptual purposes, we will not take into account all possible experimental details, but work with an optimistic best case scenario of $\theta_e \sim \SI{1}{\milli\radian}$. \label{fn:resolution}}

\paragraph{Photon selections/veto.}
The photon veto in the experiment proposals is realized with an ECAL placed centrally in the beam direction behind the target and recoil trackers.
The LDMX ECAL is a silicon-tungsten sampling calorimeter, which is assembled from seven hexagonal modules and uses technology and designs developed for the CMS High Granularity Calorimeter \cite{LDMX:2025pkp}.
For the purpose of this study, we estimate that it roughly covers a cone with a polar opening angle of $\sim \SI{0.6}{\radian}$.
For \textsc{Lohengrin}, a silicon-tungsten sampling
calorimeter based on a prototype by the CALICE collaboration (Ref.~\cite{Poschl:2022lra}) has been used in the preliminary simulation studies.
In the first candidate design, it is placed $\SI{3.5}{\meter}$ behind the target.
With a baseline choice of a $48\times\SI{ 48}{\centi\meter\squared}$ large ECAL, it is thus estimated to cover a cone with a polar opening angle of $\sim \SI{0.07}{\radian}$.

\paragraph{Hadronic selections/veto.}
\label{sec:phasespace_nuclear_stability}
From these estimations, so far without any guidance from theory, we find a set of minimal final-state selections for the recoiling electron.
In general, the missing momentum search strategy aims to detect electrons with low energy and at the same time veto on events accompanied by additional radiation (photons or hadronic activity).
To achieve this, the detectors must also veto on hadronic activity, and here we also make some rough statements about the appearance of hadronic final 
states in the detectors. This can be relevant in the case of diffractive scattering where quasi-free nucleons may be ejected from the target nucleus above a certain energy threshold.

We estimate this threshold by requiring that the nucleon kinetic energy exceeds the binding energy, which is defined as the energy required to remove one nucleon from the nucleus. 
From $E_{N,\text{kin}} \approx E_B = \SI{8.005078\pm 0.000004 }{\mega\electronvolt}$  \cite{Wang_2021}, we find for tungsten a momentum threshold of,
\begin{equation}
	|\vec{P}^\prime|_\text{thr} = \sqrt{(E_B + m_N)^2 - m_N^2} \simeq \SI{123}{\mega\electronvolt}\,.
\end{equation}

We emphasize that this simple threshold estimate should be interpreted with caution, given the highly complex nuclear environment.

We found that our results are not very sensitive to the exact choice of momentum threshold, because low energy electrons always imply a high momentum
transfer onto the nucleons.
Specifically, this applies to the results in Secs.~\ref{sec:DP_signal_yield} (Experimental final state selections), \ref{sec:photon_background_yield} and \ref{sec:sig_bkg}. 
A more sophisticated treatment of the nuclear effects experienced by the struck nucleon and a precise simulation of the hadronic backgrounds, especially the production of 
neutrals, requires a dedicated study. 
There are a few simulation frameworks that appear particularly suitable for this purpose, for example 
\texttt{Geant4} \cite{AGOSTINELLI2003250, 1610988, ALLISON2016186}, \texttt{FLUKA} \cite{FLUKA, BOHLEN2014211}, and \texttt{GiBUU} \cite{Buss:2011mx}. \texttt{Geant4} and \texttt{FLUKA} are already in use for studying backgrounds in the \textsc{Lohengrin} experiment, but the results will be published in a dedicated study.\footnote{We especially thank Laney Klipphahn for her efforts in this direction.}

\paragraph{Summary tables of fiducial selections.}
In Tables \ref{tab:LDMX_selections} and \ref{tab:LOH_selections} we summarize the fiducial selections that we enforce on the final state kinematics of real emission processes for the LDMX and \textsc{Lohengrin} scenarios, respectively.
The presented cuts select measurable, low energy electrons while at the same time removing final state photons and nucleons that are estimated to be directed into the ECAL and HCAL, respectively.
What remains are then solitary, low-energy electrons at the detector level.
In the LDMX case, the selections on the electron correspond roughly to a experimental trigger of $> \SI{3500}{\mega\electronvolt}$ missing energy (where we assume for simplicity a single incoming electron).
For \textsc{Lohengrin} they correspond to the signal region SR1 defined in Ref.~\cite{Bechtle:2024atq}.

\begin{table}
\setlength{\tabcolsep}{5pt}
\centering
\begin{tabular}{c>{\centering\arraybackslash}p{5cm}>{\centering\arraybackslash}p{1cm} >{\centering\arraybackslash}p{5cm}}
\toprule \toprule 
\multicolumn{4}{c}{\textsc{LDMX} scenario}\\
\midrule\midrule
\multirow{2}{2cm}[-2pt]{\centering Final state particle} & \multicolumn{3}{c}{Acceptance cuts} \\
\cmidrule(lr){2-4}
& Momentum & & Angle \\ \midrule
$e^-$ & $\SI{50}{\mega\electronvolt}< |\vec{p}'| < \SI{4.5}{\giga\electronvolt}$& & $\theta_e < \SI{0.5}{\radian}$ \\
$\varPhi^+$ & - & & - \\
$N$ & $|\vec{P}'| < \SI{123}{\mega\electronvolt}$ & OR & $\theta_N > \pi/2$ \\
$\ap$ & - & & - \\
$\gamma$ & - & & $\theta_\gamma > \theta_\text{ECAL}$ \\
\bottomrule\bottomrule
\end{tabular}
\caption{Final state selections corresponding approximately to events below the LDMX trigger with an estimated LDMX HCAL veto.}
\label{tab:LDMX_selections}
\end{table}

\begin{table}
\setlength{\tabcolsep}{5pt}
\centering
\begin{tabular}{c>{\centering\arraybackslash}p{5cm}>{\centering\arraybackslash}p{1cm} >{\centering\arraybackslash}p{5cm}}
\toprule \toprule 
\multicolumn{4}{c}{\textsc{Lohengrin} scenario}\\
\midrule\midrule
\multirow{2}{2cm}[-2pt]{\centering Final state particle} & \multicolumn{3}{c}{Acceptance cuts} \\
\cmidrule(lr){2-4}
& Momentum & & Angle \\ \midrule
$e^-$ & $\SI{25}{\mega\electronvolt}< |\vec{p}'| < \SI{75}{\mega\electronvolt}$& & $\theta_e < \SI{0.25}{\radian}$ \\
$\varPhi^+$ & - & & - \\
$N$ & $|\vec{P}'| < \SI{123}{\mega\electronvolt}$ & OR & $\theta_N > \SI{0.24}{\radian}$ \\
$\ap$ & - & & - \\
$\gamma$ & - & & $\theta_\gamma > \theta_\text{ECAL}$ \\
\bottomrule\bottomrule
\end{tabular}
\caption{Final state selections corresponding to \textsc{Lohengrin} SR1 and \textsc{Lohengrin} HCAL veto.}
\label{tab:LOH_selections}
\end{table}

\subsection{Kinematically Available Phase Space}
\label{sec:phasespace_constrains}

A set of feasible experimental selections have been summarised in Tables~\ref{tab:LDMX_selections} and \ref{tab:LOH_selections}.
To better inform potential BSM search strategies, it is useful to contrast these selections with the physically allowed regions in both $2 \to 2$ and $2 \to 3$ scattering processes, exactly accounting for particle mass effects.
To achieve that we derive the outgoing electron kinematics for all such scenarios, working in the lab frame.

\subsubsection{\texorpdfstring{$2 \to 2$}{TEXT}}
\label{sec:phasespace_constrains:2to2}

Energy-momentum conservation in the process $e^- \had \rightarrow e^-
\had$ implies a fixed relation between the 
energy fraction $\xi_e$ and the scattering angle $\theta_e$ of the outgoing 
electron in the lab frame. It reads,
\begin{align}
    \theta_{e,\,{2\to 2}}(\xi_e) = \arccos\left(\frac{\abs{\vec{p}}^2+E^2 \xi_e^2-m_e^2+m_\had^2-\left(E+m_\had-E\xi_e\right)^2}{2\abs{\vec{p}}\sqrt{E^2\xi_e^2-m_e^2}}\right)\, ,
    \label{eq:theta_2to2_bound}
\end{align}
and defines a line in $(\xi_e,\theta_e)$ phase space along 
which the scattering is elastic for a given hadronic target mass 
$m_\had$. At the boundaries $\cos\theta_e= \pm 1$ we get the minimal and 
maximal values of $\xi_e$ in any $2 \to 2$ process:
\begin{equation}
    \xi^\text{\:max}_{e,\, 2\to 2} = 1 \qquad \text{and} \qquad \xi^\text{\:min}_{e,\, 2\to 2} = \frac{1}{E}\,\frac{Em_\had^2+m_e^2\,(E+2m_\had)}{2E m_\had+m_\had^2+m_e^2}\,.
\end{equation}
With $E=\SI{4}{\giga\electronvolt}$, the minimal $\xi_e$ value in coherent scattering is still very 
close to one, due to the large nucleus mass: $\xi_{e,\:2\to 2, \text{coh}} 
^\text{min} \simeq 0.955$. In diffractive scattering this boundary is lower 
$\xi_{e,\:2\to 2, \text{dif}}^\text{min} \simeq 0.105$, meaning that this 
elastic $2\to2$ contribution can in principle also produce widely scattered electrons 
with a missing momentum of $\mathcal{O}(\SI{}{\giga\electronvolt})$, which 
potentially mimic and overlap with the real $A^\prime$ emission signal.
As Fig.~\ref{fig:phasespace01} (right) shows, this effect is not relevant, as the elastic nucleon line only bends to low electron 
energies at very large scattering angles, which are not accessible with the proposed forward detectors. Moreover, low energy electrons imply high momentum transfers, 
which, in the scattering on nucleons, potentially leads to vetoable 
hadronic final states, see Sec.~\ref{sec:phasespace_nuclear_stability}.

\subsubsection{\texorpdfstring{$2 \to 3$}{TEXT}}
\label{sec:phasespace_constrains:2to3}

Determining the exact phase space bounds in the lab frame for $e^-\had\to e^- \had R$ 
scattering is more involved.
In the center-of-mass (CM) frame, the electron retains an energy $E'^*$ in the range $m_e \leq E'^* \leq E'^*_\text{max}$. The maximal energy is given by the expression,
\begin{equation}
    E'^*_\text{max} = \sqrt{\left(\frac{\lambda^{\frac{1}{2}}(E_\text{\tiny{CM}}, m_R + m_\had, m_e)}{2E_\text{\tiny{CM}}}\right)^2 + m_e^2},
\end{equation}
where $\lambda^{\frac{1}{2}}(x,y,z) = \sqrt{x^4 + y^4 + z^4 - 2x^2y^2 - 2y^2z^2 - 2z^2x^2}$ is the square root of the K\"all\'{e}n function and $E_\text{\tiny{CM}} = \sqrt{m_e^2 + m_\had^2 + 2Em_\had}$ is the center-of-mass energy.
The scattering angles in the CM frame lie in the range $0 \leq \theta_e^* \leq \pi$.
One obtains the boundary of allowed electron kinematics in the lab frame by boosting the CM electron momenta with maximum energy $E'^*_\text{max}$ and arbitrary scattering angle $\theta_e^*$ into the lab frame.
The boost is in negative $\hat z$-direction with speed,
\begin{equation}
    \beta_\text{lab} = \frac{\lambda^{\frac{1}{2}}(E_\text{\tiny{CM}}, m_e, m_\had)/(2E_\text{\tiny{CM}})}{\sqrt{\left(\displaystyle\frac{\lambda^{\frac{1}{2}}(E_\text{\tiny{CM}}, m_e, m_\had)}{2E_\text{\tiny{CM}}}\right)^2+m_\had^2}}\, .
\end{equation}
Some algebra leads then to the following expression relating the lab frame scattering angle to the energy fraction $\xi_e$,
\begin{equation}
		\cos \theta_{e,\:2\to 3}^\text{bound} = \frac{2E^2\xi_e-2E m_\had(1-\xi_e)+m_R^2-2m_e^2 +  2m_\had m_R}{2\abs{\vec{p}}\sqrt{E^2\xi_e^2 - m_e^2}} \, .
		\label{eq:costheta_bounds_equations_e}
\end{equation}
The kinematically accessible scattering angles in the lab frame then lie in the range $0 \leq \theta_e \leq \arccos(\cos\theta_{e, 2\to 3}^\text{max})$, where the maximum value is determined through,
\begin{equation}
    \cos\theta_{e, 2\to 3}^\text{max} =
    \begin{cases}
    -1 & \text{if} \quad \cos \theta_{e,\:2\to 3}^\text{bound} < -1, \\
    \cos \theta_{e,\:2\to 3}^\text{bound} & \text{if} \quad \cos \theta_{e,\:2\to 3}^\text{bound} \geq -1,
    \end{cases}
    \label{eq:costheta_max2to3}
\end{equation}
and is implicitly understood as a function of the lab frame energy fraction $\xi_e$.
Essentially, the boost stretches the electron momenta for $\theta_e^* < \pi/2$ and contracts those for $\theta_e^* > \pi/2$, such that (viewed from the lab frame) with increasing scattering angles, the electron can carry away less energy.
The boost also implies that the maximum electron energy fraction is reached for $\theta_e = 0$ and has the analytic form,
\begin{align}
		\xi_{e,\:2\to 3}^{\text{max}}
        =& \: \frac{1}{2 E \left(m_\had^2 + 2 E m_\had  + m_e^2\right)} \nonumber \\
        &\times \Bigg[\left(E+m_\had\right)\left(2 E m_\had - m_R^2-2 m_\had m_R + 2m_e^2\right) \nonumber\\
		& \qquad +\abs{\vec{p}}\sqrt{\left(m_R^2+2 m_\had m_R-2 E m_\had \right)^2-4 m_e^2\big(m_\had + m_R\big)^2}\;\Bigg] \label{eq:y_max}\, .
\end{align}
In electron-nucleus scattering one might apply the limit $m_e 
\to 0$ and $m_\had \to \infty$, such that Eq. \eqref{eq:y_max} reduces
to $\xi_{e,\:2\to 3}^{\text{max}} = 1-m_R/E$.

It is useful to have analytic expressions of the momentum transfers in $2\to 3$ scattering as function of the electron lab frame kinematics. In Bethe-Heitler scattering, energy-momentum 
conservation restricts the momentum transfers to lie in the range $Q^2_{\bh;\text{min}} \leq Q_\bh^2 \leq Q^2_{\bh;\text{max}}$, where the maximal and minimal values are given by,
\begin{align}
	&Q^2_{\bh;\,{\text{max,min}}} = \frac{1}{\left(E- E\xi_e+m_\had\right)^2-V^2} \nonumber \\
	& \:\:\times \Bigg[m_\had \big(E- E\xi_e+m_\had\big)\big[(p-p^\prime)^2-m_R^2\big]+2 m_\had^2 V^2 \nonumber \\
	& \qquad \pm m_\had V \Bigg\{\Big[(p-p^\prime)^2-m_R^2\Big] \Big[4m_\had(E- E\xi_e+m_\had)+(p-p^\prime)^2-m_R^2\Big]+4m_\had^2V^2\Bigg\}^{1/2}\,\Bigg]\, ,
	\label{eq:Q2limits}
\end{align}
with $V = |\vec{p}^\prime -\vec{p}|$. The minimal and maximal values are understood as functions of $\xi_e$ and $\theta_e$, $Q^2_{\bh;\,{\text{max,min}}} = Q^2_{\bh;\,{\text{max,min}}}(\xi_e,\theta_e)$.

In virtual Compton scattering, the momentum transfer is fixed to,
\begin{equation}
    \label{eq:Q2vcs_labframe}
    Q^2_\vcs = -2m_e^2 + 2 E^2 \xi_e - 2 |\vec{p}| \sqrt{E^2 \xi_e^2 - m_e^2} \cos\theta_e\,.
\end{equation}

\begin{figure}[ht!]
		\centering
		\includegraphics[width=0.49\textwidth]{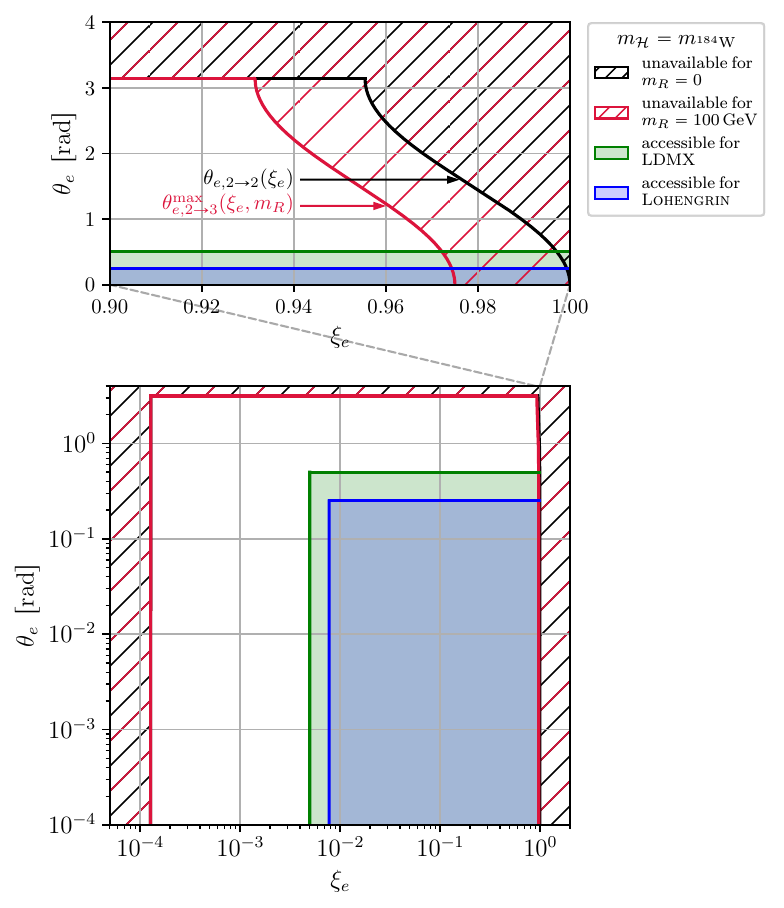}
        \includegraphics[width=0.49\textwidth]{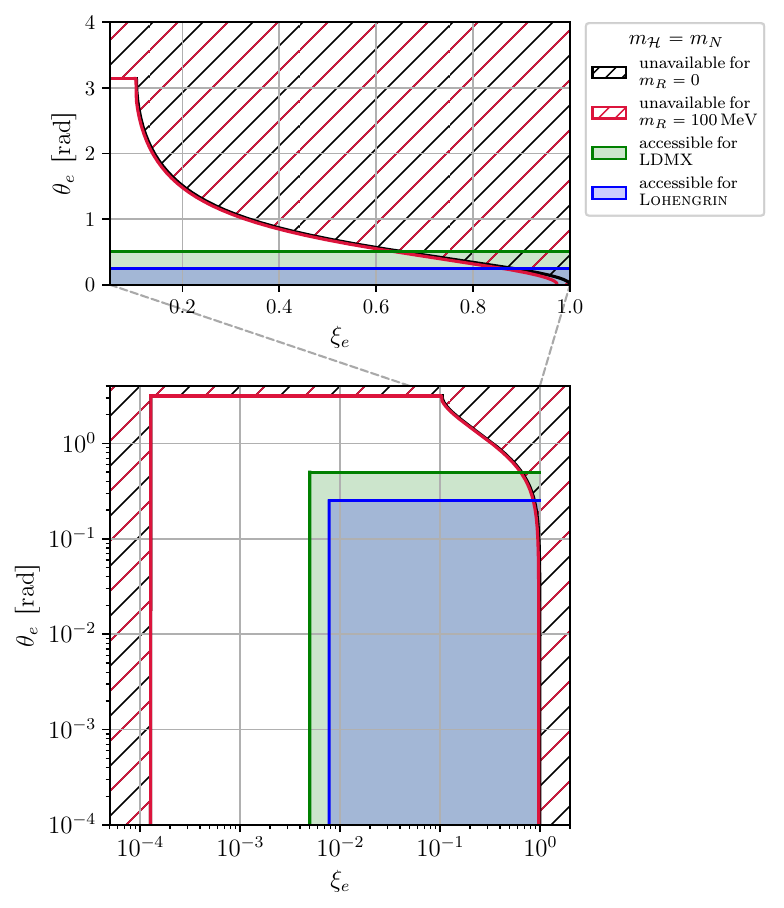}
		\caption[]{Depiction of the phase space available to the recoiling electron in a $2\to 2$ or $2\to 3$ process as in Eq. \eqref{eq:phasespace_process}. The hashed area in red (black) is kinematically forbidden for dark photon (QED photon) production. We assume a dark photon mass of $m_R = \SI{100}{\mega\electronvolt}$. The boundaries for high $\xi_e$ values are obtained from Eq.~\eqref{eq:costheta_max2to3}. The green (blue) area shows the part of phase space where we expect the electron is accessible to the LDMX (\textsc{Lohengrin}) detector setup. The top panels show a close up of the large $\xi_e$ region in linear scale. Left: electron-nucleus scattering, Right: electron-nucleon scattering.}
		\label{fig:phasespace01}
\end{figure}

A depiction of the emerging relevant phase space windows is 
presented in Fig.~\ref{fig:phasespace01} for $m_R = \SI{100} {\mega 
\electronvolt}$ and $m_R = 0$, respectively.
We show the electron phase space as a two-dimensional plane spanned 
by the energy fraction $\xi_e$ and the scattering angle $\theta_e$, 
because the fundamental processes in the target are symmetric around the 
beam axis, which renders the azimuthal angles $\phi_i$ uninteresting when 
showing cross sections.
The red (black) hashed area 
shows the phase space region which is forbidden for dark photon (QED 
photon) emission: no final state electron will populate these regions 
after $2\to 3$ scattering. On the contrary, the white area is allowed and 
thus populated by $2\to 3$ events. We notice that for $2\to 3$ scattering 
with $m_R \neq 0$, the boundary of the available phase space is shifted away 
from the region where $2 \to 2$ scattering is located (the black curve on the
right). However, for the dark photon masses in question, this is only a small
effect. For $2\to 3$ scattering with $m_R = 0$ 
(corresponding kinematically to QED bremsstrahlung), the boundary derived in 
Eq.~\eqref{eq:costheta_bounds_equations_e} coincides exactly with the $2\to 2$ 
elastic line.
The blue and green boxes in Fig.~\ref{fig:phasespace01} show the 
phase space portion available to the experiments, which is further explained 
in Sec.~\ref{sec:phasespace_measurable}.

\subsection{Theoretical implications of the final-state selections}

Based on the previous discussions of detector capabilities and kinematically accessible phase space of the $2\to 2$ or $2\to 3$ scattering processes, we broadly define two types of possible BSM search strategies.
The first would aim to measure the scattering rate close to the elastic scattering line at large $\xi_e$ values, as kinematically accessible in the coherent $2\to 2$ scattering processes for example, and detect small deviations with respect to the SM expectation.
The second is the usual missing momentum strategy which aims to identify electrons with low $\xi_e$ values in the absence of visible radiation.
In line with these two strategies, we can thus motivate a minimal set of theory selections which are summarised in Table~\ref{tab:theory_selections}.
The first `Minimal Coulomb-safe selection' applies only an angular selection on the outgoing electron and does not exclude the elastic contribution from the $2\to 2$ process. 
In contrast, the `minimal real emission selections' require $\xi_e < 0.9$, $\theta_e < \pi/2$ and $p_{T, e} < \SI{0.5}{\giga\electronvolt}$ and are designed to remove all $2\to 2$ contributions to the cross section in a benchmark scenario with a $\SI{4}{\giga\electronvolt}$ beam, while at the same time preserving a large portion of the kinematically available real emission phase space.

In the following we discuss the reliability of the theoretical predictions for these scenarios.

\paragraph{The large $\xi_e$ region (the elastic line).}
In the first scenario we must consider the contributions arising from the $2\to 2$ scattering process (\textit{i.e.} that occupy the elastic line).
The perturbative prediction for such scattering process in both SM and BSM processes involve amplitudes that arise from $t$-channel photon exchange.
As such, these amplitudes (and hence the cross-section prediction) suffer from a divergence as $Q^2 \rightarrow 0$ encoded in the photon propagator.
The limit $Q^2 \rightarrow 0$ is equivalent to $\xi_e \to 1$ or $\theta_e \rightarrow 0$ in the lab frame, \textit{i.e.} for forward electrons.
Without a direct restriction on the electron phase-space (\textit{e.g.} through $\xi_e$ or $\theta_e$) the cross-section is therefore not integrable\footnote{The application of more advanced/non-perturbative methods would be required to overcome the presence of this Coulomb divergence, for example those presented in Ref. \cite{FuentesZamoro:2025exp}.}.
In practice we consider a minimal $\theta_e$ requirement of $\theta_e^{\rm min} = \SI{1}{\milli\radian}$ when considering this scenario, thus avoiding the $Q^2\to0$ region. This selection can also be found in Table \ref{tab:theory_selections} as `Minimal Coulomb-safe selection'.

\paragraph{Implications for the missing momentum strategy.}
The missing momentum strategy instead aims to measure scattered electrons with roughly $\xi_e \lesssim 0.5$ and scattering angles maximally of $\mathcal{O}(\SI{0.1}{\radian})$.
When considering the dominant coherent scattering process, such a constraint cannot be kinematically satisfied by the $2\to 2$ scattering process.
For the $2\to 3$ scattering process this can be achieved if the radiated particle (massive or massless) carries a significant fraction of the initial energy.
In the following, we consider the case of Bethe-Heiler scattering, but the statements translate analogously to virtual Compton scattering.

For the SM process, the ECAL of the experiment is used to veto the presence of hard photons. 
The coverage of the ECAL is however not complete and it is possible that processes with a hard wide-angled photon (which is not detected by the ECAL) mimic the signal process.
Due to the presence of such a hard photon emission, these contributions do not probe the region of small $Q^2_\bh$, and the tree-level prediction is numerically finite at $\mathcal{O}(\alpha^3)$.
At the same time, since one is interested in kinematics with $\xi_e \lesssim 0.5$, the emitted photon cannot be soft.

The situation for the BSM signal is more complicated, as there are no selections on the kinematics of the emitted dark photon (it is invisible).
However, the presence of a non-zero dark photon mass does ensure that the mass of the combined outgoing electron and dark photon system always satisfies the relation $\left(p^{\prime} + k\right)^2 > m_e^2$. 
This implies that the $t$-channel singularity related to $Q^2_\bh  \rightarrow 0$ is never encountered, and the cross-section is integrable over the phase space provided.
In practice, we typically consider the mass range of $m_{A'}\approx \mathcal{O} (\SI{1}{\mega\electronvolt}) - \mathcal{O}(\SI{1}{\giga\electronvolt})$, and the condition is always satisfied.
Instead, if one considers a minimal requirement on $\theta_e$, the region of $Q^2_\bh \rightarrow 0$ is kinematically avoided and the cross-section remains integrable even in the limits $m_\ap \rightarrow 0$ and $m_e \rightarrow 0$.
\begin{table}
\setlength{\tabcolsep}{5pt}
\centering
\begin{tabular}{ccc>{\centering\arraybackslash}p{4cm} >{\centering\arraybackslash}p{4cm}}
\toprule \toprule 
\multicolumn{4}{c}{Theory selections}\\
\midrule\midrule
\multirow{2}{2cm}[-2pt]{\centering Final state particle} & \multicolumn{3}{c}{Minimal Coulomb-safe selections} \\
\cmidrule(lr){2-4}
& Energy & Transverse momentum & Angle \\ \midrule
$e^-$ & - & - & $\theta_e > \SI{e-3}{\radian}$ \\
$\varPhi^+, N, \ap, \gamma $ & - & - & - \\
\midrule\midrule
\multirow{2}{2cm}[-2pt]{\centering Final state particle} & \multicolumn{3}{c}{Minimal real emission selections} \\
\cmidrule(lr){2-4}
& Energy & Transverse momentum & Angle \\ \midrule
$e^-$ & $\xi_e < 0.9$ & $p_{T,e} < \SI{0.5}{\giga\electronvolt}$ & $\theta_e < \pi/2$ \\
$\varPhi^+, N, \ap, \gamma $ & - & - & - \\
\bottomrule\bottomrule
\end{tabular}
\caption{Final state selections motivated by theoretical considerations.}
\label{tab:theory_selections}
\end{table}

When considering values of $m_{A'}$ closer to the $\SI{1}{\mega\electronvolt}$ range, the relative size of the signal contribution arising from the small $\theta_e$ values increases (these contributions receive an enhancement at small $Q^2$ values).
To improve the experimental sensitivity to the BSM process, it can then be favourable to not place any minimal $\theta_e$ requirement.
Under such conditions, for $m_{A'} \sim \SI{1}{\mega\electronvolt}$, values of the variable $Q_\bh^2$ as small as $\SI{e-13}{} \sim \SI{e-14} {\giga \electronvolt}^2$ are sampled (depending also on the beam energy, \textit{cf.} Eq.~\eqref{eq:Q2limits}).
To achieve numerically stable results in this kinematic region, we found that it was necessary to perform the computation in \texttt{long double} precision.
Essentially, the numerical manipulations involving the (squared) ratio of $Q_\bh^2$ as compared to other mass scales (such as the tungsten nucleus) were not under control without higher numerical precision.
Our Monte Carlo program was thus extended to facilitate this feature.

\section{Virtual Dark Photons: \texorpdfstring{$2\to 2$}{TEXT} Scattering Processes}
\label{sec:2to2}
In this section, we look at the BSM contributions populating the elastic 
lines. While the leading contribution for a vanilla dark photon 
(Secs.~\ref{sec:2to2_amplitudes:leptophilic} and 
\ref{sec:2to2_results_vanilla}) should be interpreted in the context of 
a hypothetical precision measurement of the elastic lines, the 1-loop 
leptophilic $A'$ contributions are relevant for the correct 
interpretation of the inclusive $A'$ production in the experimental 
proposals of LDMX, DarkSHINE and \textsc{Lohengrin}.

\subsection{(Squared) Amplitudes}
\label{sec:2to2_amplitudes}
We consider both QED and dark photon contributions in the scattering 
process 
\begin{equation}
    e^- + \had \rightarrow e^- + \had\,.
\end{equation}
In case of a vanilla dark photon as outlined in 
Sec.~\ref{sec:dark_photon_basics}, the leading BSM contribution will 
arise from interference of tree-level diagrams. We also consider the case
of a leptophilic dark photon, which dominantly contributes through 
interferences involving one-loop graphs.

\subsubsection{Tree Level: Vanilla Dark Photon}
\label{sec:2to2_amplitudes:vanilla}
The leading, pure SM contribution at $\mathcal{O} (\alpha)$ in 
the amplitude is
\begin{equation}
        i \mathcal{M}(e^- \had \to e^- \had; \alpha) ~ = ~
        \begin{tikzpicture}[baseline=15]
        \setlength{\feynhandblobsize}{3mm}
        \setlength{\feynhandarrowsize}{5pt}
        \begin{feynhand}
        \vertex (ei) at (-1.25, 1.2) {$e^-$};
        \vertex (eo) at (1.25, 1.2) {$e^-$};
        \vertex (veg) at (0,1.2);
        \vertex [grayblob] (vhg) at (0,0) {};
        \vertex (hi) at (-1.25,0) {$\had$};
        \vertex (ho) at (1.25,0) {$\had$};
        \propag [fermion] (ei) to (veg);
        \propag [fermion] (veg) to (eo);
        \propag [photon, revmom = $q$] (veg) to [edge label' = $\gamma$] (vhg);
        \propag [double,double distance=0.2ex,thick] (hi) to (vhg);
        \propag [double,double distance=0.2ex,thick] (vhg) to (ho);
        \end{feynhand}
        \end{tikzpicture}
        ~ ,
\end{equation}
while at $\mathcal{O}(\alpha \eps^2)$, the following virtual dark photon exchange contributes:
\begin{equation}
        i \mathcal{M}(e^- \had \to e^- \had; \alpha\eps^2) ~ = ~
        \begin{tikzpicture}[baseline=15]
        \setlength{\feynhandblobsize}{3mm}
        \setlength{\feynhandarrowsize}{5pt}
        \begin{feynhand}
        \vertex (ei) at (-1.25, 1.2) {$e^-$};
        \vertex (eo) at (1.25, 1.2) {$e^-$};
        \vertex (veg) at (0,1.2);
        \vertex [grayblob] (vhg) at (0,0) {};
        \vertex (hi) at (-1.25,0) {$\had$};
        \vertex (ho) at (1.25,0) {$\had$};
        \propag [fermion] (ei) to (veg);
        \propag [fermion] (veg) to (eo);
        \propag [photon, color = crimson, revmom ={[arrow style = black]$q$}] (veg) to [edge label' = $\ap$] (vhg);
        \propag [double,double distance=0.2ex,thick] (hi) to (vhg);
        \propag [double,double distance=0.2ex,thick] (vhg) to (ho);
        \end{feynhand}
        \end{tikzpicture}
        ~ .
\end{equation}
Here, $\begin{tikzpicture}[baseline=-2]
        \setlength{\feynhandblobsize}{3mm}
        \begin{feynhand}
        \vertex (veg) at (0,0.5);
        \vertex [grayblob] (vhg) at (0,0) {};
        \vertex (hi) at (-0.4,0);
        \vertex (ho) at (0.4,0);
        \propag [photon] (veg) to (vhg);
        \propag [double,double distance=0.2ex,thick] (hi) to (vhg);
        \propag [double,double distance=0.2ex,thick] (vhg) to (ho);
        \end{feynhand}
        \end{tikzpicture}$
and
$\begin{tikzpicture}[baseline=-2]
        \setlength{\feynhandblobsize}{3mm}
        \begin{feynhand}
        \vertex (veg) at (0,0.5);
        \vertex [grayblob] (vhg) at (0,0) {};
        \vertex (hi) at (-0.4,0);
        \vertex (ho) at (0.4,0);
        \propag [photon, color = crimson] (veg) to (vhg);
        \propag [double,double distance=0.2ex,thick] (hi) to (vhg);
        \propag [double,double distance=0.2ex,thick] (vhg) to (ho);
        \end{feynhand}
        \end{tikzpicture}$
represent the interactions of the photons with the hadronic systems, 
which were discussed in Secs.~\ref{sec:amplitudes_regimes_coherent} and 
\ref{sec:amplitudes_regimes_diffractive}.

We collect the squared amplitudes up to lowest nontrivial order in the 
kinetic mixing parameter, \textit{i.e.} $\mathcal{O}(\eps^2)$. The 
leading contribution is purely from QED,
\begin{equation}
        \abs{\mathcal{M}}^2(e^- \had \to e^- \had ; \alpha^{2})
        ~ = ~
        \abs{~
        \begin{tikzpicture}[baseline=15]
        \setlength{\feynhandblobsize}{3mm}
        \setlength{\feynhandarrowsize}{5pt}
        \begin{feynhand}
        \vertex (ei) at (-1, 1.2);
        \vertex (eo) at (1, 1.2);
        \vertex (veg) at (0,1.2);
        \vertex [grayblob] (vhg) at (0,0) {};
        \vertex (hi) at (-1,0);
        \vertex (ho) at (1,0);
        \propag [fermion] (ei) to (veg);
        \propag [fermion] (veg) to (eo);
        \propag [photon] (veg) to (vhg);
        \propag [double,double distance=0.2ex,thick] (hi) to (vhg);
        \propag [double,double distance=0.2ex,thick] (vhg) to (ho);
        \end{feynhand}
        \end{tikzpicture}
        ~ }^2 ~ ,
\end{equation}
which is followed by the tree-level interference of the QED photon 
exchange with the dark photon exchange:
\begin{equation}
    \label{eq:ME2_a3e2}
        \abs{\mathcal{M}}^2(e^- \had \to e^- \had ; \alpha^{2}\eps^2)
        ~ = ~ 2 \Re \left\{~
        \begin{tikzpicture}[baseline=15]
        \setlength{\feynhandblobsize}{3mm}
        \setlength{\feynhandarrowsize}{5pt}
        \begin{feynhand}
        \vertex (ei) at (-1, 1.2);
        \vertex (eo) at (1, 1.2);
        \vertex (veg) at (0,1.2);
        \vertex [grayblob] (vhg) at (0,0) {};
        \vertex (hi) at (-1,0);
        \vertex (ho) at (1,0);
        \propag [fermion] (ei) to (veg);
        \propag [fermion] (veg) to (eo);
        \propag [photon] (veg) to (vhg);
        \propag [double,double distance=0.2ex,thick] (hi) to (vhg);
        \propag [double,double distance=0.2ex,thick] (vhg) to (ho);
        \end{feynhand}
        \end{tikzpicture}
        ~ ~
        \begin{tikzpicture}[baseline=15]
        \setlength{\feynhandblobsize}{3mm}
        \setlength{\feynhandarrowsize}{5pt}
        \begin{feynhand}
        \vertex (ei) at (-1, 1.2);
        \vertex (eo) at (1, 1.2);
        \vertex (veg) at (0,1.2);
        \vertex [grayblob] (vhg) at (0,0) {};
        \vertex (hi) at (-1,0);
        \vertex (ho) at (1,0);
        \propag [fermion] (ei) to (veg);
        \propag [fermion] (veg) to (eo);
        \propag [photon, color=crimson] (veg) to (vhg);
        \propag [double,double distance=0.2ex,thick] (hi) to (vhg);
        \propag [double,double distance=0.2ex,thick] (vhg) to (ho);
        \end{feynhand}
        \end{tikzpicture}^{\scaleto{\dagger}{10pt}}
        \right\}~ .
\end{equation}  

\subsubsection{One-Loop Level: Leptophilic Dark Photon}
\label{sec:2to2_amplitudes:leptophilic}

For a leptophilic dark photon, \textit{i.e.} one that does not couple to hadrons
but only to leptons,
the BSM contributions appear in the following loop graphs:
\begin{equation}
i \mathcal{M}(e^- \had \to e^- \had; \alpha^2 \eps^2) ~ = ~
        \begin{tikzpicture}[baseline=15]
        \setlength{\feynhandblobsize}{3mm}
        \setlength{\feynhandarrowsize}{5pt}
        \begin{feynhand}
        \vertex (ei) at (-1, 1.2);
        \vertex (eo) at (1, 1.2);
        \vertex (veg) at (0,0.6);
        \vertex (vedp1) at (-0.4,1.2);
        \vertex (vedp2) at (0.4,1.2);
        \vertex [grayblob] (vhg) at (0,0) {};
        \vertex (hi) at (-1,0);
        \vertex (ho) at (1,0);
        \propag [fermion] (ei) to (vedp1);
        \propag [fermion] (vedp1) to (veg);
        \propag [fermion] (veg) to (vedp2);
        \propag [fermion] (vedp2) to (eo);
        \propag [photon] (veg) to (vhg);
        \propag [photon, color = crimson] (vedp1) to (vedp2);
        \propag [double,double distance=0.2ex,thick] (hi) to (vhg);
        \propag [double,double distance=0.2ex,thick] (vhg) to (ho);
        \end{feynhand}
        \end{tikzpicture}
        ~+~
        \begin{tikzpicture}[baseline=15]
        \setlength{\feynhandblobsize}{3mm}
        \setlength{\feynhandarrowsize}{5pt}
        \begin{feynhand}
        \vertex (ei) at (-1, 1.2);
        \vertex (eo) at (1, 1.2);
        \vertex [crossdot] (veg) at (0,1.2) {};
        \vertex [grayblob] (vhg) at (0,0) {};
        \vertex (hi) at (-1,0);
        \vertex (ho) at (1,0);
        \propag [fermion] (ei) to (veg);
        \propag [fermion] (veg) to (eo);
        \propag [photon] (veg) to (vhg);
        \propag [double,double distance=0.2ex,thick] (hi) to (vhg);
        \propag [double,double distance=0.2ex,thick] (vhg) to (ho);
        \end{feynhand}
        \end{tikzpicture} 
        ~+~
        \begin{tikzpicture}[baseline=15]
        \setlength{\feynhandblobsize}{3mm}
        \setlength{\feynhandarrowsize}{5pt}
        \begin{feynhand}
        \vertex (ei) at (-1, 1.2);
        \vertex (eo) at (1, 1.2);
        \vertex (veg) at (0,  1.2);
        \vertex [grayblob] (vhg) at (0,0) {};
        \vertex (hi) at (-1,0);
        \vertex (ho) at (1,0);
        \vertex (vbubble1) at (0, 0.85);
        \vertex (vbubble2) at (0, 0.45);
        \propag [fermion] (ei) to (veg);
        \propag [fermion] (veg) to (eo);
        \propag [fermion] (vbubble1) to [in=180, out=180, looseness=1.7] (vbubble2);
        \propag [fermion] (vbubble2) to [in=0, out=0, looseness=1.7] (vbubble1);
        \propag [photon, color = crimson] (veg) to (vbubble1);
        \propag [photon] (vbubble2) to (vhg);
        \propag [double,double distance=0.2ex,thick] (hi) to (vhg);
        \propag [double,double distance=0.2ex,thick] (vhg) to (ho);
        \end{feynhand}
        \end{tikzpicture}
        ~+~
        \begin{tikzpicture}[baseline=15]
        \setlength{\feynhandblobsize}{3mm}
        \setlength{\feynhandarrowsize}{5pt}
        \begin{feynhand}
        \vertex (ei) at (-1, 1.2);
        \vertex (eo) at (1, 1.2);
        \vertex (veg) at (0,  1.2);
        \vertex [grayblob] (vhg) at (0,0) {};
        \vertex (hi) at (-1,0);
        \vertex (ho) at (1,0);
        \vertex [crossdot] (vct) at (0, 0.6) {};
        \propag [fermion] (ei) to (veg);
        \propag [fermion] (veg) to (eo);
        \propag [photon, color = crimson] (veg) to (vct);
        \propag [photon] (vct) to (vhg);
        \propag [double,double distance=0.3ex,thick] (hi) to (vhg);
        \propag [double,double distance=0.3ex,thick] (vhg) to (ho);
        \end{feynhand}
        \end{tikzpicture}
        ~ ,
        \label{eq:NLO_1-loop_amp}
\end{equation}
where the crossdots ($\begin{tikzpicture}[baseline=-2]
	\setlength{\feynhandblobsize}{4mm}
	\begin{feynhand}
		\vertex [crossdot] (v) at (0,0.05) {};
	\end{feynhand}
\end{tikzpicture}$)
mark the renormalization counterterms which cancel the UV divergences potentially arising from the loop.
They are further discussed in App.~\ref{app:renormalization_counterterms}.
The loops contribute in the squared amplitude through interference with the QED tree-level graph:
\begin{align}
    \abs{\mathcal{M}}^2(e^- \had \to e^- \had ; \alpha^{3}\eps^2) 
    &= 2 \Re \left\{~
        \left(
        \begin{tikzpicture}[baseline=15]
        \setlength{\feynhandblobsize}{3mm}
        \setlength{\feynhandarrowsize}{5pt}
        \begin{feynhand}
        \vertex (ei) at (-1, 1.2);
        \vertex (eo) at (1, 1.2);
        \vertex (veg) at (0,0.6);
        \vertex (vedp1) at (-0.4,1.2);
        \vertex (vedp2) at (0.4,1.2);
        \vertex [grayblob] (vhg) at (0,0) {};
        \vertex (hi) at (-1,0);
        \vertex (ho) at (1,0);
        \propag [fermion] (ei) to (vedp1);
        \propag [fermion] (vedp1) to (veg);
        \propag [fermion] (veg) to (vedp2);
        \propag [fermion] (vedp2) to (eo);
        \propag [photon] (veg) to (vhg);
        \propag [photon, color = crimson] (vedp1) to (vedp2);
        \propag [double,double distance=0.2ex,thick] (hi) to (vhg);
        \propag [double,double distance=0.2ex,thick] (vhg) to (ho);
        \end{feynhand}
        \end{tikzpicture}
        ~+~
        \begin{tikzpicture}[baseline=15]
        \setlength{\feynhandblobsize}{3mm}
        \setlength{\feynhandarrowsize}{5pt}
        \begin{feynhand}
        \vertex (ei) at (-1, 1.2);
        \vertex (eo) at (1, 1.2);
        \vertex [crossdot] (veg) at (0,1.2) {};
        \vertex [grayblob] (vhg) at (0,0) {};
        \vertex (hi) at (-1,0);
        \vertex (ho) at (1,0);
        \propag [fermion] (ei) to (veg);
        \propag [fermion] (veg) to (eo);
        \propag [photon] (veg) to (vhg);
        \propag [double,double distance=0.2ex,thick] (hi) to (vhg);
        \propag [double,double distance=0.2ex,thick] (vhg) to (ho);
        \end{feynhand}
        \end{tikzpicture}
        ~\right)
        ~~
        \begin{tikzpicture}[baseline=15]
        \setlength{\feynhandblobsize}{3mm}
        \setlength{\feynhandarrowsize}{5pt}
        \begin{feynhand}
        \vertex (ei) at (-1, 1.2);
        \vertex (eo) at (1, 1.2);
        \vertex (veg) at (0,1.2);
        \vertex [grayblob] (vhg) at (0,0) {};
        \vertex (hi) at (-1,0);
        \vertex (ho) at (1,0);
        \propag [fermion] (ei) to (veg);
        \propag [fermion] (veg) to (eo);
        \propag [photon] (veg) to (vhg);
        \propag [double,double distance=0.2ex,thick] (hi) to (vhg);
        \propag [double,double distance=0.2ex,thick] (vhg) to (ho);
        \end{feynhand}
        \end{tikzpicture}^{\scaleto{\dagger}{10pt}}
    ~\right\} \nonumber \\
    & \quad + 2 \Re \left\{~
    
        \left(
        \begin{tikzpicture}[baseline=15]
        \setlength{\feynhandblobsize}{3mm}
        \setlength{\feynhandarrowsize}{5pt}
        \begin{feynhand}
        \vertex (ei) at (-1, 1.2);
        \vertex (eo) at (1, 1.2);
        \vertex (veg) at (0,  1.2);
        \vertex [grayblob] (vhg) at (0,0) {};
        \vertex (hi) at (-1,0);
        \vertex (ho) at (1,0);
        \vertex (vbubble1) at (0, 0.85);
        \vertex (vbubble2) at (0, 0.45);
        \propag [fermion] (ei) to (veg);
        \propag [fermion] (veg) to (eo);
        \propag [fermion] (vbubble1) to [in=180, out=180, looseness=1.7] (vbubble2);
        \propag [fermion] (vbubble2) to [in=0, out=0, looseness=1.7] (vbubble1);
        \propag [photon, color = crimson] (veg) to (vbubble1);
        \propag [photon] (vbubble2) to (vhg);
        \propag [double,double distance=0.2ex,thick] (hi) to (vhg);
        \propag [double,double distance=0.2ex,thick] (vhg) to (ho);
        \end{feynhand}
        \end{tikzpicture}
        ~+~
        \begin{tikzpicture}[baseline=15]
        \setlength{\feynhandblobsize}{3mm}
        \setlength{\feynhandarrowsize}{5pt}
        \begin{feynhand}
        \vertex (ei) at (-1, 1.2);
        \vertex (eo) at (1, 1.2);
        \vertex (veg) at (0,  1.2);
        \vertex [grayblob] (vhg) at (0,0) {};
        \vertex (hi) at (-1,0);
        \vertex (ho) at (1,0);
        \vertex [crossdot] (vct) at (0, 0.6) {};
        \propag [fermion] (ei) to (veg);
        \propag [fermion] (veg) to (eo);
        \propag [photon, color = crimson] (veg) to (vct);
        \propag [photon] (vct) to (vhg);
        \propag [double,double distance=0.3ex,thick] (hi) to (vhg);
        \propag [double,double distance=0.3ex,thick] (vhg) to (ho);
        \end{feynhand}
        \end{tikzpicture} 
        ~\right)
        ~~
        \begin{tikzpicture}[baseline=15]
        \setlength{\feynhandblobsize}{3mm}
        \setlength{\feynhandarrowsize}{5pt}
        \begin{feynhand}
        \vertex (ei) at (-1, 1.2);
        \vertex (eo) at (1, 1.2);
        \vertex (veg) at (0,1.2);
        \vertex [grayblob] (vhg) at (0,0) {};
        \vertex (hi) at (-1,0);
        \vertex (ho) at (1,0);
        \propag [fermion] (ei) to (veg);
        \propag [fermion] (veg) to (eo);
        \propag [photon] (veg) to (vhg);
        \propag [double,double distance=0.2ex,thick] (hi) to (vhg);
        \propag [double,double distance=0.2ex,thick] (vhg) to (ho);
        \end{feynhand}
        \end{tikzpicture}^{\scaleto{\dagger}{10pt}}
        
    ~\right\}
    ~ .
    \label{eq:ME2_2to2_NLO}
\end{align}
The interference terms containing the loop corrections are of the same order in 
perturbation theory as the real emission graphs, see 
Eq.~\eqref{eq:ME2_2to3_DarkPhoton} below.
The virtual corrections must be added to the real emission corrections involving the dark photon to provide a consistent
cross section prediction at order $\mathcal{O} (\alpha^3 \eps^2)$ (in the lepton charge).
We discuss this in further detail in 
Sec.~\ref{sec:2to2_results_leptophilic}.

\subsection{Results}
\label{sec:2to2_results}
\subsubsection{Tree Level: Vanilla Dark Photon}
\label{sec:2to2_results_vanilla}

The contributions to the $2\to 2$ cross section populate the 
elastic lines specified in Eq.~\eqref{eq:theta_2to2_bound}. In case of a 
vanilla dark photon, having couplings to both leptons and hadrons, the 
leading BSM contribution is the tree-level interference in 
Eq.~\eqref{eq:ME2_a3e2}. In coherent scattering, the interference term takes 
the simple form,
\begin{align}
        &\overline{\abs{\mathcal{M}}^2}(e^- \varPhi^+ \to e^- \varPhi^+ ; \alpha^{2}\eps^2) \nonumber \\
        &= 128\alpha^2 \varepsilon^2 \: |F(Q^2)|^2 \frac{2 m_{\varPhi}^2 m_e^2 - s_{13}(m_{\varPhi}^2 + s_{12}) + 2m_e^2 s_{12} + s_{12}^2}{(s_{13} - 2 m_e^2 )(s_{13} - 2 m_e^2 + m_{A^\prime}^2)}\,,  
\end{align}
where $s_{13}=2 p \cdot p' = 2 ( E E^\prime - |\vec{p}||\vec{p}^\prime|
\cos\theta_e) $ and $s_{12}=2 p \cdot P = 2 E m_{\varPhi}$.
Compared to the LO QED piece, there are no extra terms in the numerator, 
since the longitudinal term in the photon propagator, $q_\rho q_\sigma /
m_{A^\prime}^2$, vanishes when being contracted with the hadronic current. The 
squared amplitude contains,
\begin{equation}
   \overline{|\mathcal{M}|^2} \supset ...\: q_\sigma  H^{\ast\,\sigma}_{\bh, \text{coh}; \, A^\prime} \,,
\end{equation}
where the ellipsis summarizes all other factors.
Since $q_\sigma (P+P^\prime)^\sigma = (P^2 - P^{\prime\,2}) = 0$ for 
on-shell external $\varPhi^+$, the longitudinal term vanishes. It also 
vanishes in diffractive scattering; this is easy to see when looking at 
the full expression of the interference term for an individual nucleon. 
We have,
\begin{equation}
   \overline{|\mathcal{M}|^2} = 2 \Re \left\{ \frac{1}{4} \sum_{\text{spins}} L^\mu \frac{-ig_{\mu\nu}}{q^2} H^\nu_{\bh, N;\,\gamma} \,  (-\eps)L^{\ast\,\rho} \frac{i \left( g_{\rho\sigma} - \frac{q_\rho q_\sigma }{m_{A^\prime}^2}\right)}{q^2 - m_{A^\prime}^2} H^{\ast\,\sigma}_{\bh, N;\, A^\prime} \right\}\,,
\end{equation}
where $L^\mu$ is the QED leptonic current. After performing the spin 
sums, the hadronic piece combines to,
\begin{align}
    &\frac{1}{2} \sum_{\text{spins}} H^\nu_{\bh, N;\, \gamma}\: H^{\ast\, \sigma}_{\bh, N;\, \ap} \nonumber \\
    &= e^2 \varepsilon^2 \left[ H_1^N(Q^2) (2P^\nu - q^\nu)(2P^\sigma - q^\sigma) - H_2^N(Q^2) (Q^2 g^{\nu\sigma} + q^\nu q^\sigma )\right]\,, \label{eq:diff_hadronic_tensor_gammaA}
\end{align}
where the $H_{1,2}^N(Q^2)$ are linear combinations of the $F^N_{1,2}(Q^2)$,
\begin{equation}
    H_1^N(Q^2) = |F_1^N(Q^2)|^2 + \tau |F_2^N(Q^2)|^2, \quad \text{and} \quad H_2^N(Q^2) = |F_1^N(Q^2) + F_2^N(Q^2)|^2,
\end{equation}
with $\tau = Q^2/4m_N^2$, \textit{cf.} Refs.~\cite{Ballett:2018uuc, Kniehl:1990iv}.
The longitudinal part of the dark photon propagator vanishes on 
Eq.~\eqref{eq:diff_hadronic_tensor_gammaA}:
\begin{align}
    \overline{|\mathcal{M}|^2} &\supset 2\Re\Bigg\{ \frac{  \frac{1}{2} \sum_{\text{spins}} L_\nu L^{\ast\,\rho}}{q^2(q^2 - m_{A^\prime}^2)}  \frac{q_\rho q_\sigma}{m_{A^\prime}^2} e^2 \varepsilon^2 \Big[ H_1^N(Q^2) (2P^\nu - q^\nu)(2P^\sigma - q^\sigma) \nonumber \\
    &\qquad \qquad \qquad \qquad \qquad \qquad \qquad \qquad - H_2^N(Q^2) (Q^2 g^{\nu\sigma} + q^\nu q^\sigma ) \Big] \Bigg\} \\
    &=  2\Re\Bigg\{ \frac{  \frac{1}{2} \sum_{\text{spins}} L_\nu L^{\ast\,\rho}}{q^2(q^2 - m_{A^\prime}^2)}  \frac{q_\rho}{m_{A^\prime}^2} e^2 \varepsilon^2 \Big[ H_1^N(Q^2) (2P^\nu - q^\nu)(P^2 - P^{\prime\,2}) \nonumber \\
    &\qquad \qquad \qquad \qquad \qquad \qquad \qquad \qquad - H_2^N(Q^2) (Q^2 q^\nu - Q^2 q^\nu ) \Big] \Bigg\} \\
    &= 0\,.
\end{align}
Thus, compared to the LO QED term, the contribution of the
interference term to the cross section is rescaled according to,
\begin{equation}
    \label{eq:2to_2alpha2eps2_reldev}
    \d \sigma (e^- \had \to e^- \had ; \alpha^{2}\eps^2) = 2\varepsilon^2\,\frac{ Q^2}{Q^2 + m_{A^\prime}^2} \: \d \sigma (e^- \had \to e^- \had ; \alpha^{2})\,.
\end{equation}
This implies two limiting cases. For $Q^2 \ll m_{A^\prime}^2$, the 
the $\mathcal{O}(\alpha^2 \eps^2)$ contribution to the
differential cross section is suppressed by $2\eps^2 Q^2/m_{A^\prime}^{2}$ 
compared to the $\mathcal{O}(\alpha^2)$ contribution
and its scaling in terms of kinematic variables is altered by a factor of
$Q^2$. For $Q^2 \gg m_{A^\prime}^2$, the interference term behaves just 
as the LO QED piece times a factor of $2\eps^2$. We quantify this by 
defining the relative deviation of the two contributions:
\begin{equation}
    \varDelta_{\alpha^2, \, \alpha^2 \eps^2} \equiv \frac{\d \sigma (e^- \had \to e^- \had ; \alpha^{2}\eps^2) - \d \sigma (e^- \had \to e^- \had ; \alpha^{2})}{\d \sigma (e^- \had \to e^- \had ; \alpha^{2})} = \frac{(2\eps^2 - 1 ) Q^2 - m_{A^\prime}^2}{Q^2 + m_{A^\prime}^2}\,,
\end{equation}
where the last equality follows from 
Eq.~\eqref{eq:2to_2alpha2eps2_reldev}. The limiting cases are then 
explicitly,
\begin{equation}
    \varDelta_{\alpha^2, \, \alpha^2 \eps^2} \rightarrow 
    \begin{cases}
    -1 + 2 \eps^2 Q^2/m_{A'}^2 &\quad \text{for} \quad Q^2 \ll m_{A^\prime}^2\,, \\
    -1+2\eps^2 &\quad \text{for} \quad Q^2 \gg m_{A^\prime}^2\,. \\
    \end{cases}
\end{equation}
\begin{figure}[ht]
    \centering
    \includegraphics[width=0.7\linewidth]{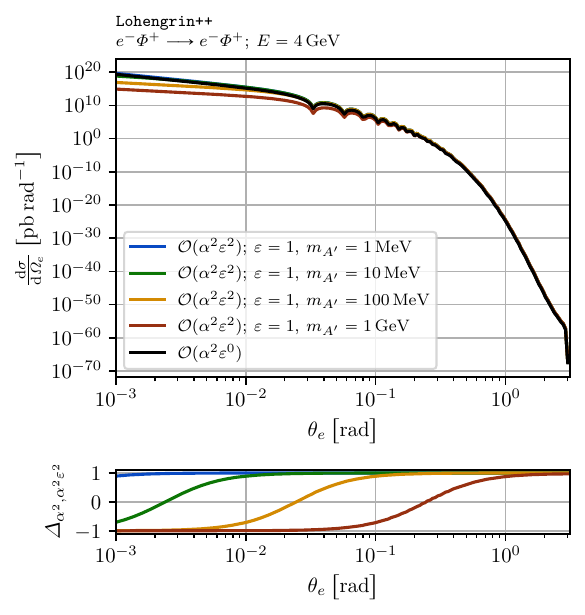}
    \caption{Top: Cross sections arising from $e^- \varPhi^+ \to e^- \varPhi^+$ at $\mathcal{O}(\alpha^2)$ and $\mathcal{O}(\alpha^2\eps^2)$ as function of the electron scattering angle $\theta_e$. Bottom: Relative deviation of the BSM contribution from the LO QED piece. We pick four different $m_{A^\prime}$ and set $\eps = 1$. Results presented here (and hereafter) were generated using \texttt{Lohengrin++}, \textit{cf.} Sec.~\ref{sec:amplitudes_tools}, as indicated in the top left corner.}
    \label{fig:dcs_2to2_interf_coh}
\end{figure}

\begin{figure}[ht]
    \centering
    \includegraphics[width=\linewidth]{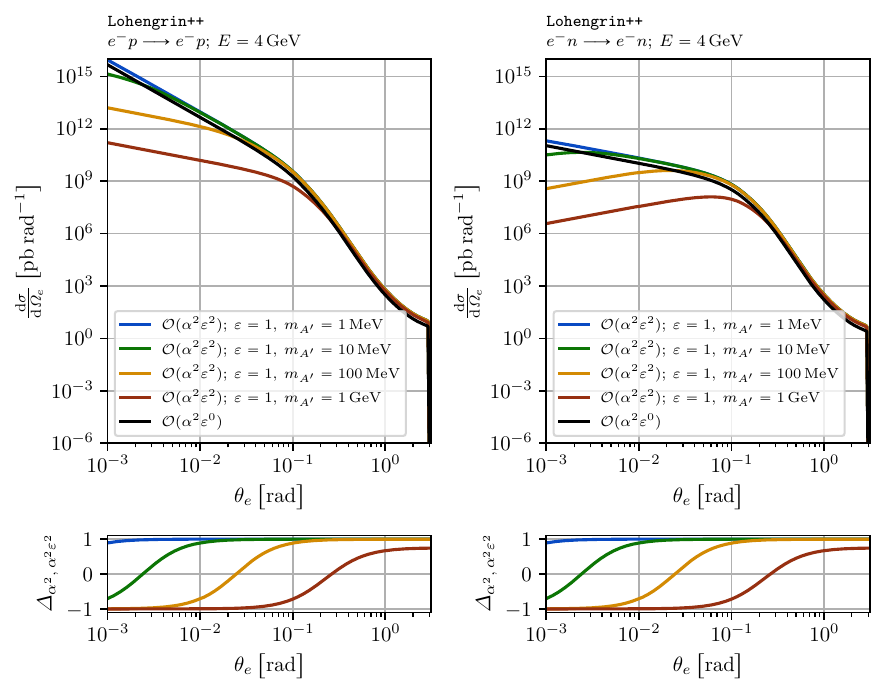}
    \caption{Top left (right): Cross sections arising from $e^- p \to e^- p$ ($e^- n \to e^- n$) at $\mathcal{O}(\alpha^2)$ and $\mathcal{O}(\alpha^2\eps^2)$ as function of the electron scattering angle $\theta_e$. Bottom left (right): Relative deviation of the BSM contribution from the LO QED proton (neutron) contribution. We consider four different $m_{A^\prime}$ and set $\eps = 1$.}
    \label{fig:dcs_2to2_interf_dif}
\end{figure}

Fig. \ref{fig:dcs_2to2_interf_coh} (Fig. \ref{fig:dcs_2to2_interf_dif}) 
shows the differential cross sections and $\varDelta_{\alpha^2, \, 
\alpha^2 \eps^2}$ for coherent (diffractive) scattering on tungsten 184, 
as a function of the electron scattering angle 
$\theta_e$. We picked a benchmark beam energy of $\SI{4}{\giga 
\electronvolt}$, representative for $\SI{}{\giga\electronvolt}$-scale 
experiments and show the dark photon interference term for several 
$m_{A^\prime}$ and $\eps = 1$. The qualitative behaviour, which we have deduced analytically, is also reproduced numerically, see especially the lower panels in 
Figs.~\ref{fig:dcs_2to2_interf_coh} and \ref{fig:dcs_2to2_interf_dif}.
We observe that for small angles, the LO cross section contribution scales as a power law, $\propto \theta_e^k$.
In this regime, the momentum transfer scales as $Q^2 \propto \theta_e^2$, such that for $Q^2 \ll m_{A^\prime}^2$, the interference term scales as $\propto \theta_e^{k+2}$.

For coherent scattering, we have $k \approx -4.19$, until the oscillating behaviour of the form factor shows for larger angles (which corresponds to larger $Q^2$). In diffractive scattering, the proton contribution has $k \approx -3$ and the neutron contribution $k \approx - 1$.
This implies that, in neutron scattering, the interference term does not monotonically decrease at small angles; instead, it increases proportionally to $\propto \theta_e$.

Consequently, the leading dark photon contribution would in fact cause a small bump in the angular spectrum that shows up for scattering angles corresponding to $Q^2 \approx m_{A^\prime}^{2}$.
The BSM contribution is of course still suppressed by $\eps^2$. To find this effect it would require a precision measurement of the elastic line as well as the knowledge of higher order QED corrections to control the SM background to the level of the numerical size of the leading dark photon contribution.
Dark photons in the mass range $\SIrange[]{10}{100}{\mega\electronvolt}$ 
are excluded by NA48/2 \cite{NA482:2015wmo} and BaBar \cite{BaBar:2014zli}, with an upper bound on the kinetic mixing parameter in the range $\eps \lesssim \SIrange{2e-4}{e-3}{}$ \cite{Fabbrichesi:2020wbt}.
These limits imply that the QED contribution must be known to an order of $n \approx 5$.
The $\text{LO}$ SM prediction begins at the level of $n = 2$ implying that a SM calculation at the $\text{N}^3\text{LO}$ level would be required have control over terms which, numerically, are expected to be of comparable size to the BSM contribution on the elastic line.

\subsubsection{One-Loop Level: Leptophilic Dark Photon}
\label{sec:2to2_results_leptophilic}

\begin{figure}
    \centering
    \includegraphics[width=0.75\linewidth]{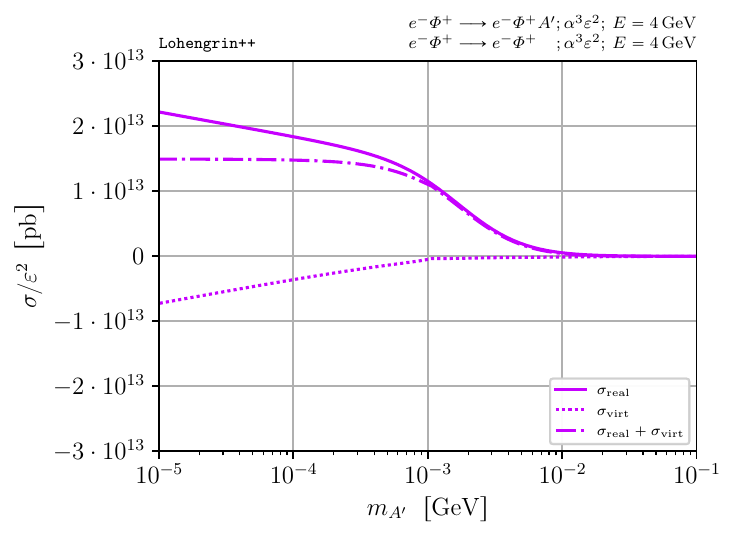}
    \caption{Real emission and virtual interference cross sections for a leptophilic dark photon $A^\prime$.}
    \label{fig:cs_mAD_real_virt}
\end{figure}

In the previous subsection, the discussion was focused on sensitivity of BSM searches to the $\mathcal{O}(\alpha^2 \eps^2)$ contribution in which a dark photon can couple to both leptonic and hadronic sectors.
At the next perturbative order in the electromagnetic coupling, \textit{i.e.} at $\mathcal{O}(\alpha^3 \eps^2)$, there are contributions to the cross-section which arise from both virtual and real-emission diagrams.

The discussion of the real-emission contributions which result in the signature of missing momentum will be the focus of the following Section.
However, it is also theoretically interesting to study the interplay between real and virtual corrections for the scattering processes when rather inclusive selections are applied---such as those outlined in Table~\ref{tab:theory_selections}. 
Under such inclusive selections (\textit{e.g.} the `Minimal Coulomb-safe selection' selections) both real and virtual corrections contribute to the cross-section, and it is necessary to include both to predict the expected event yield.

To perform such a study we have carried out the NLO calculation in the simplified case in which the dark photon is leptophilic and couples only to the electron line---essentially one has to consider the NLO corrections to the Bethe-Heitler component of the cross-section. 
The real emission contributions at $\mathcal{O}(\alpha^3 \eps^2)$ arise from the subset of diagrams displayed in Eq.\eqref{eq:ME2_2to3_DarkPhoton} where the dark photon couples directly to the lepton line, while the corresponding virtual corrections are given by the interference terms displayed in Eq.~\eqref{eq:ME2_2to2_NLO}.
Details on the renormalisation procedure required to perform the virtual correction are given in Appendix~\ref{app:renormalization_counterterms}.

To compare the impact of the real and virtual contributions, we consider the cross-section obtained after phase-space integration when applying only a minimal selection $\theta_e~>~\SI{1}{\milli\radian}$ on the outgoing electron.
The result of this comparison at a beam energy of $E=\SI{4}{\giga \electronvolt}$ is presented in Fig.~\ref{fig:cs_mAD_real_virt}, where both real and virtual contributions for coherent Bethe-Heitler scattering are shown. 
The predicted cross-sections are shown as a function of $m_{A^\prime}$, and in particular we take the limit $m_{A^\prime} \rightarrow 0$.

For values of $m_{A'} \lesssim \SI{1}{\mega\electronvolt}$ we observe a large positive (negative) logarithmic growth of the real (virtual) corrections, resulting from emissions of pseudo soft and/or collinear dark photon emissions.
The predicted cross-section at this order, obtained from summing both virtual and real corrections, remains almost constant in the limit of $m_{A^\prime} \rightarrow 0$ and the logarithmic corrections which depend on the dark photon mass cancel between the real and virtual contributions.
This (expected) cancellation  provides an important numerical check of these two parts of the calculation and their implementation in the Monte Carlo framework.
Furthermore, the calculation also indicates that for values of $m_{A^{\prime}}\lesssim \SI{1}{\mega \electronvolt}$ incorrect predictions for the event yield would be obtained if only the real emission corrections are considered.
Instead for larger mass values, as considered in~\cite{Izaguirre:2014bca,LDMX:2018cma}, the impact of the virtual corrections is relatively small compared to those of the real emission process.
It is also worth re-iterating that we have applied the `Minimal Couloumb-safe selections' as indicated in Table~\ref{tab:theory_selections} for this study. 
The presence of the minimal selection of the scattering angle is required to ensure that the cross-section prediction is integrable.
Finally, note that the physical electron mass is always kept throughout the calculations, such that the electron mass acts as a regulator for (pseudo) collinear dark photon emissions in the $m_{A^\prime} \rightarrow 0$ limit.

In the following Section we will focus on the kinematic selections summarised in \Cref{tab:theory_selections,tab:LDMX_selections,tab:LOH_selections} that include an upper bound on the electron energy.
Such a restriction forms the basis of the missing momentum search scenarios.

\section{Real (Dark) Photons: \texorpdfstring{$2 \to 3$}{TEXT} Processes}
\label{sec:2to3}

Various cross section contributions arise from the real $2\to 3$ 
scattering (dark) photon production when considering both the 
coherent and diffractive regimes. In this section, we show the most 
prominent features of the individual contributions and discuss their 
experimental interpretation in terms of signal and background.

\subsection{(Squared) Amplitudes}
\label{sec:2to3_amplitudes}
The following amplitudes contribute to real emission processes:
\begin{equation}
    i \mathcal{M}(e^- \had \to e^- \had \gamma ; \alpha^{3/2}) ~ = ~
        \begin{tikzpicture}[baseline=12.5]
        \setlength{\feynhandblobsize}{3mm}
        \setlength{\feynhandarrowsize}{5pt}
        \begin{feynhand}
        \vertex (ei) at (-1, 1);
        \vertex (eo) at (1, 1);
        \vertex (veg) at (0,1);
        \vertex (vedp) at (-0.5, 1);
        \vertex (dpo) at (0,1.4);
        \vertex [grayblob] (vhg) at (0,0) {};
        \vertex (hi) at (-1,0);
        \vertex (ho) at (1,0);
        \propag [fermion] (ei) to (vedp);
        \propag [fermion] (vedp) to (veg);
        \propag [fermion] (veg) to (eo);
        \propag [photon] (veg) to (vhg);
        \propag [photon] (vedp) to (dpo);
        \propag [double,double distance=0.3ex,thick] (hi) to (vhg);
        \propag [double,double distance=0.3ex,thick] (vhg) to (ho);
        \end{feynhand}
        \end{tikzpicture}
        ~+~
        \begin{tikzpicture}[baseline=12.5]
        \setlength{\feynhandblobsize}{3mm}
        \setlength{\feynhandarrowsize}{5pt}
        \begin{feynhand}
        \vertex (ei) at (-1, 1);
        \vertex (eo) at (1, 1);
        \vertex (veg) at (0,1);
        \vertex (vedp) at (0.5,1);
        \vertex (dpo) at (1,1.4);
        \vertex [grayblob] (vhg) at (0,0) {};
        \vertex (hi) at (-1,0);
        \vertex (ho) at (1,0);
        \propag [fermion] (ei) to (veg);
        \propag [fermion] (veg) to (vedp);
        \propag [fermion] (vedp) to (eo);
        \propag [photon] (veg) to (vhg);
        \propag [photon] (vedp) to (dpo);
        \propag [double,double distance=0.3ex,thick] (hi) to (vhg);
        \propag [double,double distance=0.3ex,thick] (vhg) to (ho);
        \end{feynhand}
        \end{tikzpicture}
        ~+~
        \begin{tikzpicture}[baseline=12.5]
        \setlength{\feynhandblobsize}{5mm}
        \setlength{\feynhandarrowsize}{5pt}
        \begin{feynhand}
        \vertex (ei) at (-1, 1);
        \vertex (eo) at (1, 1);
        \vertex (veg) at (0,1);
        \vertex (dpo) at (0.7,-0.4);
        \vertex [NEblob] (vhg) at (0,0) {};
        \vertex (hi) at (-1,0);
        \vertex (ho) at (1,0);
        \propag [fermion] (ei) to (veg);
        \propag [fermion] (veg) to (eo);
        \propag [photon] (veg) to (vhg);
        \propag [photon] (vhg) to (dpo);
        \propag [double,double distance=0.3ex,thick] (hi) to (vhg);
        \propag [double,double distance=0.3ex,thick] (vhg) to (ho);
        \end{feynhand}
        \end{tikzpicture}
        ~ ,
\end{equation}

\begin{equation}
    i \mathcal{M}(e^- \had \to e^- \had \ap ; \alpha^{3/2}\eps) ~ = ~
        \begin{tikzpicture}[baseline=12.5]
        \setlength{\feynhandblobsize}{3mm}
        \setlength{\feynhandarrowsize}{5pt}
        \begin{feynhand}
        \vertex (ei) at (-1, 1);
        \vertex (eo) at (1, 1);
        \vertex (veg) at (0,1);
        \vertex (vedp) at (-0.5, 1);
        \vertex (dpo) at (0,1.4);
        \vertex [grayblob] (vhg) at (0,0) {};
        \vertex (hi) at (-1,0);
        \vertex (ho) at (1,0);
        \propag [fermion] (ei) to (vedp);
        \propag [fermion] (vedp) to (veg);
        \propag [fermion] (veg) to (eo);
        \propag [photon] (veg) to (vhg);
        \propag [photon, color = crimson] (vedp) to (dpo);
        \propag [double,double distance=0.3ex,thick] (hi) to (vhg);
        \propag [double,double distance=0.3ex,thick] (vhg) to (ho);
        \end{feynhand}
        \end{tikzpicture}
        ~+~
        \begin{tikzpicture}[baseline=12.5]
        \setlength{\feynhandblobsize}{3mm}
        \setlength{\feynhandarrowsize}{5pt}
        \begin{feynhand}
        \vertex (ei) at (-1, 1);
        \vertex (eo) at (1, 1);
        \vertex (veg) at (0,1);
        \vertex (vedp) at (0.5,1);
        \vertex (dpo) at (1,1.4);
        \vertex [grayblob] (vhg) at (0,0) {};
        \vertex (hi) at (-1,0);
        \vertex (ho) at (1,0);
        \propag [fermion] (ei) to (veg);
        \propag [fermion] (veg) to (vedp);
        \propag [fermion] (vedp) to (eo);
        \propag [photon] (veg) to (vhg);
        \propag [photon, color = crimson] (vedp) to (dpo);
        \propag [double,double distance=0.3ex,thick] (hi) to (vhg);
        \propag [double,double distance=0.3ex,thick] (vhg) to (ho);
        \end{feynhand}
        \end{tikzpicture}
        ~+~
        \begin{tikzpicture}[baseline=12.5]
        \setlength{\feynhandblobsize}{5mm}
        \setlength{\feynhandarrowsize}{5pt}
        \begin{feynhand}
        \vertex (ei) at (-1, 1);
        \vertex (eo) at (1, 1);
        \vertex (veg) at (0,1);
        \vertex (dpo) at (0.7,-0.4);
        \vertex [NEblob] (vhg) at (0,0) {};
        \vertex (hi) at (-1,0);
        \vertex (ho) at (1,0);
        \propag [fermion] (ei) to (veg);
        \propag [fermion] (veg) to (eo);
        \propag [photon] (veg) to (vhg);
        \propag [photon, color=crimson] (vhg) to (dpo);
        \propag [double,double distance=0.3ex,thick] (hi) to (vhg);
        \propag [double,double distance=0.3ex,thick] (vhg) to (ho);
        \end{feynhand}
        \end{tikzpicture}
        ~ ,
\end{equation}

\begin{equation}
    i \mathcal{M}(e^- \had \to e^- \had \gamma ; \alpha^{3/2}\eps^2) ~ = ~
        \begin{tikzpicture}[baseline=12.5]
        \setlength{\feynhandblobsize}{3mm}
        \setlength{\feynhandarrowsize}{5pt}
        \begin{feynhand}
        \vertex (ei) at (-1, 1);
        \vertex (eo) at (1, 1);
        \vertex (veg) at (0,1);
        \vertex (vedp) at (-0.5, 1);
        \vertex (dpo) at (0,1.4);
        \vertex [grayblob] (vhg) at (0,0) {};
        \vertex (hi) at (-1,0);
        \vertex (ho) at (1,0);
        \propag [fermion] (ei) to (vedp);
        \propag [fermion] (vedp) to (veg);
        \propag [fermion] (veg) to (eo);
        \propag [photon, color = crimson] (veg) to (vhg);
        \propag [photon] (vedp) to (dpo);
        \propag [double,double distance=0.3ex,thick] (hi) to (vhg);
        \propag [double,double distance=0.3ex,thick] (vhg) to (ho);
        \end{feynhand}
        \end{tikzpicture}
        ~+~
        \begin{tikzpicture}[baseline=12.5]
        \setlength{\feynhandblobsize}{3mm}
        \setlength{\feynhandarrowsize}{5pt}
        \begin{feynhand}
        \vertex (ei) at (-1, 1);
        \vertex (eo) at (1, 1);
        \vertex (veg) at (0,1);
        \vertex (vedp) at (0.5,1);
        \vertex (dpo) at (1,1.4);
        \vertex [grayblob] (vhg) at (0,0) {};
        \vertex (hi) at (-1,0);
        \vertex (ho) at (1,0);
        \propag [fermion] (ei) to (veg);
        \propag [fermion] (veg) to (vedp);
        \propag [fermion] (vedp) to (eo);
        \propag [photon, color = crimson] (veg) to (vhg);
        \propag [photon] (vedp) to (dpo);
        \propag [double,double distance=0.3ex,thick] (hi) to (vhg);
        \propag [double,double distance=0.3ex,thick] (vhg) to (ho);
        \end{feynhand}
        \end{tikzpicture}
        ~+~
        \begin{tikzpicture}[baseline=12.5]
        \setlength{\feynhandblobsize}{5mm}
        \setlength{\feynhandarrowsize}{5pt}
        \begin{feynhand}
        \vertex (ei) at (-1, 1);
        \vertex (eo) at (1, 1);
        \vertex (veg) at (0,1);
        \vertex (dpo) at (0.7,-0.4);
        \vertex [NEblob] (vhg) at (0,0) {};
        \vertex (hi) at (-1,0);
        \vertex (ho) at (1,0);
        \propag [fermion] (ei) to (veg);
        \propag [fermion] (veg) to (eo);
        \propag [photon, color=crimson] (veg) to (vhg);
        \propag [photon] (vhg) to (dpo);
        \propag [double,double distance=0.3ex,thick] (hi) to (vhg);
        \propag [double,double distance=0.3ex,thick] (vhg) to (ho);
        \end{feynhand}
        \end{tikzpicture}
        ~ .
\end{equation}
The first two terms in each amplitude are also referred to as Bethe-Heitler diagrams, while the third is the VCS contribution.
The 4-point interactions $\begin{tikzpicture}[baseline=-2]
        \setlength{\feynhandblobsize}{5mm}
        \begin{feynhand}
        \vertex (veg) at (-0.4,0.4);
        \vertex (veg2) at (+0.4,0.4);
        \vertex [NEblob] (vhg) at (0,0) {};
        \vertex (hi) at (-0.4,0);
        \vertex (ho) at (0.4,0);
        \propag [photon] (veg) to (vhg);
        \propag [photon] (veg2) to (vhg);
        \propag [double,double distance=0.2ex,thick] (hi) to (vhg);
        \propag [double,double distance=0.2ex,thick] (vhg) to (ho);
        \end{feynhand}
        \end{tikzpicture}$
and
$\begin{tikzpicture}[baseline=-2]
        \setlength{\feynhandblobsize}{5mm}
        \begin{feynhand}
        \vertex (veg) at (-0.4,0.4);
        \vertex (veg2) at (+0.4,0.4);
        \vertex [NEblob] (vhg) at (0,0) {};
        \vertex (hi) at (-0.4,0);
        \vertex (ho) at (0.4,0);
        \propag [photon] (veg) to (vhg);
        \propag [photon, color=crimson] (veg2) to (vhg);
        \propag [double,double distance=0.2ex,thick] (hi) to (vhg);
        \propag [double,double distance=0.2ex,thick] (vhg) to (ho);
        \end{feynhand}
\end{tikzpicture}$
can be further expanded; the form of the VCS amplitude depends on the hadronic target and we restrict ourselves to the lowest order (Born) contributions reviewed in Sec.~\ref{sec:amplitudes_regimes}.

The squared amplitudes contributing to the real emission of a QED 
photon are:
\begin{equation}
    \abs{\mathcal{M}}^2(e^- \had \to e^- \had \gamma ; \alpha^{3}) = 
    \abs{~
    \begin{tikzpicture}[baseline=12.5]
        \setlength{\feynhandblobsize}{3mm}
        \setlength{\feynhandarrowsize}{5pt}
        \begin{feynhand}
        \vertex (ei) at (-1, 1);
        \vertex (eo) at (1, 1);
        \vertex (veg) at (0,1);
        \vertex (vedp) at (-0.5, 1);
        \vertex (dpo) at (0,1.4);
        \vertex [grayblob] (vhg) at (0,0) {};
        \vertex (hi) at (-1,0);
        \vertex (ho) at (1,0);
        \propag [fermion] (ei) to (vedp);
        \propag [fermion] (vedp) to (veg);
        \propag [fermion] (veg) to (eo);
        \propag [photon] (veg) to (vhg);
        \propag [photon] (vedp) to (dpo);
        \propag [double,double distance=0.3ex,thick] (hi) to (vhg);
        \propag [double,double distance=0.3ex,thick] (vhg) to (ho);
        \end{feynhand}
        \end{tikzpicture}
        ~+~
        \begin{tikzpicture}[baseline=12.5]
        \setlength{\feynhandblobsize}{3mm}
        \setlength{\feynhandarrowsize}{5pt}
        \begin{feynhand}
        \vertex (ei) at (-1, 1);
        \vertex (eo) at (1, 1);
        \vertex (veg) at (0,1);
        \vertex (vedp) at (0.5,1);
        \vertex (dpo) at (1,1.4);
        \vertex [grayblob] (vhg) at (0,0) {};
        \vertex (hi) at (-1,0);
        \vertex (ho) at (1,0);
        \propag [fermion] (ei) to (veg);
        \propag [fermion] (veg) to (vedp);
        \propag [fermion] (vedp) to (eo);
        \propag [photon] (veg) to (vhg);
        \propag [photon] (vedp) to (dpo);
        \propag [double,double distance=0.3ex,thick] (hi) to (vhg);
        \propag [double,double distance=0.3ex,thick] (vhg) to (ho);
        \end{feynhand}
        \end{tikzpicture}
        ~+~
        \begin{tikzpicture}[baseline=12.5]
        	\setlength{\feynhandblobsize}{5mm}
        	\setlength{\feynhandarrowsize}{5pt}
        	\begin{feynhand}
        		\vertex (ei) at (-1, 1);
        		\vertex (eo) at (1, 1);
        		\vertex (veg) at (0,1);
        		\vertex (dpo) at (0.7,-0.4);
        		\vertex [NEblob] (vhg) at (0,0) {};
        		\vertex (hi) at (-1,0);
        		\vertex (ho) at (1,0);
        		\propag [fermion] (ei) to (veg);
        		\propag [fermion] (veg) to (eo);
        		\propag [photon] (veg) to (vhg);
        		\propag [photon] (vhg) to (dpo);
        		\propag [double,double distance=0.3ex,thick] (hi) to (vhg);
        		\propag [double,double distance=0.3ex,thick] (vhg) to (ho);
        	\end{feynhand}
        \end{tikzpicture}
    ~}^2
        ~ ,
        \label{eq:ME2_2to3_Photon_QED}
\end{equation}
as well as,
\begin{align}
&\abs{\mathcal{M}}^2(e^- \had \to e^- \had \gamma ; \alpha^{3}\eps^2) \nonumber \\ 
&\qquad \qquad = 2\Re \left\{  \left(
\begin{tikzpicture}[baseline=12.5]
        \setlength{\feynhandblobsize}{3mm}
        \setlength{\feynhandarrowsize}{5pt}
        \begin{feynhand}
        \vertex (ei) at (-1, 1);
        \vertex (eo) at (1, 1);
        \vertex (veg) at (0,1);
        \vertex (vedp) at (-0.5, 1);
        \vertex (dpo) at (0,1.4);
        \vertex [grayblob] (vhg) at (0,0) {};
        \vertex (hi) at (-1,0);
        \vertex (ho) at (1,0);
        \propag [fermion] (ei) to (vedp);
        \propag [fermion] (vedp) to (veg);
        \propag [fermion] (veg) to (eo);
        \propag [photon] (veg) to (vhg);
        \propag [photon] (vedp) to (dpo);
        \propag [double,double distance=0.3ex,thick] (hi) to (vhg);
        \propag [double,double distance=0.3ex,thick] (vhg) to (ho);
        \end{feynhand}
        \end{tikzpicture}
        ~+~
        \begin{tikzpicture}[baseline=12.5]
        \setlength{\feynhandblobsize}{3mm}
        \setlength{\feynhandarrowsize}{5pt}
        \begin{feynhand}
        \vertex (ei) at (-1, 1);
        \vertex (eo) at (1, 1);
        \vertex (veg) at (0,1);
        \vertex (vedp) at (0.5,1);
        \vertex (dpo) at (1,1.4);
        \vertex [grayblob] (vhg) at (0,0) {};
        \vertex (hi) at (-1,0);
        \vertex (ho) at (1,0);
        \propag [fermion] (ei) to (veg);
        \propag [fermion] (veg) to (vedp);
        \propag [fermion] (vedp) to (eo);
        \propag [photon] (veg) to (vhg);
        \propag [photon] (vedp) to (dpo);
        \propag [double,double distance=0.3ex,thick] (hi) to (vhg);
        \propag [double,double distance=0.3ex,thick] (vhg) to (ho);
        \end{feynhand}
        \end{tikzpicture}
        ~+~
        \begin{tikzpicture}[baseline=12.5]
        \setlength{\feynhandblobsize}{5mm}
        \setlength{\feynhandarrowsize}{5pt}
        \begin{feynhand}
        \vertex (ei) at (-1, 1);
        \vertex (eo) at (1, 1);
        \vertex (veg) at (0,1);
        \vertex (dpo) at (0.7,-0.4);
        \vertex [NEblob] (vhg) at (0,0) {};
        \vertex (hi) at (-1,0);
        \vertex (ho) at (1,0);
        \propag [fermion] (ei) to (veg);
        \propag [fermion] (veg) to (eo);
        \propag [photon] (veg) to (vhg);
        \propag [photon] (vhg) to (dpo);
        \propag [double,double distance=0.3ex,thick] (hi) to (vhg);
        \propag [double,double distance=0.3ex,thick] (vhg) to (ho);
        \end{feynhand}
        \end{tikzpicture}
        \right)  \right. \nonumber \\
        &\qquad \qquad \qquad \qquad \times  \left. \left(
        \begin{tikzpicture}[baseline=12.5]
        \setlength{\feynhandblobsize}{3mm}
        \setlength{\feynhandarrowsize}{5pt}
        \begin{feynhand}
        \vertex (ei) at (-1, 1);
        \vertex (eo) at (1, 1);
        \vertex (veg) at (0,1);
        \vertex (vedp) at (-0.5, 1);
        \vertex (dpo) at (0,1.4);
        \vertex [grayblob] (vhg) at (0,0) {};
        \vertex (hi) at (-1,0);
        \vertex (ho) at (1,0);
        \propag [fermion] (ei) to (vedp);
        \propag [fermion] (vedp) to (veg);
        \propag [fermion] (veg) to (eo);
        \propag [photon, color=crimson] (veg) to (vhg);
        \propag [photon] (vedp) to (dpo);
        \propag [double,double distance=0.3ex,thick] (hi) to (vhg);
        \propag [double,double distance=0.3ex,thick] (vhg) to (ho);
        \end{feynhand}
        \end{tikzpicture}
        ~+~
        \begin{tikzpicture}[baseline=12.5]
        \setlength{\feynhandblobsize}{3mm}
        \setlength{\feynhandarrowsize}{5pt}
        \begin{feynhand}
        \vertex (ei) at (-1, 1);
        \vertex (eo) at (1, 1);
        \vertex (veg) at (0,1);
        \vertex (vedp) at (0.5,1);
        \vertex (dpo) at (1,1.4);
        \vertex [grayblob] (vhg) at (0,0) {};
        \vertex (hi) at (-1,0);
        \vertex (ho) at (1,0);
        \propag [fermion] (ei) to (veg);
        \propag [fermion] (veg) to (vedp);
        \propag [fermion] (vedp) to (eo);
        \propag [photon, color=crimson] (veg) to (vhg);
        \propag [photon] (vedp) to (dpo);
        \propag [double,double distance=0.3ex,thick] (hi) to (vhg);
        \propag [double,double distance=0.3ex,thick] (vhg) to (ho);
        \end{feynhand}
        \end{tikzpicture}
        ~+~
        \begin{tikzpicture}[baseline=12.5]
        \setlength{\feynhandblobsize}{5mm}
        \setlength{\feynhandarrowsize}{5pt}
        \begin{feynhand}
        \vertex (ei) at (-1, 1);
        \vertex (eo) at (1, 1);
        \vertex (veg) at (0,1);
        \vertex (dpo) at (0.7,-0.4);
        \vertex [NEblob] (vhg) at (0,0) {};
        \vertex (hi) at (-1,0);
        \vertex (ho) at (1,0);
        \propag [fermion] (ei) to (veg);
        \propag [fermion] (veg) to (eo);
        \propag [photon, color=crimson] (veg) to (vhg);
        \propag [photon] (vhg) to (dpo);
        \propag [double,double distance=0.3ex,thick] (hi) to (vhg);
        \propag [double,double distance=0.3ex,thick] (vhg) to (ho);
        \end{feynhand}
        \end{tikzpicture}
        \right)^{\scaleto{\dagger}{10pt}}
        \right\}
        ~ .
        \label{eq:ME2_2to3_Photon_QED+D}
\end{align}
The first order contribution to Dark photon production is described by the contribution:
\begin{equation}
    \abs{\mathcal{M}}^2(e^- \had \to e^- \had \ap ; \alpha^{3}\eps^2) = 
    \abs{~
    \begin{tikzpicture}[baseline=12.5]
        \setlength{\feynhandblobsize}{3mm}
        \setlength{\feynhandarrowsize}{5pt}
        \begin{feynhand}
        \vertex (ei) at (-1, 1);
        \vertex (eo) at (1, 1);
        \vertex (veg) at (0,1);
        \vertex (vedp) at (-0.5, 1);
        \vertex (dpo) at (0,1.4);
        \vertex [grayblob] (vhg) at (0,0) {};
        \vertex (hi) at (-1,0);
        \vertex (ho) at (1,0);
        \propag [fermion] (ei) to (vedp);
        \propag [fermion] (vedp) to (veg);
        \propag [fermion] (veg) to (eo);
        \propag [photon] (veg) to (vhg);
        \propag [photon, color=crimson] (vedp) to (dpo);
        \propag [double,double distance=0.3ex,thick] (hi) to (vhg);
        \propag [double,double distance=0.3ex,thick] (vhg) to (ho);
        \end{feynhand}
        \end{tikzpicture}
        ~+~
        \begin{tikzpicture}[baseline=12.5]
        \setlength{\feynhandblobsize}{3mm}
        \setlength{\feynhandarrowsize}{5pt}
        \begin{feynhand}
        \vertex (ei) at (-1, 1);
        \vertex (eo) at (1, 1);
        \vertex (veg) at (0,1);
        \vertex (vedp) at (0.5,1);
        \vertex (dpo) at (1,1.4);
        \vertex [grayblob] (vhg) at (0,0) {};
        \vertex (hi) at (-1,0);
        \vertex (ho) at (1,0);
        \propag [fermion] (ei) to (veg);
        \propag [fermion] (veg) to (vedp);
        \propag [fermion] (vedp) to (eo);
        \propag [photon] (veg) to (vhg);
        \propag [photon, color=crimson] (vedp) to (dpo);
        \propag [double,double distance=0.3ex,thick] (hi) to (vhg);
        \propag [double,double distance=0.3ex,thick] (vhg) to (ho);
        \end{feynhand}
        \end{tikzpicture}
        ~+~
        \begin{tikzpicture}[baseline=12.5]
        \setlength{\feynhandblobsize}{5mm}
        \setlength{\feynhandarrowsize}{5pt}
        \begin{feynhand}
        \vertex (ei) at (-1, 1);
        \vertex (eo) at (1, 1);
        \vertex (veg) at (0,1);
        \vertex (dpo) at (0.7,-0.4);
        \vertex [NEblob] (vhg) at (0,0) {};
        \vertex (hi) at (-1,0);
        \vertex (ho) at (1,0);
        \propag [fermion] (ei) to (veg);
        \propag [fermion] (veg) to (eo);
        \propag [photon] (veg) to (vhg);
        \propag [photon, color=crimson] (vhg) to (dpo);
        \propag [double,double distance=0.3ex,thick] (hi) to (vhg);
        \propag [double,double distance=0.3ex,thick] (vhg) to (ho);
        \end{feynhand}
        \end{tikzpicture}
    ~}^2
        ~ .
        \label{eq:ME2_2to3_DarkPhoton}
\end{equation}
Note that the additional terms in Eq. \eqref{eq:ME2_2to3_Photon_QED+D} contribute to on-shell QED photon production, but are of the same order in perturbation theory as the squared amplitude describing real dark photon production, Eq.~\eqref{eq:ME2_2to3_DarkPhoton}.

\subsection{Dark Photon Production}
\label{sec:DP_production}
Real dark photons are produced at $\mathcal{O}(\alpha^3\eps^2)$ from the squared amplitude in Eq.~\eqref{eq:ME2_2to3_DarkPhoton}.
In the following we discuss the total cross section and kinematic distributions of the missing momentum signal induced by real dark photon emission in the different scattering regimes.
We consider two experimental scenarios, namely LDMX at $\SI{8}{\giga\electronvolt}$ and \textsc{Lohengrin} at $\SI{3.2}{\giga\electronvolt}$.

\subsubsection{Integrated Signal Yield}
\label{sec:DP_signal_yield}
\begin{figure}[ht]
    \centering
    \includegraphics[width=0.8\linewidth]{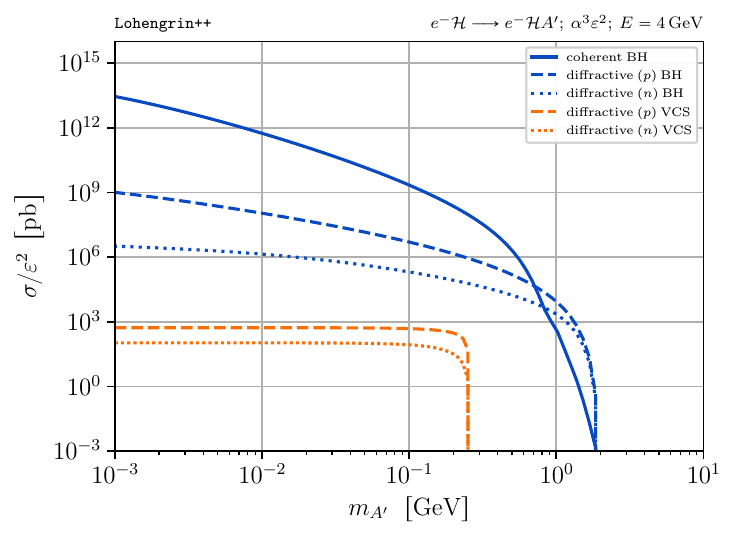}
    \caption{Real dark photon production cross section as function of the dark photon mass for the different contributing hadronic currents. We integrate over the ``minimal selection" phase space specified in Table~\ref{tab:theory_selections} and show for diffractive scattering the BHVCS and VCS results independently.}
    \label{fig:eH>eHAp_minimal_selection}
\end{figure}

\paragraph{Minimal Real Emission Selections.}
Fig.~\ref{fig:eH>eHAp_minimal_selection} shows the dark photon production cross section as a function of the ${A'}$ mass for a beam energy of $\SI{4}{\giga\electronvolt}$, representative for the planned $\mathcal{O}(\SI{}{\giga\electronvolt})$ experiments.
We consider the minimal real emission selections of the final state electrons, displayed in Table~\ref{tab:theory_selections}, that remove all virtual $\mathcal{O}(\alpha^3\eps^2)$ contributions which would otherwise be required for a physically consistent inclusive observable in the missing momentum search approach.
Although this minimal selection is not directly relevant for the experimental proposals, we present the results in order to illustrate and discuss the general properties of the cross sections.

We present the BH (blue curves) and VCS contributions (orange curves) individually.
The combined cross section result would in general also include interference terms.

At low $m_{A'}$, the VCS contributions are small compared to the BH scattering, for a number of reasons.
In general, the relevant photon propagator at the amplitude level scales as $\propto 1/Q^2_\vcs$, which favors phase-space regions with low $Q^2_\vcs$, \textit{i.e.} configurations where the electron remains close to the beam direction and carries high energy. 
However, this portion of phase space is not important for missing momentum searches and is already cut away by the minimal real emission selections, causing a general suppression of the VCS contributions.
This effect will get even stronger when probing phase space regions with very small electron momenta exclusively, which is the primary target of the missing momentum search strategy.
In the dark photon mass range of interest for the missing momentum search approach, $\SI{1}{\mega\electronvolt} \lesssim m_{A'} \lesssim \SI{100}{\mega\electronvolt}$, we always find a general photon propagator suppression of the VCS contributions compared to BH scattering, since $(Q^2_{\vcs})^2 \gg (Q^2_{\bh; \text{min}})^2$ in the phase space portion characterized by small electron momenta, \textit{cf.} Eqs. \eqref{eq:Q2limits} and \eqref{eq:Q2vcs_labframe}.

In the displayed dark photon mass range, the coherent VCS contributions are vanishing due to the resonance cutoff in Eq.~\eqref{eq:coh_res_cut}.
As we mentioned earlier, probing larger virtualities for a recoiling nucleus is associated with a lot of uncertainties and unknowns, especially in the required form factors.
Therefore, we cannot provide any reliable cross section prediction for contributions in and beyond the resonance regime.

In diffractive scattering, the VCS contributions are also small.
In addition to the reduction caused by the photon propagator, the nucleon propagator introduces inverse powers of the nucleon mass, leading to an additional suppression of the VCS contribution relative to the BH terms.
The VCS results presented here are furthermore limited by the resonance cutoff (
\textit{cf.}~Sec.~\ref{sec:physics_input}), which also constrains the VCS contribution to dark photon masses $\lesssim \SI{260}{\mega\electronvolt}$.

We can briefly comment on the behavior of the diffractive cross section if we also allow for large virtualities, \textit{i.e.} if the restriction in Eq.~\eqref{eq:dif_res_cut} is not enforced.
The VCS contribution is then of $\mathcal{O}(\SI{e6}{\pico\barn})$ for dark photons in the mass range relevant for the missing momentum search strategy, \textit{i.e.} $\SIrange{1}{100}{\mega\electronvolt}$.
The cross section would receive further damping through a phenomenological form factor, which suppresses large virtualities, but comes with model uncertainties.
Compared to BH scattering, the VCS contribution is then still highly suppressed by at least three orders of magnitude.
Including the excitation of nucleon resonances would in general lead to an enhancement of the cross section, but such contributions would produce much more complicated final states, for example through the production of pions.
A proper classification of signal and background events in the presence of nucleon resonances would require a dedicated detailed study.
For larger dark photon masses in the range $\SIrange{0.1}{1}{\giga\electronvolt}$, the cross section is enhanced by the vector meson resonances in the timelike form factors $F_{1,2}^{N}(m_{A'}^2)$.
However, we found that these contributions are dominated by configurations with high outgoing nucleon momenta of $\mathcal{O}(\SI{100}{\mega\electronvolt})$ in the final state.
We expect that such high momenta cause nucleon ejections and thus additional radiation in the detectors, rendering such events incompatible with a solitary electron signature.
Nevertheless, a dedicated and detailed study of the large virtuality regime in nucleon bremsstrahlung in the context of LDMX or \textsc{Lohengrin} could lead to the development of new search strategies within the newly proposed experimental setups, especially for larger dark photon masses $\gtrsim \SI{100}{\mega\electronvolt}$.

In the following, we want to focus only on the missing momentum search strategy and thus we will be considering only Bethe-Heitler scattering in the remaining sections.
So, let us briefly comment on the general behaviour of the coherent and diffractive contributions in this case.

In the BH case, the coherent contribution dominates the dark photon production except for
large dark photon masses. Above $m_{A'} \approx \SI{700}{\mega 
\electronvolt}$ the diffractive contribution takes over, since the 
production of heavy dark photons requires large momentum transfers, and 
in this regime the diffractive contribution dominates over the coherent 
one due to the form factors. In diffractive scattering we observe a clear
hierarchy: the proton contribution is roughly two orders of magnitude 
larger than the neutron contribution, because the former receives most 
contributions from small momentum transfers, which are suppressed in the 
neutron case due to the form factors. The difference diminishes for 
heavier dark photons, since (as stated before) their production requires 
larger momentum transfers, where the neutron form factors become numerically comparable to the proton ones.

\begin{figure}[ht]
    \centering
    \includegraphics[width=0.49\linewidth]{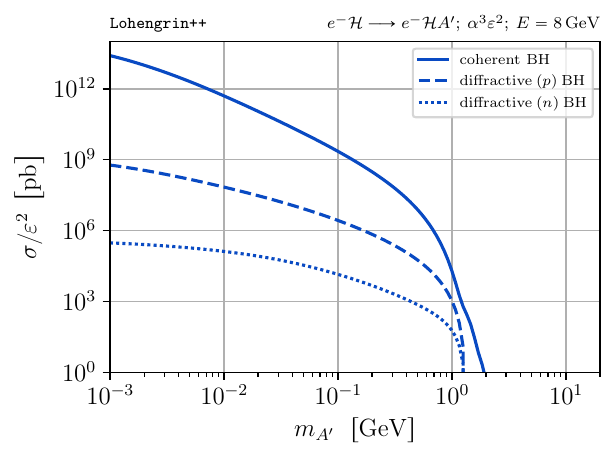}
    \includegraphics[width=0.49\linewidth]{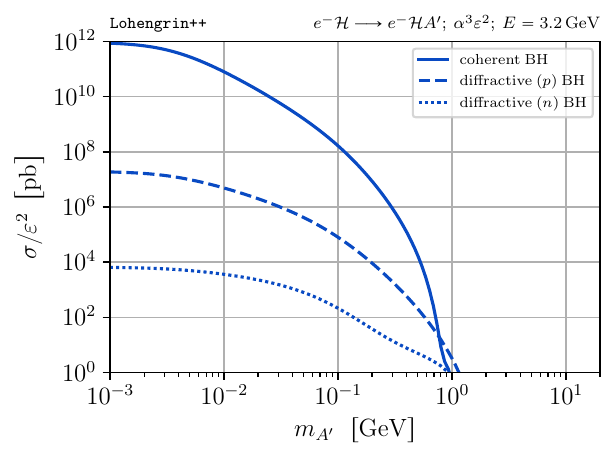}
    \caption{Real dark photon production cross sections as function of the dark photon mass for the different contributing hadronic currents. We apply final state selections corresponding to the LDMX trigger with LDMX HCAL veto (left) and to \textsc{Lohengrin} SR1 with \textsc{Lohengrin} HCAL veto (right). See 
    Tables~\ref{tab:LDMX_selections} and \ref{tab:LOH_selections} for details on the selections.}
    \label{fig:eH>eHAp_LDMX+Loh}
\end{figure}

\paragraph{Experimental Final State Selections.}
{}{In} Fig.~\ref{fig:eH>eHAp_LDMX+Loh} we present again the 
dark photon yield, but for different experimental setups and final state
selections. The left panel shows the dark photon yield for the LDMX 
$\SI{8}{\giga\electronvolt}$ scenario. We apply the final state selections in \Cref{tab:LDMX_selections}, such that we integrate over measurable electrons below a missing momentum trigger.
In addition to the electron selections, we consider (in diffractive scattering) a selection rejecting nucleons with high momenta directed into the HCAL volume, \textit{cf.} Sec. \ref{sec:phasespace_measurable}, such that we only integrate over solitary electrons.
Practically, since the electrons are 
required to have a relatively low momentum, the momentum transfers to 
the hadronic system are always large, such that the cross section results are not very sensitive to the exact choice of momentum threshold.
We found no significant change in 
the relevant cross sections when varying the threshold between 
$\mathcal{O}(\SI{10}{\mega\electronvolt})$ and $\mathcal{O}(\SI{100} 
{\mega\electronvolt})$.
In coherent scattering, we make the 
natural assumption that no hadronic activity is produced.

In this approach we assume that the HCAL is used solely as a veto to discard events with any hadronic activity in the detector.
Alternatively, an accurate reconstruction of the energy deposit in the HCAL would allow for a missing momentum analysis based on the total four-momentum of all final state particles.
However, such an approach would require a reliable simulation of all possible hadronic final states (including also additional signal contributions from virtual Compton scattering at large virtuality or invisible decays of produced pions, $\pi^0 \rightarrow \gamma A'$), which is beyond the scope of this work.
So we restrict ourselves to a `clean' signal, \textit{i.e.} solitary electrons.

As Fig.~\ref{fig:eH>eHAp_LDMX+Loh} (left) shows, the nucleon contributions are negligible in the LDMX scenario with the described assumptions.
The main reason for the suppression of the diffractive contributions is the wide-angle HCAL veto.

For \textsc{Lohengrin}, we consider the setup proposed
in Ref.~\cite{Bechtle:2024atq} with a $\SI{3.2}{\giga 
\electronvolt}$ beam.
The final state selections are summarized in 
Table~\ref{tab:LOH_selections}.
We require the electron to be located in
the proposed signal region SR1 and treat the diffractive contributions as in the LDMX case described above.
Fig.~\ref{fig:eH>eHAp_LDMX+Loh} (right) shows 
the results for the \textsc{Lohengrin} scenario: for relatively heavy dark photons, $m_{A'} 
\approx \SI{1}{\giga\electronvolt}$, we observe a slight increase of 
missing momentum events in the signal region due to the diffractive contribution.

Note that the left and right panels of Fig.~\ref{fig:eH>eHAp_LDMX+Loh} 
are not directly comparable. For LDMX, we only integrated over events below 
the missing energy trigger.
For \textsc{Lohengrin}, a simple cut and count analysis was used in the 
first design steps, such that we can straightforwardly study the impact of any extra contribution, \textit{e.g.}
diffractive dark photon production, in the proposed signal region.

\subsubsection{Differential Signal Distributions}
\label{sec:DP_distributions}
\begin{figure}[ht]
	\centering
	\includegraphics[width=0.98\textwidth]{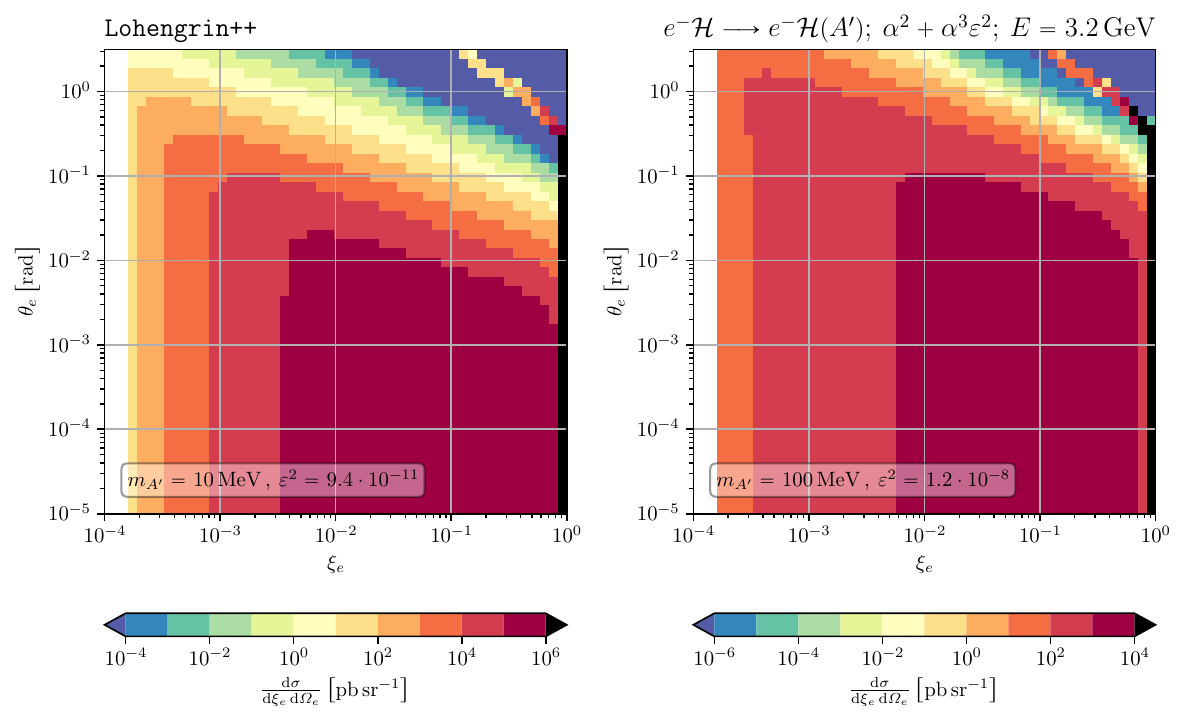}
	\caption[]{Double differential cross section w.r.t. the energy fraction $\xi_e$ and solid angle $\varOmega_e$ of the recoiling electron taking into account $\mathcal{O}(\alpha^2)$ LO and $\mathcal{O}(\alpha^3\eps^2)$ NLO contributions in coherent and diffractive scattering. We set a beam energy of $E = \SI{3.2}{\giga\electronvolt}$ and show two benchmark points defined by different choices of dark photon mass $m_{A^\prime}$ and squared kinetic mixing strength $\varepsilon$.}
	\label{fig:ddcs_sph_cohdif}
\end{figure}

In Fig.~\ref{fig:ddcs_sph_cohdif} we show the double differential 
cross section for the reaction $e^- 
\had \rightarrow e^- \had (A')$, w.r.t.~the energy fraction $\xi_e$ and
scattering angle $\theta_e$ of the outgoing electron. We consider only 
the dominant case of Bethe-Heitler scattering (with a leptophilic dark 
photon). We take into account both $A'$ real emission and one-loop 
interference effects. In total we display the differential quantity,
\begin{align}
    \frac{\d^2 \sigma}{\d\xi_e \d\varOmega_e} &= \frac{\d^2\sigma_\text{LO}}{\d\xi_e \d\varOmega_e}(e^- \had \rightarrow e^- \had; \alpha^2) \nonumber \\
    &\quad + \frac{\d^2\sigma_\text{NLO, real}}{\d\xi_e \d\varOmega_e}(e^- \had \rightarrow e^- \had A'; \alpha^3\eps^2) + \frac{\d^2\sigma_\text{NLO, virt}}{\d\xi_e \d\varOmega_e}(e^- \had \rightarrow e^- \had; \alpha^3\eps^2).
\end{align}
\begin{table}
    \setlength{\tabcolsep}{5pt}
    \centering
    \begin{tabular}{ccc}
        \toprule \toprule 
        $m_{A'}$ & $\eps^2$  & Comment \\ \midrule
        $\SI{10}{\mega\electronvolt}$ & $\sim 9.4\cdot 10^{-11}$ & \multirow{2}{*}{Scalar thermal relic target, \textit{cf.}~Ref.~\cite{Battaglieri:2017aum}}\\ 
        $\SI{100}{\mega\electronvolt}$ & $\sim 1.2\cdot 10^{-8}$ & \\ 
        \bottomrule\bottomrule
    \end{tabular}
    \caption{Minimal values of $\eps^2$ motivated by the scalar thermal relic targets for $m_{A'} = \SI{10}{\mega\electronvolt}$ and $\SI{100}{\mega\electronvolt}$.}
    \label{tab:eps2}
\end{table}

We set the beam energy to $\SI{3.2}{\giga\electronvolt}$, as 
in the \textsc{Lohengrin} scenario, and consider two
dark photon masses, $m_{A^\prime} = \SI{10}{\mega 
\electronvolt}$ and $\SI{100}{\mega 
\electronvolt}$. Since we include the 
LO piece, we also fix physically motivated values for $\eps^2$. We use the minimal values suggested 
by the scalar thermal relic target, which we obtain from Ref.~\cite{Battaglieri:2017aum}.
This assumes a benchmark model for the
dark sector: the dark matter current $J^\mu_{\text{\tiny{D}}}$ consists
of a scalar field $\chi$ with the mass ratio $m_{\chi}/ 
m_{A'}=1/3$ and the dark fine structure constant is set to $\alpha_ 
{\text{\tiny{D}}} = g_\text{\tiny{D}}^2/4\pi = 0.5$.
The values for $\eps^2$ are also listed in Table~\ref{tab:eps2}.

The distributions presented in Fig.~\ref{fig:ddcs_sph_cohdif} show the predicted behaviour of the signal electrons in a detailed and general form, since we bin in both the energy fraction and scattering angle of the outgoing electron.
In general, we observe the trend in the double differential distributions
that the maximum of the cross section shifts from high energies and 
narrow angles to low energies and wide angles when increasing the dark 
photon mass from $\SI{1}{\mega\electronvolt}$ to $\SI{1}{\giga\electronvolt}$. In consequence, the differential cross section is maximal 
for roughly the same values of transverse momentum $p'_{T}\equiv 
\sqrt{(p'_1)^2 + (p'_2)^2}$ of the recoiling electron, since lines of constant $p'_{T}$ appear 
approximately as diagonals in the presented double logarithmic phase 
space. We 
found that the differential cross section peaks at $p'_T \sim \SI{1}{\mega\electronvolt}$ if the dark photon has a mass in the range $\sim \SIrange{1}{50}{\mega\electronvolt}$.
The described behaviour of the real emission cross section as function of the lab frame variables $\xi_e, \theta_e$ is mostly determined by the behaviour of $Q^2_{\bh;\text{min}}(\xi_e, \theta_e)$, \textit{cf.} Eq.~\eqref{eq:Q2limits}, since the cross section receives most contributions from small momentum transfers as the amplitude scales as $\propto 1/Q^2_{\bh}$ due to the photon propagator.

Besides adding the diffractive elastic line, the
scattering on the nuclear constituents only mildly alters the
distributions in the large $\theta_e$ regime for heavy masses, here visible for $m_{A'} = \SI{100} 
{\mega\electronvolt}$. The same distribution taking only the coherent scattering into account can be found in Fig.~\ref{fig:ddcs_sph_coh} in 
App.~\ref{app:supp_figures} for comparison.

\subsection{QED Photon Production}
\label{sec:photon_production}
A detailed understanding of the scattering processes that result in the emission of one or more real QED photons is also critical for the missing momentum search strategy.
The leading-order prediction for this process can be obtained from the squared amplitude Eq.~\eqref{eq:ME2_2to3_Photon_QED} and at $\mathcal{O} 
(\alpha^3\eps^2)$ by the interference terms in Eq.~\eqref{eq:ME2_2to3_Photon_QED+D}.
Before providing numerical results for this scattering process, let us make some kinematic statements about this process.

Generally, this scattering process will be dominated by kinematic configurations that involve the emission of either a soft or a (pseudo)collinear photon with respect to the incoming/outgoing electron.
Additionally, the scattering rate is also enhanced for small momentum transfers (\textit{i.e.} $Q^2_\bh\to0$) due to the presence of the $t$-channel photon exchange.
The dominant contribution arises from the BH scattering subprocess, and typically the electron-photon system is highly boosted (in the lab frame) leading to a signature in which a hard photon is emitted in the forward direction.

The kinematic requirement that the scattered electron carries only a small energy fraction (\textit{e.g.} $\xi_e \lesssim 0.5$) in combination with a veto on the presence of hard photons (as measured by an ECAL) would ensure that this background can be entirely neglected.
However, in practice, the kinematic coverage of an implemented ECAL will not be complete. 
This means that the process $e^- \had \rightarrow e^- \had \gamma$ can still mimic the missing momentum signal if the emitted photon is hard (carries a substantial energy fraction) and avoids detection---\textit{e.g.} if the photon is emitted at wide-angle.
As such, the BH subprocess is a design-driving background process for the novel missing momentum experiments.

While missing QED photons produced at $\mathcal{O}(\alpha^3)$ through Eq.~\eqref{eq:ME2_2to3_Photon_QED} contribute to background events, the contribution at $\mathcal{O}(\alpha^3\eps^2)$, 
Eq.~\eqref{eq:ME2_2to3_Photon_QED+D}, should in fact be treated as signal, because it increases the number of missing momentum events at the same order in perturbation theory as the real emission process $e^- \had \rightarrow e^- \had A^\prime$.
In theory, it should be possible to disentangle both missing momentum contributions sourced by 
Eqs.~\eqref{eq:ME2_2to3_Photon_QED+D}--\eqref{eq:ME2_2to3_DarkPhoton} from the pure QED background through their distribution in phase space.
This is of course straightforward for real $A^\prime$ emission, since it causes a bump in the electron spectrum at low momenta, as demonstrated in Sec.~\ref{sec:DP_distributions}.
On the contrary, the scaling of the interference term w.r.t. kinematic variables, relative to the pure QED piece, is determined by the behaviour of the dark photon propagator w.r.t. $Q^2_\bh$, in a similar sense as in Sec. \ref{sec:2to2_results_vanilla}.
Consequently, we expect the dark photon interference effects to appear in a much more subtle way as compared to the dark photon real emission.

In the following, we discuss the background and signal originating from single-photon emission by presenting integrated cross sections and differential distributions.
The geometric parameter that controls the veto power -- specifically, the sizes of missing momentum cross sections -- is the solid angle covered by the ECAL.
For simplicity, we assume an ECAL that is symmetric around the beam axis, with its solid angle coverage defined solely by its polar opening angle, $\theta_\text{ECAL}$, which we treat as a free parameter.

\subsubsection{Integrated Background Yield}
\label{sec:photon_background_yield}
\begin{figure}[ht]
    \centering
    \includegraphics[width=0.49\linewidth]{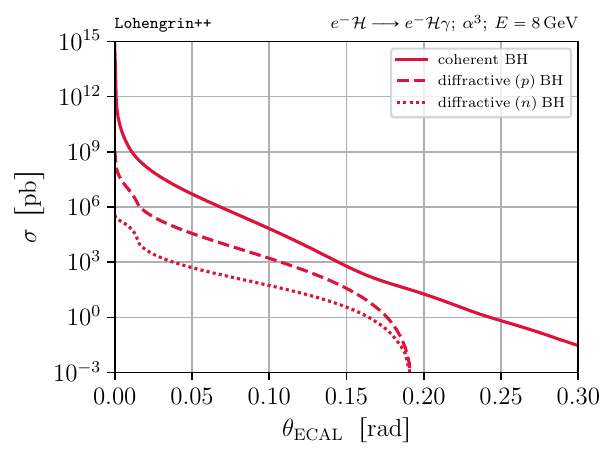}
    \includegraphics[width=0.49\linewidth]{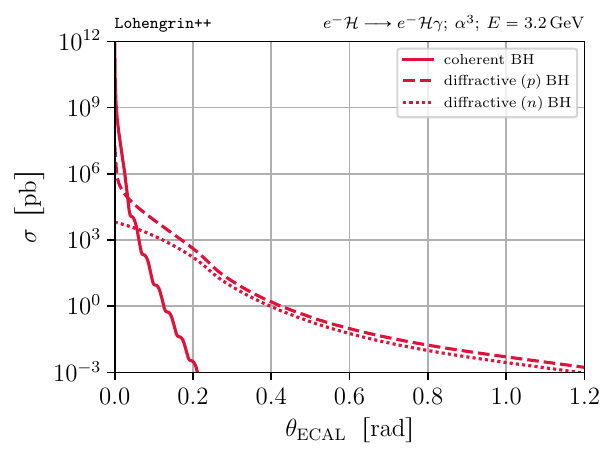}
    \caption{Real QED photon production cross sections as function of the polar ECAL opening angle $\theta_\text{ECAL}$ for the different contributing hadronic currents. We apply final state selections corresponding to LDMX (left), see Table \ref{tab:LDMX_selections}, and \textsc{Lohengrin} (right), see Table~\ref{tab:LOH_selections}.
    }
    \label{fig:eH>eHg_LDMX+Loh}
\end{figure}

Fig.~\ref{fig:eH>eHg_LDMX+Loh} shows the total background yield in terms
of the polar angle $\theta_\text{ECAL}$ covered by the ECAL. The left 
panel of Fig.~\ref{fig:eH>eHg_LDMX+Loh} shows the LDMX $\SI{8}{\giga 
\electronvolt}$ scenario with the selections in 
Table~\ref{tab:LDMX_selections} (such that we again only integrate over solitary, low-energy electrons).
Compared to the coherent contribution, the diffractive channel does not make a significant impact, again because of the aggressive HCAL veto.

The right panel of Fig.~\ref{fig:eH>eHg_LDMX+Loh} shows the 
\textsc{Lohengrin} $\SI{3.2}{\giga\electronvolt}$ scenario with 
selections as given in Table~\ref{tab:LOH_selections}. Let us first 
try to understand the oscillation behaviour 
of the coherent contribution. In the \textsc{Lohengrin} signal region 
the outgoing electron has a very small momentum, such that phase space regions with relatively high 
momentum transfers are probed. Since the BH momentum transfer, $Q^2_\bh = 
-(p - p^\prime -k)^2$, is sensitive to both the final state selections 
of the electron and photon, increasing $\theta_\text{ECAL}$ raises the 
lower boundary of the contributing $Q^2_\bh$ range, causing the window 
of allowed momentum transfers to shrink. Hence, the missing momentum cross
section as a function of $\theta_\text{ECAL}$ is sensitive to the 
behavior of the tail of the sampled $Q^2_\bh$ distribution. As a result,
the coherent contribution drops rapidly with $\theta_\text{ECAL}$ and 
exhibits the oscillatory behavior of the elastic form factor, which 
is manifest in the high $Q^2_\bh$ regime.

In the \textsc{Lohengrin} scenario, the diffractive scattering does
not drop as rapidly as in the LDMX case.
We find that the nucleons are dominantly scattered to wide angles that are not covered by an HCAL with the proposed geometry, such that the cross section suppression due to the HCAL veto is not as strong as in the LDMX case.
The diffractive scattering dominates over the
coherent scattering for $\theta_\text{ECAL} \gtrsim \SI{0.03}{\radian}$, because (as stated previously) for increasing $\theta_\text{ECAL}$, the high momentum transfer regime is probed, where the diffractive scattering dominates.

The polar angle extension of the proposed 
\textsc{Lohengrin} ECAL corresponds 
to a relatively small solid angle coverage of $\theta_\text{ECAL}
\approx \SI{0.07}{\radian}$, since then the primary beam electrons are 
bent away from it by the magnetic field. At this opening 
angle, we observe that the diffractive contribution increases the 
missing momentum background significantly (\textit{cf.}~Fig.~\ref{fig:eH>eHg_LDMX+Loh} (right)). We found that this can be 
circumvented by extending the coverage of the HCAL, since then more 
nucleon ejections can be vetoed. An independent study for the 
\textsc{Lohengrin} project using the simulation program 
\texttt{FLUKA} also implies that it is necessary to 
extend the HCAL. \texttt{FLUKA} provides a much more
accurate estimate of hadronic activity than our naive nucleon-ejection 
estimate with a momentum threshold, but it leads to a 
similar conclusion as our simple treatment.
The results of the \texttt{FLUKA} study will be published at a later date.

\subsubsection{Integrated Signal Yield}
\label{sec:photon_signal_yield}
As mentioned earlier, the missing momentum cross section at $\mathcal{O}(\alpha^3\eps^2)$ is not only sourced by $e^- \had \rightarrow e^- \had A^\prime$, but also by interference terms contributing to $e^- \had \rightarrow e^- \had \gamma$ when the photon escapes detection.

While real $A^\prime$ emission is kinematically impossible for dark photon masses 
$m_{A^\prime} \geq E_{\text{\tiny{CM}}}-m_e- m_\had$,
virtual dark photons of much higher mass can contribute to the  missing momentum cross section via the 
interference terms. We do not give explicit numbers here for dark 
photon masses approaching the EW scale, because this would imply 
significant EW mixing effects, see 
Secs.~\ref{sec:dark_photon_basics} and 
\ref{sec:amplitudes_regimes_diffractive}. This would cause an alteration
of the vertex functions and form factors used in the calculations.
However, we include the case $m_\ap = \SI{10}{\giga\electronvolt}$ as the condition for suppressed EW mixing effects, $m_\ap^2\approx m_{A_\text{D}}^2 \ll m_Z^2$, is still fulfilled.

\begin{figure}[ht]
    \centering
    \includegraphics[width=0.7\linewidth]{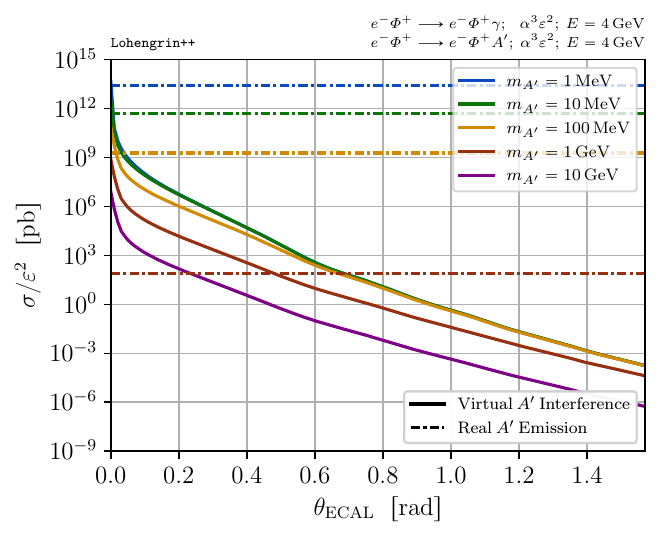}
    \caption{Missing energy cross sections arising from $e^- \varPhi^+ \to e^- \varPhi^+ A^\prime$ (real $A^\prime$ emission) and $e^- \varPhi^+ \to e^- \varPhi^+ \gamma$ (virtual $A^\prime$ interference) at $\mathcal{O}(\alpha^3 \eps^2)$ as a function of the polar opening angle of the ECAL, $\theta_\text{ECAL}$. We integrate over the minimal real emission selections shown in Table~\ref{tab:theory_selections}.}
    \label{fig:cs_thg_BSM_coh}
\end{figure}

Fig.~\ref{fig:cs_thg_BSM_coh} shows, for coherent scattering, the dependence of the
missing momentum cross sections at $\mathcal{O}(\alpha^3\eps^2)$ for 
different dark photon masses as function of the polar ECAL opening angle
$\theta_\text{ECAL}$. We choose a representative beam energy of $\SI{4} 
{\giga\electronvolt}$ and integrate over the minimal selection phase 
space specified in Table~\ref{tab:theory_selections}.
Clearly, the real emission cross section is independent of the ECAL size and thus appears as constant.
For small $\theta_\text{ECAL}$, the interference terms are in fact larger than the real $A^\prime$ emission piece, because the latter receives a phase space suppression due to the nonzero dark photon mass.
For relatively heavy dark photons, here for example $m_{A^\prime} = \SI{1}{\giga\electronvolt}$, the phase space suppression is strongest such that the interference terms dominate over the real emission terms until a sizable $\theta_\text{ECAL} \approx \SI{0.48}{\radian}$.
In general, the interference terms drop rapidly with $\theta_\text{ECAL}$, since the final state photons are focused into a forward cone, as in the pure QED case.

\begin{figure}[ht]
    \centering
    \includegraphics[width=0.49\linewidth]{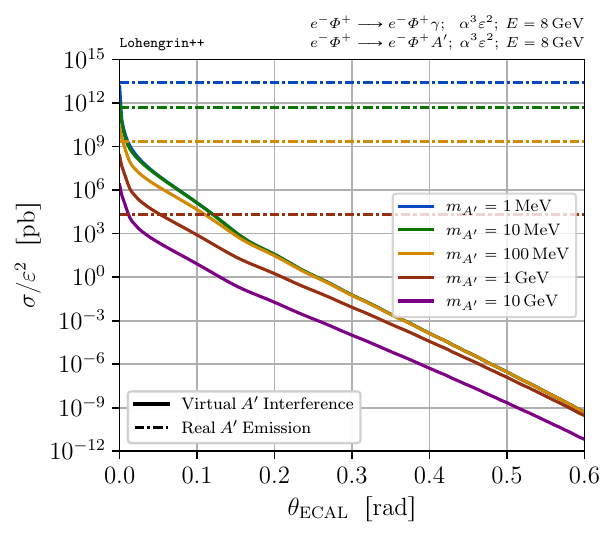}
    \includegraphics[width=0.49\linewidth]{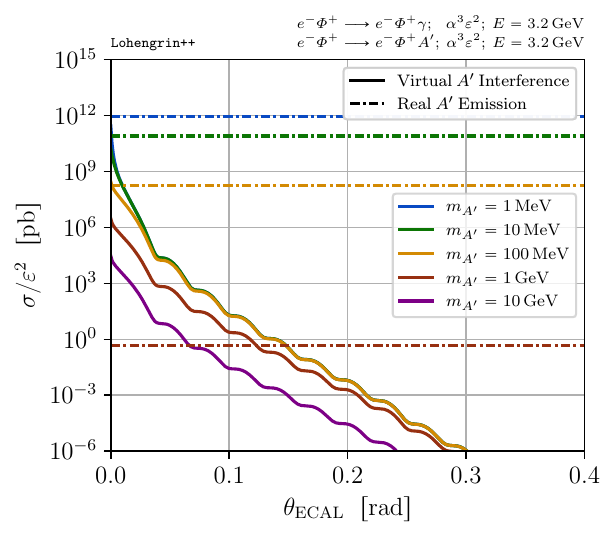}
    \caption{Same as Fig.~\ref{fig:cs_thg_BSM_coh}, but we include only electrons in the final state selections of LDMX (left) and the \textsc{Lohengrin} signal region SR1 (right), see 
    Tables~\ref{tab:LDMX_selections} and \ref{tab:LOH_selections}.}
    \label{fig:cs_thg_BSM_coh_LDMX+Loh}
\end{figure}

Fig.~\ref{fig:cs_thg_BSM_coh_LDMX+Loh} presents the same comparison of missing momentum cross sections but with final state electron selections corresponding to LDMX (left) and \textsc{Lohengrin} (right) scenarios.
The selection criteria are again the same as in 
Tables~\ref{tab:LDMX_selections} and \ref{tab:LOH_selections}.

In both cases, the interference terms fall off even more rapidly than in Fig.~\ref{fig:cs_thg_BSM_coh} due to the additional cuts on the electron momentum.  
In general, the \textsc{Lohengrin} cross sections are lower because of the reduced beam energy and more stringent electron selection criteria.  
The interference terms also reveal once again the oscillatory behavior 
of the elastic nuclear form factor, since the numerical evaluation of the squared amplitudes is forced to
the regime of large $Q^2_\bh$. The LDMX ECAL covers a polar angle of 
$\theta_\text{ECAL} \approx \SI{0.6}{\radian}$, making the new missing 
momentum contribution irrelevant for such a detector setup. Meanwhile, the
\textsc{Lohengrin} ECAL would cover $\theta_\text{ECAL}\approx \SI{0.07}
{\radian}$. For the $m_{A^\prime} = \SI{1}{\giga\electronvolt}$ case, we
then observe that the $\mathcal{O}(\alpha^3 \eps^2)$ missing momentum
contribution from virtual $A^\prime$ interference is actually about 
$\sim70$ times larger than the contribution from real $A^\prime$ 
emission. Again, this reversed hierarchy arises due to the severe phase 
space suppression of the real emission terms for heavy dark photons.

We do not explicitly show the interference contributions for diffractive scattering,
because they vanish at the proposed ECAL sizes when 
imposing forward HCAL vetos.

\subsection{Signal vs. Background}
\label{sec:sig_bkg}
Let us finally compile all previously discussed signal and background 
contributions into a single figure for the \textsc{Lohengrin} scenario.
We present all results as differential distributions in the momentum 
component perpendicular to the magnetic field, $p^\prime_{\perp B}$, 
since this is the quantity the recoil tracking system is sensitive to if
the magnetic field is homogeneous.
We do not consider a distribution in $\theta_e$, since the scattering angle itself does not discriminate between the dark photon signal and QED photon background as well as $p^\prime_{\perp B}$:
for kinematic configurations where the photon escapes detection by missing the ECAL, momentum conservation forces the electron scattering angle also to be wide.
This is especially true since the dominant process, coherent Bethe-Heitler scattering, features only small nuclear recoil.
The angular electron spectrum thus shows a similar feature as the real emission signal, \textit{i.e.} a bump at wide angles.
As a consequence, we consider only the variable $p^\prime_{\perp B}$.

We apply the selections given
in Table~\ref{tab:LOH_selections}, but omit the cuts on $|\vec{p}'|$, since we explicitly bin in $p^\prime_{\perp B}$ now.
Furthermore, we set $\theta_\text{ECAL}=0.07$ and additionally assume an HCAL veto in complete forward direction to avoid the background from nucleon knockouts that we found
in Sec.~\ref{sec:photon_background_yield}.
Again note that independent
studies based on \texttt{FLUKA} also support a
spatial extension of the \textsc{Lohengrin} HCAL.

\begin{figure}[ht]
    \centering
    \includegraphics[width=0.49\linewidth]{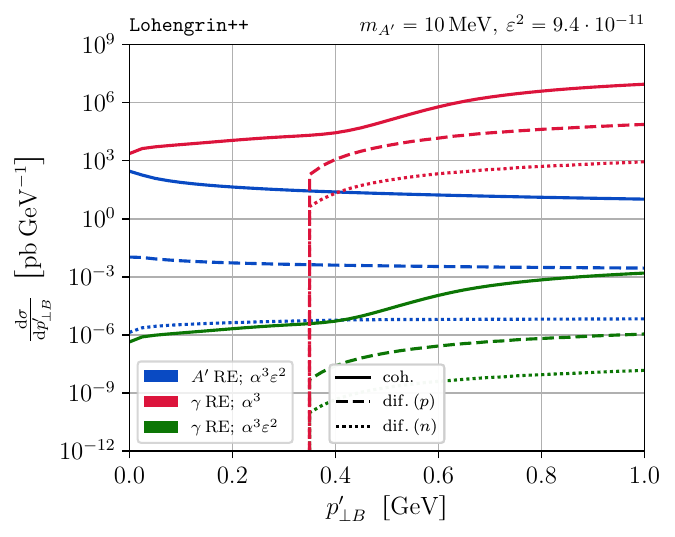}
    \includegraphics[width=0.49\linewidth]{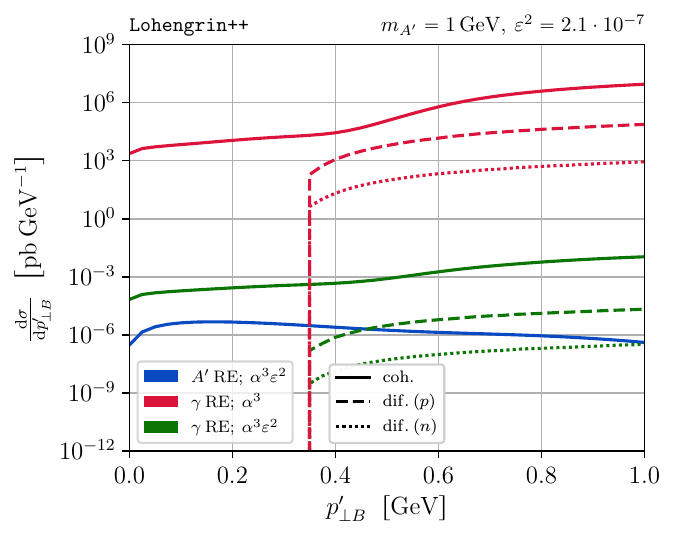}
    \includegraphics[width=0.49\linewidth]{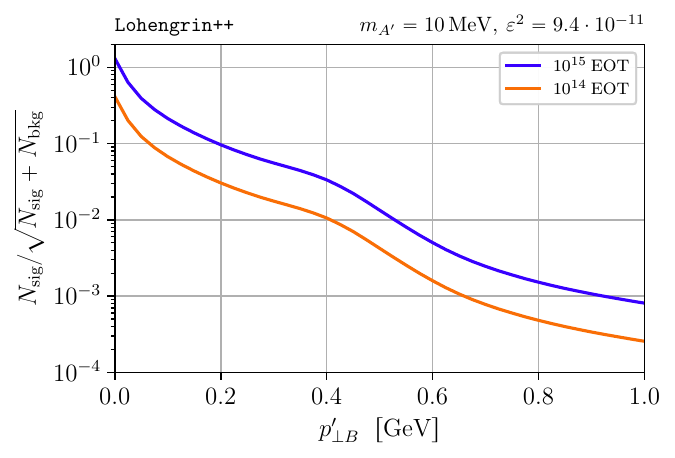}
    \includegraphics[width=0.49\linewidth]{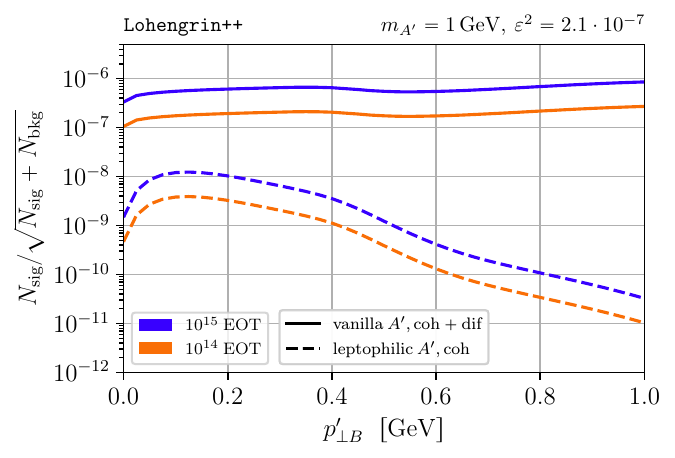}
    \caption{Top: Differential cross sections in $p^{\prime}_{\perp B}$ of all contributions at $\mathcal{O}(\alpha^3)$ and $\mathcal{O}(\alpha^3 \eps^2)$ for two representative dark photon masses. Bottom: Signal significance taking into account all contributions. We apply the selections given in Table~\ref{tab:LOH_selections}, but omit the cuts on $|\protect\vec{p}'|$, since we explicitly bin in $p^\prime_{\perp B}$. Furthermore, we set $\theta_\text{ECAL}=0.07$ and assume an HCAL veto in complete forward direction.}
    \label{fig:cs_pB_signal_significance}
\end{figure}

Fig.~\ref{fig:cs_pB_signal_significance} (top left) shows all 
contributions as a differential cross section in $p'_{\perp B}$ for a 
$\SI{10}{\mega\electronvolt}$ dark photon with a kinetic mixing 
parameter $\varepsilon^2 = 9.4\cdot 10^{-11}$, which is 
the minimal value suggested by the scalar thermal relic target, see 
Table~\ref{tab:eps2}. It is worth noting that real $\gamma$ emissions in
diffractive scattering are now significantly suppressed by the extended 
HCAL veto and do not populate the low momentum bins.
On the other hand, $A^\prime$ emission from diffractive scattering still contributes to low $p'_{\perp B}$ bins, mainly because there are no selections placed on $A'$.
However, this contribution is suppressed by several orders of magnitude compared to the coherent contribution.
Fig.~\ref{fig:cs_pB_signal_significance} (bottom left) shows the resulting signal significance,
\begin{equation}
    N_\text{sig} / \sqrt{N_\text{sig} + N_\text{bkg}},
\end{equation}
where $N_\text{sig}$ and $N_\text{bkg}$ are the signal and background events assuming $10^{14}$ and $10^{15}$ electrons on target (EOT).
We sum all contributions (and neglect detector efficiencies).
In the low-momentum bins, the signal significance reaches a value of $\sim 1.3$ for a benchmark exposure of $10^{15}$ EOT; an extended runtime and/or further expected optimizations in the experimental setup and analysis will enhance the significance further, demonstrating that the \textsc{Lohengrin} experiment can achieve exclusion and discovery sensitivity once hadronic backgrounds are better understood and under control.
The signal is well described by real dark photon production in coherent scattering, and additional signal effects from diffractive scattering and/or interference terms are negligible.

The top and bottom right panels of Fig.~\ref{fig:cs_pB_signal_significance} show
the same for a $\SI{1}{\giga\electronvolt}$ dark photon with kinetic 
mixing parameter $\eps^2 = 2.1\cdot 10^{-7}$, which is again 
motivated by the scalar thermal relic target and taken from Ref.~\cite{Battaglieri:2017aum}.
Previously, with the selections used in
Sec.~\ref{sec:DP_signal_yield}, the diffractive real $A^\prime$ emission
caused a slight increase of the real emission signal cross section around $m_\ap \approx \SI{1}{\giga\electronvolt}$, but 
this contribution now vanishes due to the extended HCAL veto (similar to
the LDMX case in Fig.~\ref{fig:eH>eHAp_LDMX+Loh} (left)). In contrast, 
the coherent virtual $A^\prime$ interference contributing to real $\gamma$ emission 
dominates over real $A^\prime$ emission for dark photons of this mass. 
As expected, it behaves in a slightly different way as compared to the 
leading, pure QED contribution, since its scaling with $Q^2_\bh$ is 
altered by the dark photon mass in the propagator.
Fig.~\ref{fig:cs_pB_signal_significance} (bottom right) shows the signal
significance, and we explicitly distinguish between the purely 
leptophilic, coherent real $A^\prime$ emission case and the scenario 
including all additional contributions, where we treat the interference 
contribution as signal. Although the signal 
significance is greatly increased, the interference term causes it to be
smeared out, removing the distinct maximum in the low electron momentum 
bins. It also still does not offer discovery potential, as the signal significance remains too small, namely $< 10^{-6}$ for $10^{15}$ EOT, which ELSA would deliver for \textsc{Lohengrin} within roughly one year.
Moreover, the interference term is still overpowered by higher-order QED contributions with unknown, but likely similar (\textit{i.e.} in the electron momentum monotonically decreasing) phase space distributions.
So, it seems experimentally impossible to measure this effect, even if all QED contributions up to $\mathcal{O}(\alpha^n) \sim \mathcal{O}(\alpha^3 \eps^2)$ would be known.

However, in the LDMX scenario, it was found that hard bremsstrahlung 
photons can produce vector mesons in the ECAL, see Refs.~\cite{Schuster:2021mlr, Akesson:2022vza}, which may subsequently 
convert into a dark photon. This process represents an additional source
of missing momentum events and can significantly enhance sensitivity to 
relatively heavy dark photons, \textit{i.e.}, $\mathcal{O} (\SI{100} 
{\mega\electronvolt} - \SI{1}{\giga\electronvolt})$. It would be worthwhile to study this contribution also for \textsc{Lohengrin} to increase sensitivity 
for heavy dark photons, but we save this for a dedicated study.

\section{Conclusions}
\label{sec:conclusions}
Novel missing momentum experiments such as LDMX, \textsc{Lohengrin}, and DarkSHINE have the potential to probe large portions of the parameter space of dark photon models, particularly the thermal relic targets.
To contribute to the general understanding of the fundamental signal and background processes, we have systematically calculated all real emission contributions up to $\mathcal{O}(\alpha^3 \varepsilon^2)$ arising from the scattering of an electron off a target nucleus.
In doing so, we incorporated both coherent and diffractive scattering regimes and allowed the dark photon to also couple to composite hadronic systems.
The later generalization ultimately introduces new contributions to the emitted radiation through virtual Compton scattering as well as the appearance of the dark photon as virtual exchange particle.
In addition to real emission processes, we also studied the leading contributions of virtual dark photons appearing in elastic scattering events.
These calculations led to the development of a dedicated Monte Carlo code, \texttt{Lohengrin++}, which enables detailed studies of real dark photon production in both coherent and diffractive scattering, as well as virtual dark photon interference effects.

Within our framework, we found that the missing momentum signal caused by the existence of a dark photon with mass $\SI{1}{\mega\electronvolt} \lesssim m_\ap \lesssim \SI{100}{\mega\electronvolt}$ is indeed well described by real dark photon production in coherent Bethe-Heitler scattering.
Additional contributions were found to be subdominant in this mass range.
In the context of virtual Compton scattering, this conclusion assumes that large virtualities of intermediate nuclei in coherent scattering receive a sufficiently large damping.

For heavier dark photons with masses $m_\ap \sim \SI{1}{\giga\electronvolt}$, vector meson resonances in the timelike nucleon form factors will enhance the dark photon production in virtual Compton scattering significantly, but for a reliable description, a more detailed understanding of the large virtuality regime, including nucleon resonances and off-shell effects, is needed.
However, considering the sizes of momentum transfers required to produce heavy dark photons, we expect additional radiation in the form of ejected nucleons to appear in the final states, which potentially spoils the clean solitary-electron signature.

The size of the interference effects caused by virtual dark photons appearing in the production of SM photon was found to depend strongly on the choice of the spacial dimensions of the ECAL, since it vetos hard photons and thus controls the abundance of missing momentum events with a photon in the final state.
We found that the interference effects are not of experimental relevance.

By explicitly studying experimentally motivated final state selections, we found that \textsc{Lohengrin} would need to cover more solid angle with the HCAL (compared to the first candidate design) to veto background events accompanied by nucleon ejections.
Given such an extended HCAL veto, we find that the \textsc{Lohengrin} experiment indeed offers discovery/exclusion potential for dark photon models with masses around $m_{A'}\sim \SI{10}{\mega\electronvolt}$.
Moreover, we found that for heavy dark photons with masses $\sim \SI{1}{\giga\electronvolt}$, the missing momentum cross section at $\mathcal{O}(\alpha^3\eps^2)$ is predominantly driven by the tree-level interference terms involving virtual dark photons, which contribute to the emission of real photons.

Our calculations can still be complemented by several additional contributions.
For instance, virtual Compton scattering was only considered at the Born level, \textit{i.e.}, including terms up to linear order in the photon energy. Higher-order terms are known and could be implemented, although we do not expect them to induce any order-of-magnitude changes.
Secondly, nucleon resonances could be implemented using effective field theory methods, but achieving a comprehensive description of the resonance regime would be cumbersome. 
Instead, as already mentioned, the hadron backgrounds are currently studied for the \textsc{Lohengrin} project separately using \texttt{FLUKA}.
Additionally, deep inelastic scattering events also contribute at $\mathcal{O}(\alpha^3 \varepsilon^2)$, but we expect these to either produce significant hadronic activity -- thus not yielding a clean electron signal -- or to be subdominant compared to the coherent and diffractive cross sections.
Therefore, we do not expect these extra contributions to significantly impact the signal significance or the core design of the experiments.

We hope that the presented discussion of effects beyond leptophilic dark photon production in Bethe-Heitler scattering contributes to the broader understanding of these novel missing momentum experiments and proves useful to the community.

\section{Acknowledgements}
\label{sed:acknowledgements}
We thank the \textsc{Lohengrin} study group for long and fruitful discussions: 
Markus Gruber, Laney Klipphahn, Patrick Schw\"abig, Tobias Schiffer, and 
especially Philipp Bechtle, Klaus Desch, Matthias Hamer, and Jan-Eric 
Heinrichs. Furthermore, we thank Howie Haber, Felix Kling, Dominik K\"ohler, 
Saurabh Nangia, Rhitaja Sengupta, and Apoorva Shah for discussions.
We also thank Felix Kahlhoefer for discussions about MeV mass Dark Photon candidates, and sources of astrophysical constraints on such particles.
HKD would like to thank the Universiteit van Amsterdam and the Nikhef Theory Group for 
hospitality while part of this work was completed.

\newpage 
\appendix 

\section{Phase Space Implementation}
\label{app:phasespace_integral}
The differential cross section for $2\to n$ scattering, with initial momenta $p$ and $P$, reads
\begin{align}
    \d\sigma_{2\rightarrow n} &= \frac{1}{F} \overline{\sum} |\mathcal{M}|^2 \left( \prod_{f=1}^n \frac{\d^3 \vec{p}_f}{(2\pi)^3 2E_f} \right) (2\pi)^4 \delta^{(4)} \left(p + P - \textstyle\sum\displaystyle  p_f \right)  \\
    &= \frac{1}{F} \:\overline{\sum}|\mathcal{M}|^2 \: \d\varPhi_n , \label{eq:dsigma2ton}
\end{align}
where $F$ is the flux factor.
The integral over the $n$-body Lorentz invariant phase space $\d\varPhi_n$ can be conveniently split up into subsequent evaluations of 2-body phase space integrals, see Ref.~\cite{Kersevan:2004yh}.
For a system of squared invariant mass $M_n^2$ that produces final states with masses $m_1, ..., m_n$, the recursion relation reads,
\begin{align}
    \int \d\varPhi_n (M_n^2, m_1, ..., m_n) &= \int\displaylimits_{\left(\sum_{i=1}^{n-1}m_i\right)^2}^{(M_n-m_n)^2} \d M_{n-1}^2 \int\d\varPhi_2 (M_n^2, M_{n-1},m_n) \nonumber \\
    &\qquad \times  \int\d\varPhi_{n-1} (M_{n-1}^2, m_1, ..., m_{n-1}) \, . \label{eq:nbody_recursion}
\end{align}
The two-body phase space is given by
\begin{equation}
    \int\d\varPhi_2(M_i^2, m_j, m_i) = \frac{\lambda^{\frac{1}{2}}(M_i, m_j, m_i)}{8M_i^2} \int \d \varOmega_{i,j}^* \, ,
\end{equation}
where $\lambda^{\frac{1}{2}}(x,y,z) = \sqrt{x^4+y^4+z^4-2x^2y^2-2y^2z^2-2z^2x^2}$ is the square root of the K\"all\'{e}n function.
The superscript $*$ indicates that the solid angle integration is each time performed in the COM frame of the $ij$ system.

For the $2\to 2$ processes, the phase space integral is straightforward.
Let us be more explicit about the $2\to 3$ case: here, assign $m_1=m_e$, $m_2=m_R$, $m_3 = m_\had$, such that $M_2$ is the invariant mass of the electron-boson system, $M_2^2 = m_{e R}^2$.
$M_3^2$ is the center of mass energy, $M_3^2 = E_\text{\tiny{CM}}^2 = (p+P)^2 = m_e^2 + m_\had^2 + 2 E m_\had$.
With Eq.~\eqref{eq:nbody_recursion}, the total cross section, \textit{e.g.} the integral of Eq.~\eqref{eq:dsigma2ton}, becomes
\begin{align}
    \sigma_{2\to 3} &= \frac{(2\pi)^{-5}}{2 \lambda^{\frac{1}{2}}(E_\text{\tiny{CM}}, m_e, m_\had)} \int\displaylimits_{(m_e+m_R)^2}^{(E_\text{\tiny{CM}}-m_\had)^2} \d m_{e R}^2 
    \frac{\lambda^{\frac{1}{2}}(E_\text{\tiny{CM}}, m_{e R}, m_\had)}{8E_\text{\tiny{CM}}^2} \int \d \varOmega_{\had,e R}^*  \nonumber \\
    & \qquad \times \frac{\lambda^{\frac{1}{2}}(m_{e R}, m_e, m_R)}{8m_{e R}^2} \int \d \varOmega_{R,e}^* \: \overline{\sum}|\mathcal{M}|^2 ,
\end{align}
where we expressed the flux through the K\"all\'{e}n function, $F = \lambda^{\frac{1}{2}}(E_\text{\tiny{CM}}, m_e, m_\had) / 2$.
We further rewrite the solid angle integral $\int \d \varOmega_{\had,e R}^*$ in terms of the Bethe-Heitler momentum transfer $Q_\bh^2$. To that end, note
\begin{equation}
    Q_\bh^2 = - (p - p_{e R} )^2 = -m_e^2 - m_{e R}^2 + 2 \big( E^* E_{e R}^* - |\vec{p}^*| |\vec{p}^*_{e R}| \cos\theta_{\had, e R} \big)\, ,
\end{equation}
such that $\d \cos\theta_{\had, e R} = - \d Q^2_\bh /(2|\vec{p}^*| |\vec{p}_{e R}^*|)$, where $\theta_{\had, e R}$ is the angle between the outgoing hadronic system and the electron-boson system.
The integral over the azimuthal angle yields trivially $2\pi$, such that the integral becomes
\begin{equation}
    \int\d\varOmega_{\had,e R}^* = \int_{-1}^1 \d \cos\theta_{\had, e R} \int^{2\pi}_0 \d\phi_{\had, e R} = \frac{2\pi}{2|\vec{p}^*| |\vec{p}^*_{e R}|}  \int_{Q^2_{\bh;\text{min}}}^{Q^2_{\bh;\text{max}}} \d Q_\bh^2 \, ,
\end{equation}
where the limits are given by
\begin{align}
    Q^2_{\bh;\text{min}} &= -m_e^2 - m_{e R}^2 + 2 \big( E^* E_{e R}^* - |\vec{p}^*| |\vec{p}^*_{e R}| \big) \, ,\\
    Q^2_{\bh;\text{max}} &= -m_e^2 - m_{e R}^2 + 2 \big( E^* E^*_{e R} + |\vec{p}^*| |\vec{p}^*_{e R}| \big) \, .
\end{align}
To obtain differential distributions in any frame-dependent kinematic variable $\zeta$, we consider the cross section in a bin of size $\Delta\zeta = \zeta_\text{max} - \zeta_\text{min}$. The differential distribution is then,
\begin{align}
    \frac{\Delta\sigma_{2\to 3}}{\Delta \zeta} &= \frac{(2\pi)^{-5}}{2 \lambda^{\frac{1}{2}}(E_\text{\tiny{CM}}, m_e, m_\had)} \int\displaylimits_{(m_e+m_R)^2}^{(E_\text{\tiny{CM}}-m_\had)^2} \d m_{e R}^2 
    \frac{\lambda^{\frac{1}{2}}(E_\text{\tiny{CM}}, m_{e R}, m_\had)}{8E_\text{\tiny{CM}}^2} \int \d \varOmega_{\had,e R}^*  \nonumber \\
    & \qquad \times \frac{\lambda^{\frac{1}{2}}(m_{e R}, m_e, m_R)}{8m_{e R}^2} \int \d \varOmega_{X,e}^* \: \overline{\sum}|\mathcal{M}|^2 \:\: \frac{\varTheta(\zeta - \zeta_\text{min})\,\varTheta(\zeta_\text{max}-\zeta)}{\zeta_\text{max}-\zeta_\text{min}}\, ,
\end{align}
where $\varTheta$ is the Heaviside step function with $\varTheta(0) = 0$.
We are ultimately interested in variables $\zeta$ in the lab frame, and their values are thus determined by the lab frame momenta $p^\prime$, $P^\prime$ and $k$. 
In each Monte Carlo sampling step, the lab frame momenta are obtained in the following way:
first, boost the momenta in the $R,e$ frame to the $\had, e R$ frame with the boost vector $\vec{\beta} = -\vec{p}_{e R}/E_{e R}$.
Then, boost the momenta in the $\had, e R$ frame to the lab frame with a boost along the $\hat{z}$-axis with velocity
\begin{equation}
    \beta_\text{lab} = \frac{\lambda^{\frac{1}{2}}(E_\text{\tiny{CM}}, m_e, m_\had)/(2E_\text{\tiny{CM}})}{\sqrt{\left(\displaystyle\frac{\lambda^{\frac{1}{2}}(E_\text{\tiny{CM}}, m_e, m_\had)}{2E_\text{\tiny{CM}}}\right)^2+m_\had^2}}\, .
\end{equation}

\section{Dark Photon Compton Tensor in Coherent Scattering}
\label{app:coherent_compton_tensor}
In this appendix, we will outline how the Compton tensors with one dark photon leg are obtained for coherent scattering.
Electromagnetically interacting scalar particles with finite size and charge $Ze$ can be described by an effective Lagrangian, given in Refs. \cite{Fearing:1996gs, Moinester:2019sew}.
Although these references are mainly focused on pions, the discussion holds in general, so also for an 
effective field theory of scalar nuclei. 
The Lagrangian reads,
\begin{equation}
    \lgr = (D_\mu^f \varPhi^+)^\dagger D^\mu_f \varPhi^+ - m_{\varPhi}^2 \varPhi^- \varPhi^+,
\end{equation}
where the noncanonical covariant derivative is given by,
\begin{equation}
    D_\mu^f \varPhi^+ = \{ \del_\mu + i eZ A_\mu +ie [f(-\dalembertian) \del^\nu F_{\mu\nu}] \} \varPhi^+ .
\end{equation}
The function $f$ acts on the field strength tensor.
In particular, $f(-\dalembertian)$ acting on a plane wave yields $f(-\dalembertian)e^{iqx} = f(q^2) e^{iqx}$.
Moreover, $f(q^2)$ is connected to the form factor $F(q^2)$ by,
\begin{equation}
    f(q^2) = \frac{F(q^2) - Z}{q^2}.
\end{equation}
Expanding the Lagrangian gives:
\begin{align}
    \lgr_0 &= (D_\mu \varPhi^+)^\dagger D^\mu \varPhi^+ - m_{\varPhi}^2 \varPhi^- \varPhi^+ , \label{eq:L_scalarQED}\\
    \lgr_1 &= ie(D_\mu \varPhi^+)^\dagger \varPhi^+ [f(-\dalembertian)\del_\nu F^{\mu\nu}] -ie \varPhi^- D_\mu \varPhi^+ [f(-\dalembertian)\del_\nu F^{\mu\nu}]^\dagger , \\
    \lgr_2 &= e^2 \varPhi^- \varPhi^+ [f(-\dalembertian)\del^\nu F_{\mu\nu}] [f(-\dalembertian)\del_\rho F^{\mu\rho}]^\dagger ,
\end{align}
with the canonical covariant derivative $D_\mu \varPhi^+ = (\del_\mu + i eZ A_\mu)\varPhi^+$.
While $\lgr_0$ is the usual scalar QED Lagrangian, $\lgr_1$ and $\lgr_2$ encode the finite size effects.
In particular, $\lgr_0$ and $\lgr_1$ together yield the electromagnetic vertex,
\begin{equation}
\begin{tikzpicture}[baseline=-3.6]
    	\setlength{\feynhandblobsize}{3mm}
    \begin{feynhand}
    		\vertex (veg) at (0,1.2) {$\nu$};
    		\vertex [grayblob] (vhg) at (0,0) {};
    		\vertex (hi) at (-1.3,0) {$\varPhi^+$};
    		\vertex (ho) at (1.3,0) {$\varPhi^+$};
    		\propag [photon, revmom'={$q$} ] (veg) to (vhg);
    		\propag [scalar, mom'={$P$} ] (hi) to (vhg);
    		\propag [scalar, mom'={$P^\prime$}] (vhg) to (ho);
    	\end{feynhand}
    \end{tikzpicture}
    = ie F(q^2) \left(2P^\nu - q^\nu\right) - ie q^\mu (P^2 - {P^\prime}^2) f(q^2),
\end{equation}
which satisfies the Ward-Takahashi identity. For on-shell external $\varPhi^+$, this reduces to Eq.~\eqref{eq:H_nu_BH_gamma}.
Now, expanding out the QED interactions and performing the substitution $A_\mu \mapsto A_\mu^\text{\tiny{SM}} \approx A_\mu -\eps A_\mu^\prime$ will yield the effective low-energy dark photon interactions.

The interaction terms linear in the photon field $A_\mu^\text{\tiny{SM}}$ are straightforward. They yield the same interactions involving $A^\prime$, but with the coupling rescaled by $-\eps$, thus giving Eq.~\eqref{eq:H_nu_BH_Ap}.

The interaction terms quadratic $A_\mu^\text{\tiny{SM}}$ are more involved.
Imposing $A_\mu^\text{\tiny{SM}} \approx A_\mu -\eps A_\mu^\prime$ results in terms $\propto A_\mu A^{\mu}$,  $\propto A_\mu A^{\prime\, \mu}$ and $\propto A^\prime_\mu A^{\prime\, \mu}$.
While the $A_\mu A^\mu$ terms contribute to the QED Compton tensor with a symmetry factor of $2$, we find the mixed terms $\propto A_\mu A^{\prime\, \mu}$ to appear always twice.
Thus, the contribution from $\lgr_0 + \lgr_1$ to the Compton tensor with one dark photon leg is identical to the pure QED Compton tensor, except for the $-\eps$ rescaling.
The contribution to the QED Compton tensor from $\lgr_2$ vanishes if at least one photon is real, satisfying $\del_\nu F^{\mu\nu} = 0$.
This is not true in the case of the dark photon, since it satisfies the Proca equation, $\del_\nu F^{\prime \, \mu\nu} + m_{A'}^2 A^{\prime\,\mu} = 0$.
In presence of a dark photon, $\lgr_2$ becomes:
\begin{align}
    \lgr_2 &= e^2 \varPhi^- \varPhi^+ [f(-\dalembertian) \del^\nu F_{\mu\nu}] [f(-\dalembertian) \del_\rho F^{\mu\rho}]^\dagger \nonumber \\
    &\quad -\eps e^2 \varPhi^- \varPhi^+ [f(-\dalembertian) \del^\nu F^\prime_{\mu\nu}] [f(-\dalembertian) \del_\rho F^{\mu\rho}]^\dagger \nonumber \\
    & \quad -\eps e^2 \varPhi^- \varPhi^+ [f(-\dalembertian) \del^\nu F_{\mu\nu}] [f(-\dalembertian) \del_\rho F^{\prime\, \mu\rho}]^\dagger \nonumber \\
    & \quad + \eps^2 e^2 \varPhi^- \varPhi^+ [f(-\dalembertian) \del^\nu F^\prime_{\mu\nu}] [f(-\dalembertian) \del_\rho F^{\prime\,\mu\rho}]^\dagger .
\end{align}
Taking the dark photon to be on-shell then results in the following extra contribution to the mixed Compton tensor:
\begin{equation}
    \begin{tikzpicture}[baseline=-2]
        \setlength{\feynhandblobsize}{5mm}
        \begin{feynhand}
        \vertex (veg) at (-1,1) {$\nu$};
        \vertex (veg2) at (+1,1) {$\mu$};
        \vertex [NEblob] (vhg) at (0,0) {};
        \vertex (hi) at (-1.05,-1.05) {$\varPhi^+$};
        \vertex (ho) at (1.05,-1.05) {$\varPhi^+$};
        \propag [photon, color=black, revmom'={$q$} ] (veg) to (vhg);
        \propag [photon, color=crimson, revmom'={[arrow style = black]$k$} ] (veg2) to (vhg);
        \propag [scalar, mom'={$P$} ] (hi) to (vhg);
        \propag [scalar, mom={$P^\prime$}] (vhg) to (ho);
        \end{feynhand}
    \end{tikzpicture}
    \supset 2\eps e^2 \big[F(k^2) - Z \big] \big[F(q^2) - Z \big] \left( g^{\mu\nu} - \frac{q^\mu q^\nu}{q^2} \right),
\end{equation}
such that we end up with Eq.~\eqref{eq:Cmunu_Apg} in total.

\section{Dark Goldstone Boson in \texorpdfstring{$R_\xi$}{TEXT} and Feynman Gauge}

\label{app:dark_goldstone}

Loop corrections are most commonly carried out in Feynman-t'Hooft gauge, where the Goldstone bosons associated to the massive vector bosons are propagating degrees of freedom.
To calculate the loop corrections involving the dark photon, we will thus rephrase the commonly used effective theory below the electroweak scale to explicitly contain the Goldstone boson and a gauge parameter that controls its presence.
For physics applications below the electroweak scale, it is sufficient to use kinetic mixing of the new, heavy $U(1)_\text{\tiny{D}}$ gauge boson with the QED photon.
Let the $U(1)_\text{\tiny{D}}$ gauge field be $\ad^\mu$ with field strength tensor $F_\text{\tiny{D}}^{\mu\nu} = \del^\mu \ad^\nu - \del^\nu \ad^\mu$ and mass $m_{\ad}$.
The dark sector Lagrangian in which the $U(1)_\text{\tiny{D}}$ gauge symmetry is spontaneously broken by an abelian Higgs mechanism then contains,
\begin{equation}
\label{eq:lgrD_symbr}
    \lgr_\text{\tiny{D}} \supset -\frac{1}{4} {F_\text{\tiny{D}}}_{\mu\nu} {F_\text{\tiny{D}}}^{\mu\nu} + \frac{1}{2} m_{\ad}^2 \left[ \ad^\mu + \frac{1}{m_{\ad}}\del^\mu \pi \right]^2\, ,
\end{equation}
where $\pi$ is the Goldstone boson associated to $\ad^\mu$.
The portal interaction is the kinetic mixing of $\ad^\mu$ (dark sector) and $A_{\text{\tiny{SM}}}^\mu$ (SM photon),
\begin{equation}
    \lgr_{\text{\tiny{SM}}\otimes \text{\tiny{D}}} = -\frac{\eps}{2} {\fsm}_{\mu\nu} \fd^{\mu\nu} = -\frac{\eps}{2} \: \Big[ \del_\mu {\asm}_\nu - \del_\nu {\asm}_\mu \Big] \: \left[ \del^\mu \ad^\nu - \del^\nu \ad^\mu \right] \, .
\end{equation}
Now, to achieve a consistent treatment, we will modify the field strength tensor to also contain the Goldstone field $\pi$.
In light of the form of the mass term in Eq.~\eqref{eq:lgrD_symbr}, we write
\begin{equation}
    \label{eq:kin_mixing_new}
    \lgr_{\text{\tiny{SM}} \otimes \text{\tiny{D}}} = -\frac{\eps}{2} \Big[ \del_\mu {\asm}_\mu - \del_\nu {\asm}_\mu \Big] \: \left[ \del^\mu \left( \ad^\nu + \frac{1}{m_{\ad}}\del^\nu \pi \right) - \del^\nu \left( \ad^\mu + \frac{1}{m_{\ad}}\del^\mu \pi \right) \right] \, .
\end{equation}
Note that substituting $\ad^\mu \mapsto \ad^\mu + m_{\ad}^{-1}\del^\mu \pi$ behaves here exactly like a gauge transformation:
Since the derivatives commute, the field strength tensor is unchanged and thus the kinetic mixing operator.
In the abelian Higgs mechanism, the Goldstone field is conveniently removed by performing a gauge fixing of this type.
The next step is to rotate away the kinetic mixing of the two photons.
This is achieved with the transformation,
\begin{equation}
    \begin{pmatrix}
    \ad^\mu + \frac{1}{m_{\ad}}\del^\mu \pi\\
    \asm^\mu
    \end{pmatrix}
    =
    \begin{pmatrix}
    \frac{1}{\sqrt{1-\eps^2}} & 0 \\
    -\frac{\eps}{\sqrt{1-\eps^2}} & 1
    \end{pmatrix}
    \begin{pmatrix}
    {\ap}^\mu + \frac{1}{m_{\ap}}\del^\mu \pi\\
    A^\mu
    \end{pmatrix}\, .
\end{equation}
The field strength tensors then become
\begin{align}
    \fd^{\mu\nu} &= \frac{1}{\sqrt{1-\eps^2}} {\fp}^{\mu\nu}, \\
    \fsm^{\mu\nu} &= \del^\mu \left[ \frac{-\eps}{\sqrt{1-\eps^2}}\left( A^{\prime\,\nu} + \frac{1}{m_{A^\prime}} \del^\nu \pi \right) + A^\nu \right] - \del^\nu \left[ \frac{-\eps}{\sqrt{1-\eps^2}}\left( A^{\prime\,\mu} + \frac{1}{m_{A^\prime}} \del^\mu \pi \right) + A^\mu \right] \nonumber \\
    &\quad= \frac{-\eps}{\sqrt{1-\eps^2}} F^{\prime\,\mu\nu} + F^{\mu\nu}.
\end{align}
Thus, the kinetic mixing among the photons is removed and the Lagrangian contains,
\begin{align}
    \lgr &\supset -\frac{1}{4}F^{\prime}_{\mu\nu} F^{\prime\,\mu\nu} + \frac{1}{2} \frac{m_{\ad}^2}{1-\eps^2} \left[ A^\prime_{\mu} + \frac{1}{m_{A^\prime}} \del^\mu \pi \right]^2 \nonumber \\
    &\qquad -\frac{1}{4}F_{\mu\nu}F^{\mu\nu}
    - J^\mu \left( \frac{-\eps}{\sqrt{1-\eps^2}} \left( A^\prime_\mu + \frac{1}{m_{A^\prime}} \del_\mu \pi \right) + A_\mu \right) \, . 
\end{align}
Here, we can identify $m_{A^\prime}^2 = m_{\ad}^2/(1-\eps^2)$ and obtain a coupling of the QED current $J^\mu$ to the mass eigenstate $A^\prime$, which we call `dark photon'.
As last step, we can now perform the gauge fixing for the new spin-1 field. We introduce a gauge fixing parameter $\xi^\prime$ and add the gauge-fixing term to the Lagrangian, which removes the kinetic mixing of $A^\prime$ with $\pi$,
\begin{equation}
    \lgr \supset -\frac{1}{4}F^{\prime}_{\mu\nu}F^{\prime\,\mu\nu} + \frac{1}{2} \frac{m_{\ad}^2}{1-\eps^2} \left[ A^\prime_\mu + \frac{1}{m_{A^\prime}} \del_\mu \pi \right]^2 - \frac{1}{2\xi^\prime} \left( \del_\mu A^{\prime\,\mu} - \xi^\prime m_{A^\prime} \pi \right)^2 \, .
\end{equation}
The propagators for the BSM fields thus are,
\begin{align}
        \begin{tikzpicture}[baseline=0]
        \setlength{\feynhandblobsize}{3mm}
        \setlength{\feynhandarrowsize}{5pt}
        \begin{feynhand}
        \vertex (i) at (-1., 0) {$\mu$};
        \vertex (o) at (+1, 0) {$\nu$};
        \propag [photon, color = crimson] (i) to (o);
        \end{feynhand}
        \end{tikzpicture}
        ~ &\sim ~ 
        \frac{i}{p^2 - m_{A^\prime}^2} \left( - g_{\mu\nu} + \frac{p_\mu p_\nu}{p^2 - \xi^\prime m_{A^\prime}^2} (1-\xi^\prime) \right) \, , \\
        \begin{tikzpicture}[baseline=0]
        \setlength{\feynhandblobsize}{3mm}
        \setlength{\feynhandarrowsize}{5pt}
        \begin{feynhand}
        \vertex (i) at (-1., 0) {$\hphantom{\mu}$};
        \vertex (o) at (+1, 0) {$\hphantom{\nu}$};
        \propag [scalar, color = crimson] (i) to (o);
        \end{feynhand}
        \end{tikzpicture}
        ~ &\sim ~
        \frac{i}{p^2 - \xi^\prime m_{A^\prime}^2}\, ,
\end{align}
and with $J^\mu = eQ_\ell \bar\ell \gamma^\mu \ell$ we get for the interaction vertices,
\begin{align}
        \begin{tikzpicture}[baseline=15]
        \setlength{\feynhandblobsize}{3mm}
        \setlength{\feynhandarrowsize}{5pt}
        \begin{feynhand}
        \vertex (ei) at (-1, 0) {$\ell$};
        \vertex (eo) at (1, 0) {$\ell$};
        \vertex (veg) at (0, 0.6);
        \vertex (g) at (0, 1.5) {$\mu$};
        \propag [fermion] (ei) to (veg);
        \propag [fermion] (veg) to (eo);
        \propag [photon, color = crimson] (veg) to (g);
        \end{feynhand}
        \end{tikzpicture}
        ~ &\sim ~ 
        -i e Q_\ell \gamma^\mu \frac{-\eps}{\sqrt{1- \eps^2}} ~ \approx ~ i e Q_\ell \eps \gamma^\mu \, , \\
        \begin{tikzpicture}[baseline=15]
        \setlength{\feynhandblobsize}{3mm}
        \setlength{\feynhandarrowsize}{5pt}
        \begin{feynhand}
        \vertex (ei) at (-1, 0) {$\ell$};
        \vertex (eo) at (1, 0) {$\ell$};
        \vertex (veg) at (0, 0.6);
        \vertex (g) at (0, 1.5);
        \propag [fermion] (ei) to (veg);
        \propag [fermion] (veg) to (eo);
        \propag [scalar, color = crimson, mom ={[arrow style = black]$k$}] (veg) to (g);
        \end{feynhand}
        \end{tikzpicture}
        ~ &\sim ~ 
        -i e Q_\ell \gamma^\mu \frac{-\eps}{\sqrt{1- \eps^2}} \frac{- i k_\mu}{m_{A^\prime}}  ~ \approx ~ \frac{e Q_\ell \eps \slashed{k}}{m_{A^\prime}}\, ,
\end{align}
where $e = \sqrt{4\pi \alpha}$, and the approximation $\eps \ll 1$ was applied.
In Feynman-t'Hooft gauge, $\xi^\prime \to 1$, we now indeed have the Goldstone as propagating degree of freedom.
For $\xi^\prime \to \infty$ we are back in unitary gauge, where the Goldstone boson becomes infinitely heavy and thus decouples, while the dark photon propagator gets its longitudinal term $\propto p_\mu p_\nu / m_{\ap}^2$.

\section{Renormalization and Counterterms}
\label{app:renormalization_counterterms}

In this appendix, we present the key results of the renormalization procedure necessary to evaluate the NLO interference terms involving a leptophilic dark photon, see Eq. \eqref{eq:NLO_1-loop_amp}.
We consider a collection of fermions (the SM charged leptons) which are charged under U$(1)_\text{\tiny{EM}}$ and receive an additional coupling to a massive dark photon, generated through the kinetic mixing operator.
Since LDMX-like experiments use beam energies well below the electroweak scale, we work in a theory without $W^\pm$ and $Z^0$ bosons and assume that the dark photon couples to left- and right-chiral fermions in the same way.

Our treatment is similar to the techniques presented in Ref.~\cite{Denner:1991kt}:
we rephrase the Lagrangian to include counterterms in renormalized perturbation theory and determine the counterterms in the physical on-shell scheme by imposing appropriate renormalization conditions.

\subsection{Parameter and Field Renormalization}
First, we express the bare parameters and fields in terms of renormalized quantities.
The bare quantities will carry a subscript $0$, while the renormalized ones have no subscript.
The rescaling factors are denoted by the letter $Z$ with appropiate subscripts and are then expanded in terms of counterterms $\delta Z$ at the one loop level.
The renormalized parameters are the electric charge, the dark photon mass and the fermion masses, so we write
\begin{align}
    e_0 &= Z_e e = (1+\delta Z_e) e \\
    m^2_{\ap,0} &= Z_{m_\ap} m_\ap^2 = \left(1 + \delta Z_{m_\ap} \right) m_\ap^2 = m_\ap^2 + \delta m_\ap^2 \\
    m_{\ell,i,0} &= Z_{m_{\ell,i}} m_{\ell,i} = \left(1 + \delta Z_{m_{\ell,i}} \right) m_{\ell,i} = m_{\ell,i} + \delta m_{\ell,i} \, .
\end{align}
The vector bosons mix at one loop. So we define the field strength renormalizations in the vector boson sector as,
\begin{equation}
\begin{pmatrix}
    \ap^0 \\
    A^0
\end{pmatrix}
=
\begin{pmatrix}
Z^{1/2}_{\ap\ap}& Z^{1/2}_{\ap A} \\
Z^{1/2}_{A\ap} & Z^{1/2}_{AA} 
\end{pmatrix}
\begin{pmatrix}
    \ap \\
    A
\end{pmatrix}
=
\begin{pmatrix}
1+\frac{1}{2}\delta Z_{\ap\ap}& \frac{1}{2} \delta Z_{\ap A} \\
\frac{1}{2} \delta Z_{A\ap} & 1+ \frac{1}{2}\delta Z_{AA} 
\end{pmatrix}
\begin{pmatrix}
    \ap \\
    A
\end{pmatrix} \, .
\end{equation}
and in the fermion sector simply as,
\begin{align}
    \ell_{i,0} &= \left[1 + \frac{1}{2} \delta Z_{i}^{\ell}  \right] \ell_{i} \, .
\end{align}

\subsection{Renormalization of the Vector Boson Sector}
To obtain the $\delta Z_{A' A}$ and $\delta Z_{AA'}$ counter terms, we expand the vector boson kinetic terms and the dark photon mass term in terms of the renormalized fields and the counterterms. We find,
\begin{align}
    \lgr \supset& -\frac{1}{4} F^0_{\mu\nu}F_0^{\mu\nu} - \frac{1}{4} {F^\prime}^0_{\mu\nu} {F^\prime}_0^{\mu\nu} - \frac{1}{2} m^2_{\ap,0} {A^\prime}^0_\mu {A^\prime}_0^\mu \\
    =&\: \frac{1}{2} \left( A_\mu \dalembertian A^\mu - A_\mu \del_\nu \del^\mu A^\nu\right) (1+\delta Z_{AA})
    +  \frac{1}{2} \left( {A^\prime}_\mu \dalembertian {A^\prime}^\mu - {A^\prime}_\mu \del_\nu \del^\mu {A^\prime}^\nu\right) (1 + \delta Z_{{A^\prime}{A^\prime}}) \nonumber  \\
    &+\left( A_\mu \dalembertian {A^\prime}^\mu - A_\mu \del_\nu \del^\mu {A^\prime}^\nu\right)\frac{\delta Z_{A{A^\prime}} + \delta Z_{{A^\prime}A}}{2} \nonumber  \\
    &- \frac{1}{2} m_\ap^2 {A^\prime}_\mu {A^\prime}^\mu (1 + \delta Z_{{A^\prime}{A^\prime}}) - {m_\ap}^2 {A^\prime}_\mu A^\mu \frac{1}{2}\delta Z_{{A^\prime}A} - \frac{1}{2} \delta m_\ap^2 {A^\prime}_\mu {A^\prime}^\mu  \, ,
\end{align}
where we neglected contributions quadratic in the counterterms. The relevant diagram involving the renormalisation constants $\delta Z_{A' A}$ and $\delta Z_{AA'}$ gives:
\begin{align}
\begin{tikzpicture}[baseline=(o.base)]
\begin{feynhand}
\vertex (l) at (-1.2,0) {$A, \mu$};
\vertex (o) at (0,0);
\vertex (r) at (1.2,0) {$\ap, \nu$};
\vertex [crossdot] (o) at (0,0) {};
\propag [photon] (l) to (o);
\propag [photon, color = crimson] (o) to (r);
\end{feynhand}
\end{tikzpicture}
&=
\begin{tikzpicture}[baseline=(o.base)]
\begin{feynhand}
\vertex (l) at (-1.2,0) {$\ap, \mu$};
\vertex (o) at (0,0);
\vertex (r) at (1.2,0) {$A, \nu$};
\vertex [crossdot] (o) at (0,0) {};
\propag [photon, color = crimson] (l) to (o);
\propag [photon] (o) to (r);
\end{feynhand}
\end{tikzpicture} \nonumber \\
&= -i \frac{\delta Z_{A\ap} + \delta Z_{\ap A}}{2} \left( p^2 g_{\mu\nu} -p_\mu p_\nu \right) + ig_{\mu\nu} \frac{\delta Z_{\ap A}}{2} m_\ap^2 .
\label{eq:FR_ct_VB_mixing}
\end{align}
As a renormalization condition, we demand that the propagator matrix (derived from the vector boson self-energies and their one-loop mixing) is diagonal at both the photon and dark photon poles.
Explicitly, we then get for the two relevant counterterms:
\begin{align}
    \delta Z_{AA^\prime} &= \sum_{i} \frac{2 \alpha \eps Q_{\ell_i}^2}{9 \pi m_\ap^2} \Bigg\{ m_\ap^2 - 6 m_{\ell_i}^2 + 6 A_0(m_{\ell_i}^2) \nonumber \\
    & \qquad \qquad \qquad \quad - 3 \left(m_\ap^2 + 2 m_{\ell_i}^2 \right) B_0(m_\ap^2; m_{\ell_i}^2, m_{\ell_i}^2)  \Bigg\} \, ,\\
    \delta Z_{A^\prime A} &= 0 \, ,
\end{align}
where $A_0(m^2)$ and $B_0(p^2; m_a^2, m_b^2)$ are the usual Passarino-Veltman loop functions.

\subsection{Vertex Renormalization}
The fermion-photon vertex counterterm $\delta Z_1^i$ is formally defined through the interaction terms.
Reexpressing the unrenormalized interaction Lagrangian in terms of renormalized quantities and counterterms gives,
\begin{align}
    \lgr \supset& -e_0 Q_\ell \bar{\ell}_{i,0} \gamma^\mu \ell_{i,0} A_\mu^0 - e_0 Q_\ell \eps \bar{\ell}_{i,0} \gamma^\mu \ell_{i,0} {\ap}^0_\mu \\
    \supset& -\left(1 + \delta Z_e + \delta Z_{i}^{\ell} + \frac{1}{2} \delta Z_{AA} + \eps \frac{1}{2} \delta Z_{\ap A} \right) e Q_\ell \bar{\ell}_{i} \gamma^\mu \ell_i  A_\mu \\
    \equiv& -\left(1 + \delta Z_1^i \right) e Q_\ell \bar{\ell}_{i} \gamma^\mu \ell_i  A_\mu ,
\end{align}
where we again neglected contributions quadratic in the counterterms and implcitly defined $\delta Z_1^i$ in the last step. So the counterterm Feynman rule simply is,
\begin{equation}
    \label{eq:FR_ct_llA}
    \begin{tikzpicture}[baseline=(o.base)]
    \begin{feynhand}
    \vertex (l1) at (-1.2,0) {$\ell_i$};
    \vertex (l2) at (1.2,0) {$\ell_i$};
    \vertex (r) at (0,1.2) {$A, \mu$};
    \vertex [crossdot] (o) at (0,0) {};
    \propag [fermion] (l1) to (o);
    \propag [anti fermion] (l2) to (o);
    \propag [photon] (r) to (o);
    \end{feynhand}
    \end{tikzpicture}
    \quad \sim \quad
    -ieQ_\ell \delta Z_1^i\, \gamma^\mu .
\end{equation}
Instead of obtaining $\delta Z_1^i$ from imposing the Thomson limit on the vertex function, we found it more convenient to use the relation implied by the Ward-Takahashi identity, \textit{i.e.}
\begin{equation}
    \delta Z_1^i = \delta Z_i^\ell ,
\end{equation}
which relates the vertex counterterm with the fermion wave function renormalization counterterm.
The latter is straightforwardly derived from the fermion self-energy function. The fermionic kinetic term is expanded as,
\begin{align}
    \lgr &\supset \bar{\ell}_i^0 \left(i \slashed{\del} - m_{\ell,i}^0 \right) \ell_i^0 \\
    &= \bar{\ell}_i \left(i \slashed{\del} - m_{\ell,i} \right) \ell_i (1+\delta Z_i^\ell) - \bar{\ell}_i \delta m_{\ell,i} \ell_i \, ,
\end{align}
such that the counterterm will give:
\begin{equation}
\begin{tikzpicture}[baseline=(o.base)]
\begin{feynhand}
\vertex (l) at (-1.2,0) {$\ell_i$};
\vertex (o) at (0,0);
\vertex (r) at (1.2,0) {$\ell_i$};
\vertex [crossdot] (o) at (0,0) {};
\propag [fermion] (l) to (o);
\propag [fermion] (o) to (r);
\end{feynhand}
\end{tikzpicture}
= i \left[ \left( \slashed{p} - m_{\ell,i} \right) \delta Z_i^\ell - \delta m_{\ell,i} \right] \, .
\end{equation}
The wave function renormalization counterterm is obtained from demanding that the propagator resulting form the 1-loop self-energy function has residue $i$.
Explicitly we find for the BSM part $\propto \eps^2$,
\begin{align}
    \delta Z_i^\ell &= \frac{\alpha \eps^2 Q_{\ell_i}^2}{4 \pi m_\ap^2 m_{\ell,i}^2} \Bigg\{ \left( m_\ap^2 + m_{\ell,i}^2 \right) A_0(m_\ap^2) - m_\ap^2 A_0(m_{\ell,i}^2) - m_\ap^4 B_0(m_{\ell,i}^2;m_\ap^2,m_{\ell,i}^2) \nonumber \\
    &\qquad\qquad\qquad\quad + m_{\ell,i}^2 m_\ap^2\left[ 1+ 2 \left( m_\ap^2 + 2 m_{\ell,i}^2 \right) B_0^\prime(m_{\ell,i}^2;m_\ap^2,m_{\ell,i}^2)\right]\Bigg\} \, ,
\end{align}
where $A_0(m^2)$ and $B_0(p^2; m_a^2, m_b^2)$ are again the Passarino-Veltman loop functions and $B_0^\prime(p^2; m_a^2, m_b^2)=\frac{\partial}{\partial p^2} B_0(p^2; m_a^2, m_b^2)$.

\section{Supplementary Figures}
\label{app:supp_figures}
\begin{figure}[htbp]
	\centering
	\includegraphics[width=0.98\textwidth]{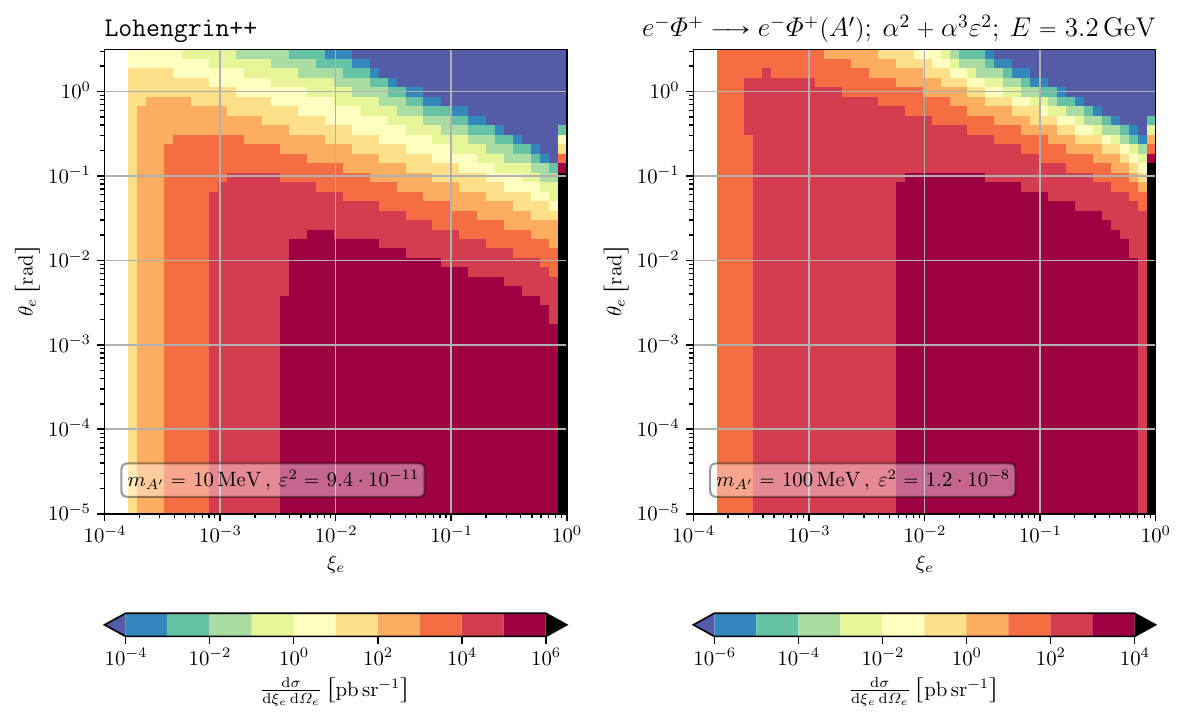}
	\caption[]{Double differential cross section w.r.t. the energy fraction $\xi_e$ and solid angle $\varOmega_e$ of the recoiling electron taking into account $\mathcal{O}(\alpha^2)$ LO and $\mathcal{O}(\alpha^3\eps^2)$ NLO contributions in coherent scattering. We set a beam energy of $E = \SI{3.2}{\giga\electronvolt}$ and show two benchmark points defined by different choices of dark photon mass $m_{A^\prime}$ and squared kinetic mixing strength $\varepsilon$.}
	\label{fig:ddcs_sph_coh}
\end{figure}

Fig.~\ref{fig:ddcs_sph_coh} shows the double differential cross section,
\begin{align}
    \frac{\d^2 \sigma}{\d\xi_e \d\varOmega_e} &= \frac{\d^2\sigma_\text{LO}}{\d\xi_e \d\varOmega_e}(e^- \varPhi^+ \rightarrow e^- \varPhi^+; \alpha^2) \nonumber \\
    &\quad + \frac{\d^2\sigma_\text{NLO, real}}{\d\xi_e \d\varOmega_e}(e^- \varPhi^+ \rightarrow e^- \varPhi^+ A'; \alpha^3\eps^2) + \frac{\d^2\sigma_\text{NLO, virt}}{\d\xi_e \d\varOmega_e}(e^- \varPhi^+ \rightarrow e^- \varPhi^+; \alpha^3\eps^2).
\end{align}
for coherent scattering with a beam energy of $E = \SI{3.2}{\giga\electronvolt}$.
This may be compared to Fig.~\ref{fig:ddcs_sph_cohdif} which shows the combination of coherent and diffractive scattering.

\section{Supplementary Tables}
\label{app:supp_tables}
The numerical evaluation of the nucleon form factors, $F_{i}^{N}(q^2)$ with $i=1,2$ and $N=p,n$, requires the input of several parameters, cf. Sec.~\ref{sec:amplitudes_regimes_diffractive}.
In this appendix, we explicitly state the parameter values used for the calculations presented in this work.

For space-like arguments, $q^2 < 0$, we use the Kelly parameterization of the Sachs form factors, \textit{cf.} Eqs.~\eqref{eq:Kelly_GK}--\eqref{eq:Kelly_NM}.
Table~\ref{tab:Kellycoefficients} shows the parameter values, taken from \cite{Kelly:2004hm}.
Furthermore, for the nucleon magnetic moments we use the values:
\begin{align}
    \mu_p &= 2.7928473446 \,\mu_N \, , \label{eq:mu_p_value} \\
    \mu_n &= -1.9130427 \,\mu_N \, , \label{eq:mu_n_value}
\end{align}
with the nuclear magneton $\mu_N = 3.15245125844$ \cite{ParticleDataGroup:2024cfk}.

For time-like arguments, $q^2 > 0$ we use the parameterization following from the vector meson dominance approach, which is reviewed in Ref.~\cite{Kling:2025udr}, \textit{cf.} also Eqs.~\eqref{eq:vmd}--\eqref{eq:bw_form}.
We choose to adopt Model 1 from Ref.~\cite{Kling:2025udr} in which the masses and widths of the highest resonances $\omega'', \rho'', \phi''$ are varied around their PDG values, while the parameters of the lower resonances $\omega, \omega', \rho, \rho', \phi, \phi'$ remain fixed to their PDG values.
Table~\ref{tab:timelikeFF_coeff} shows the parameter values entering our calculations.

\begin{table}
\setlength{\tabcolsep}{12pt}
\centering
\begin{tabular}{lcccccc}
\toprule
\toprule
Quantity\: & $a_1$ & $b_1$ & $b_2$ & $b_3$ & $A$ & $B$  \\ \midrule
$G_\text{\tiny{E}}^p$ & $-0.24$ & $10.98$ & $12.82$ & $21.97$ & - & - \\
$G_\text{\tiny{M}}^p/\mu_p$ & $0.12$ & $10.97$ & $18.86$ & $6.55$ & - & - \\
$G_\text{\tiny{M}}^n/\mu_n$ & $2.33$ & $14.72$ & $24.2$ & $84.1$ & - & - \\
$G_\text{\tiny{E}}^n$ & - & - & - & - & 1.7 & 3.3  \\
\bottomrule
\bottomrule
\end{tabular}
\caption{Parameter values used in the Kelly parametrization of the Sachs form factors (in the space-like region). Taken from Ref.~\cite{Kelly:2004hm}.}
\label{tab:Kellycoefficients}
\end{table}

\begin{table}
\setlength{\tabcolsep}{12pt}
\centering
\begin{tabular}{lcc}
\toprule
\toprule
Parameter & Value & Reference \\ \midrule
$m_{\omega_0}$ & $0.78266$ & \multirow{13}{*}{PDG, Ref.~\cite{ParticleDataGroup:2024cfk}} \\ %
$\varGamma_{\omega_0}$ & $0.00868$ &  \\
$m_{\rho_0}$ & $0.77526$ &  \\
$\varGamma_{\rho_0}$ & $0.1474$ &  \\
$m_{\phi_0}$ & $1.01946$ &  \\
$\varGamma_{\phi_0}$ & $0.004249$ &  \\
$m_{\omega_1}$ & $1.410$ &  \\
$\varGamma_{\omega_1}$ & $0.29$ &  \\
$m_{\rho_1}$ & $1.465$ &  \\
$\varGamma_{\rho_1}$ & $0.4$ &  \\
$m_{\phi_1}$ & $1.68$ &  \\
$\varGamma_{\phi_1}$ & $0.15$ &  \\
\midrule 
$m_{\omega_2}$ & $1.6106$ &  \multirow{9}{*}{Ref.~\cite{Kling:2025udr} (Model 1)} \\
$\varGamma_{\omega_2}$ & $0.2638$ &  \\
$m_{\rho_2}$ & $1.5782$ & \\
$\varGamma_{\rho_2}$ & $0.3538$ &  \\
$m_{\phi_2}$ & $2.1752$ &  \\
$\varGamma_{\phi_2}$ & $0.2384$ &  \\
$a_{1,\omega_0}$ & $0.7829$ &  \\
$a_{1,\rho_0}$ & $0.4856$ &  \\
$a_{1,\phi_0}$ & $-0.0157$ & \\
\bottomrule
\bottomrule
\end{tabular}
\caption{Parameter values used for the evaluation of the nucleon form factors in the time-like region.}
\label{tab:timelikeFF_coeff}
\end{table}

\bibliographystyle{JHEP}
\newpage 
\bibliography{bibliography}

\end{document}